\documentclass[preprint,preprintnumbers,amsmath,amssymb]{revtex4-1}

\usepackage{color}
\usepackage[]{graphicx}
\usepackage{latexsym}
\usepackage{natbib}
\usepackage{amsmath}
\usepackage[caption=false]{subfig}
\usepackage{multirow}

\usepackage[final]{feynmp}
\newcommand{\dd}{\mathrm{d}}

\newcommand{\s}{\sigma}

\newcommand{\la}{\langle}
\newcommand{\ra}{\rangle}

\newcommand{\Mia}{\hat{\mathcal{M}}}
\newcommand{\Bia}{\hat{\mathcal{B}}}

\newcommand{\be}{\begin{equation}}
\newcommand{\ee}{\end{equation}}
\newcommand{\bc}{\begin{cases}}
\newcommand{\ec}{\end{cases}}

\begin{document}

\title{Influence maximization in complex networks through optimal percolation}

\author{Flaviano Morone}
\author{Hern\'an A. Makse}

\affiliation{Levich Institute and Physics Department, City College of
  New York, New York, NY 10031}

\begin{abstract}

\bf{The whole frame of interconnections in complex networks hinges on
  a specific set of structural nodes, much smaller than the total
  size, which, if activated, would cause the spread of information to
  the whole network \cite{richardson1}; or, if immunized, would
  prevent the diffusion of a large scale epidemic
  \cite{pastor,newman-perco}.  Localizing this optimal, i.e. minimal,
  set of structural nodes, called influencers, is one of the most
  important problems in network science
  \cite{kempe,newman-book}. Despite the vast use of heuristic
  strategies to identify influential spreaders
  \cite{freeman,pagerank,kleinberg,barab,cohen,chen,gallos,
    zecchina1,zecchina2}, the problem remains unsolved.  Here, we map
  the problem onto optimal percolation in random networks to identify
  the minimal set of influencers, which arise by minimizing the energy
  of a many-body system, where the form of the interactions is fixed
  by the non-backtracking matrix \cite{hashimoto} of the network.  Big
  data analyses reveal that the set of optimal influencers is much
  smaller than the one predicted by previous heuristic
  centralities. Remarkably, a large number of previously neglected
  weakly-connected nodes emerges among the optimal influencers. These
  are topologically tagged as low-degree nodes surrounded by
  hierarchical coronas of hubs, and are uncovered only through the
  optimal collective interplay of all the influencers in the network.
  Eventually, the present theoretical framework may hold a larger
  degree of universality, being applicable to other hard optimization
  problems exhibiting a continuous transition from a known phase
  \cite{Coja}.  }

\end{abstract}

\maketitle

The optimal influence problem was initially introduced in the context
of viral marketing \cite{richardson1}, and its solution was shown to
be NP-hard \cite{kempe} for a generic class of linear threshold models
of information spreading \cite{LTM,watts}. Indeed, finding the optimal
set of influencers is a many-body problem in which the topological
interactions between them play a crucial role
\cite{zecchina1,zecchina2}.  On the other hand, there has been an
abundant production of heuristic rankings to identify influential
nodes and "superspreaders" in networks
\cite{freeman,pagerank,kleinberg,barab,cohen,chen,gallos,pei}.  The
main problem is that heuristic methods do not optimize a global
function of influence. As a consequence, there is no guarantee of
their performance.

Here we address the problem of quantifying node's influence by finding
the optimal (i.e. minimal) set of structural influencers.  After
defining a unified mathematical framework for both immunization and
spreading,
we provide its optimal solution in random networks by mapping the
problem onto optimal percolation.  In addition, we present CI (which
stands for Collective Influence), a scalable algorithm to solve the
optimization problem in large scale datasets. The thorough comparison
with competing methods (Methods Section \ref{heuristics} \cite{pei2})
ultimately establishes the major performance of our algorithm.  By
taking into account collective influence effects, our optimization
theory identifies a new class of strategic influencers, called
weak-nodes, which outrank the hubs in the network.  Thus, the top
influencers are highly counterintuitive: low degree nodes play a major
broker role in the network, and despite being weakly connected, can be
powerful influencers.

The problem of finding the minimal set of activated nodes
\cite{LTM,watts} to spread information to the whole network
\cite{kempe} or to optimally immunize a network against epidemics
\cite{chen} can be exactly mapped onto optimal percolation (see
Methods Section \ref{mapping}).
This mapping provides the mathematical support to the intuitive
relation between influence and the concept of cohesion of a network:
the most influential nodes are the ones forming the minimal set that
guarantees a global connection of the network
\cite{barab,cohen,newman-book}.  We call this minimal set the
``optimal influencers'' of the network. At a general level, the
optimal influence problem can be stated as follows: find the minimal
set of nodes which, if removed, would break down the network in many
disconnected pieces. The natural measure of influence is, therefore,
the size of the largest (giant) connected component as the influencers
are removed from the network.

We consider a network composed of $N$ nodes tied with $M$ links with
arbitrary degree distribution $P(k)$. Let us suppose we remove a
certain fraction $q$ of the total number of nodes. It is well known
from percolation theory \cite{bollobas} that, if we choose these nodes
randomly, the network undergoes a structural collapse at a certain
critical fraction where the probability of existence
of the giant connected component vanishes, $G=0$. The optimal
influence problem corresponds to finding the minimum fraction $q_c$ of
influencers to fragment the network: $q_c\ =\ \min
\{q\in[0,1]:\ G(q)\ = 0\}$.

\begin{figure}[h!]
\includegraphics[width=1\textwidth]{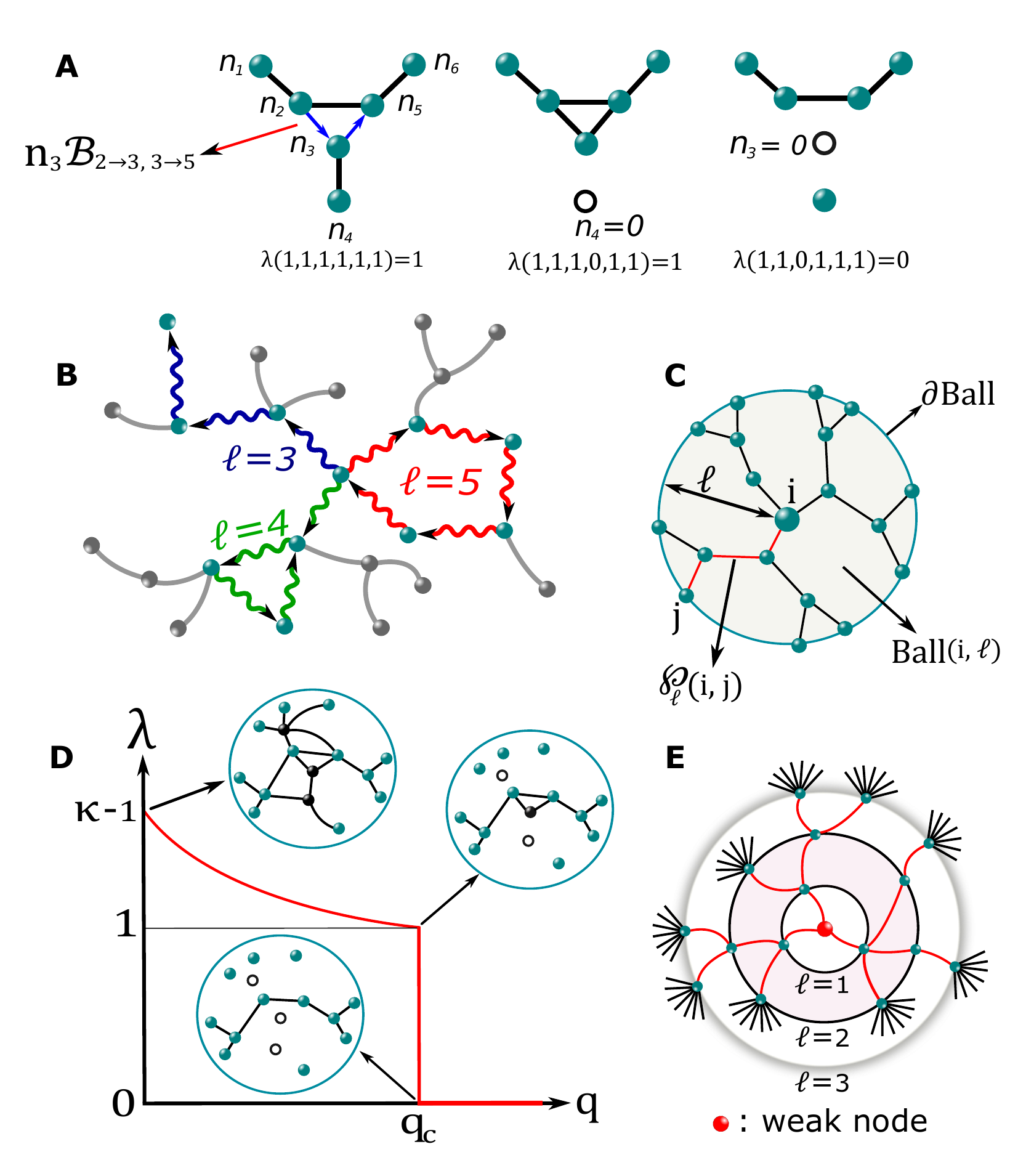} 
\caption{}
\label{fig:explanation}
\end{figure}

Let the vector $\mathbf{n}=(n_1,\dots,n_N)$ represents which node is
removed ($n_i=0$, influencer) or left $(n_i=1$, the rest) in the
network ($q=1-1/N\sum_i n_i$), and consider a link from $i \to j$.
The order parameter of the influence problem is the probability that
$i$ belongs to the giant component in a modified network where $j$ is
absent, $\nu_{i\to j}$ \cite{bianconi,lenka}.
Clearly, in the absence of a giant component we find
$\{\nu_{i\to j}=0\}$ for all $i\to j$.
The stability of the solution $\{\nu_{i\to j} = 0\}$ is controlled by
the largest eigenvalue $\lambda(\mathbf{n}; q)$ of the linear operator
$\Mia$ defined on the $2M \times 2M$ directed edges as
$\mathcal{M}_{k\to \ell, i\to j} \equiv \frac{\partial \nu_{i\to
    j}}{\partial \nu_{k\to\ell}}\Big|_{\{ \nu_{i\to j} = 0 \}}$. We
find for locally-tree like random graphs (see
Fig. \ref{fig:explanation}a and Methods Section \ref{theory}): 

\be
\mathcal{M}_{k\to \ell, i\to j} = n_i \ \mathcal{B}_{k\to \ell, i\to
  j}
\ee 
where $\mathcal{B}_{k\to \ell, i\to j}$ is the
\textit{non-backtracking} matrix of the network \cite{hashimoto,nbt}.
The matrix $\mathcal{B}_{k\to \ell, i\to j}$ has non-zero entries only
when $(k\to \ell, i\to j)$ form a pair of consecutive non-backtracking
directed edges, i.e. $(k\to \ell, \ell\to j)$ with $k\neq j$. In this
case $\mathcal{B}_{k\to \ell, \ell\to j}=1$
(Eq. S\ref{definition}). Powers of the matrix $\Bia$ count the number
of non-backtracking walks of a given length in the network
(Fig. \ref{fig:explanation}b) \cite{nbt}, much in the same way as
powers of the adjacency matrix count the number of paths
\cite{newman-book}.  Operator $\Bia$ has recently received a lot of
attention thanks to its high performance in the problem of community
detection \cite{florant,newman}.
Below, we show its
topological power in the problem of optimal percolation.

Stability of the solution $\{\nu_{i\to j}=0\}$ requires
$\lambda(\mathbf{n}; q)\leq 1$. The optimal
influence problem for a given $q$ ($\ge q_c)$ can be rephrased as
finding the optimal configuration $\mathbf{n}$ that minimizes the
largest eigenvalue $\lambda(\mathbf{n}; q)$
(Fig. \ref{fig:explanation}c).  The optimal set $\mathbf{n}^*$ of
$Nq_c$ influencers is obtained when the minimum of the largest
eigenvalue reaches the critical threshold:
\begin{equation} 
\lambda(\mathbf{n}^*; q_c) = 1.
\label{one} 
\end{equation}
The formal mathematical mapping of the optimal influence problem to
the minimization of the largest eigenvalue of the modified
non-backtracking matrix for random networks,
Eq. (\ref{one}), represents our first main result.

An example of non-optimized solution corresponds to choosing $n_i$ at
random and decoupled from the non-backtracking matrix
\cite{lenka,radicchi} (random percolation \cite{bollobas}, Methods
Section \ref{zero}). In the optimized case we seek to derandomize the
selection of the set $n_i=0$ and optimally choose them to find the
best configuration $\mathbf{n}^*$ with the lowest $q_c$ according to
Eq. (\ref{one}). The eigenvalue $\lambda(\mathbf{n})$ (from now on we
omit $q$ in $\lambda(\mathbf{n}; q)\equiv\lambda(\mathbf{n})$,
which is always kept fixed) determines the growth rate of an arbitrary
vector $\mathbf{w}_0$ with $2M$ entries after $\ell$ iterations of the
matrix $\Mia$: $ |\mathbf{w}_\ell(\mathbf{n})|\ =\ \langle
\mathbf{w}_\ell | \mathbf{w}_\ell \rangle^{1/2} = |\Mia^{\ell}
\mathbf{w}_0|\ =
\ \la\mathbf{w}_0|(\Mia^{\ell})^{\dagger}\Mia^{\ell}|\mathbf{w}_0\ra^{1/2}\sim\ e^{\ell\log\lambda(\mathbf{n})}$. The
  largest eigenvalue is then calculated by the Power Method:

\be \lambda(\mathbf{n})\ =\ \lim_{\ell\to\infty}\ \left[
  \frac{|\mathbf{w}_\ell(\mathbf{n})|}{|\mathbf{w}_0|} \right]^{1/\ell}\ .
\label{eq:growth rate}
\ee 

Equation \eqref{eq:growth rate} is the starting point of an (infinite)
perturbation series which provides the exact solution to the many-body
influence problem in random networks and therefore contains all
physical effects, including the collective influence.
In practice, we minimize the cost energy function of influence $
|\mathbf{w}_\ell(\mathbf{n})|$ in Eq. (\ref{eq:growth rate}) for a
finite $\ell$. The solution rapidly converges to the exact value at
$\ell\to\infty$; the faster the larger the spectral gap. We find for
$\ell\geq 1$ as a leading order in $1/N$ (Methods Section
\ref{mainformula}):

\be |\mathbf{w}_\ell(\mathbf{n})|^2 = \sum_{i=1}^{N} (k_i-1)
\sum_{j\in\partial\mathrm{Ball}(i,2\ell-1)}\left(\prod_{k\in\mathcal{P}_{2\ell-1}(i,j)}n_k\right)
(k_j-1)\ ,
\label{eq:Costfunction}
\ee where $\mathrm{Ball}(i,\ell)$ is the set of nodes inside a ball of
radius $\ell$ (defined as the shortest path) around node $i$,
$\partial\mathrm{Ball}(i,\ell)$ is the frontier of the ball and
$\mathcal{P}_{\ell}(i,j)$ is the shortest path of length $\ell$
connecting $i$ and $j$ (Fig. \ref{fig:explanation}d).

The first collective optimization in Eq. \eqref{eq:Costfunction} is
$\ell=1$.  We find $ |\mathbf{w}_1(\mathbf{n})|^2\ =
\ \sum_{i,j=1}^{N} A_{ij}(k_i-1)(k_j-1) n_in_j$
(Eq. S\ref{norm1}). This term is interpreted as the energy of
an antiferromagnetic Ising model with random bonds in a random
external field at fixed magnetization, which is an example of a
pair-wise NP-complete spin-glass whose solution is found in Methods
Section \ref{cavity} with the cavity method \cite{MP}.

For $\ell\ge 2$, the problem can be mapped exactly to a statistical
mechanical system with many-body interactions which can be recast in
terms of a diagrammatic expansion, Eqs.  S\ref{d1}-S\ref{d3}. For
example, $|\mathbf{w}_2(\mathbf{n})|^2$ leads to $4$-body interactions
(Eq. S\ref{4body}), and, in general, the energy cost
$|\mathbf{w}_{\ell}(\mathbf{n})|^2$ contains $2\ell$-body
interactions.
 As soon as $\ell \ge 2$, the cavity method becomes much more
 complicated to implement and we use another suitable method, called
 extremal optimization (EO) \cite{EO} (Methods Section
 \ref{sec:EO}). This method estimates the true optimal value of the
 threshold by finite size scaling following extrapolation to
 $\ell\to\infty$ (Extended Data Fig. \ref{fig:tau_pc}).  However, EO
 is not scalable to find the optimal configuration in large
 networks. Therefore, we develop an adaptive method, which performs
 excellently in practice, preserves the features of EO, and is highly
 scalable to present-day big-data.

The idea is to remove the nodes causing the biggest drop in the energy
function Eq. \eqref{eq:Costfunction}.
First, we define a ball of radius $\ell$ around every node
(Fig. \ref{fig:explanation}d). Then, we consider the nodes belonging
to the frontier $\partial\mathrm{Ball}(i,\ell)$ and assign to node $i$
the collective influence (CI) strength at level $\ell$ following
Eq. (\ref{eq:Costfunction}) (see Methods Section \ref{CI} for
implementation and Section \ref{neq} for minimizing $G(q)\neq 0$): \be
\mathrm{CI}_\ell(i)\ =\ (k_i-1)\sum_{j\in\partial\mathrm{Ball}(i,\ell)}(k_j-1)\ .
\label{eq:CI}
\ee We notice that, while Eq. \eqref{eq:Costfunction} is valid only
for odd radii of the ball, $\mathrm{CI}_\ell(i)$ is defined also for
even radii. This generalization is possible by considering an energy
function for even radii analogous to Eq. \eqref{eq:Costfunction}, as
explained in Methods Section \ref{odd}. The case of one-body
interaction with zero radius $\ell=0$ (Eq. S\ref{highdegree}) leads to
the high-degree (HD) ranking (Eq. S\ref{eq:HDaveraged}) \cite{cohen}.

The collective influence Eq. (\ref{eq:CI}) is our second and most
important result since it is the basis for the highly scalable and
optimized CI-algorithm which follows. In the beginning all the nodes
are present: $n_i=1$ for all $i$. Then, we remove node $i^*$ with
highest $\mathrm{CI}_\ell$ and set $n_{i^*}=0$. The degrees of its
neighbours are decreased by one, and the procedure is repeated to find
the new top CI node to remove. The algorithm is terminated when
the giant component is zero.  By increasing the radius $\ell$ of the
ball we obtain better and better approximations of the optimal exact
solution at $\ell\to \infty$ (for finite networks, $\ell$ does not
exceed the network diameter).

The collective influence $\mathrm{CI}_\ell$ for $\ell\ge 1$ has a rich
topological content, and consequently can tell us more about the role
played by nodes in the network than the non-interacting high-degree
hub removal strategy at $\ell=0$, CI$_0$. The augmented information
comes from the sum in the r.h.s, which is absent in the naive
high-degree rank. This sum contains the contribution of the nodes
living on the surface of the ball surrounding the central vertex $i$,
each node weighted by the factor $k_j-1$.  This means that a node
placed at the center of a corona irradiating many links--- the
structure hierarchically emerging at different $\ell$-levels as seen
in Fig. \ref{fig:explanation}e--- can have a very large collective
influence, even if it has a moderate or low degree.
Such ``weak-nodes'' can outrank nodes with larger degree that occupy
mediocre peripherical locations in the network.
The commonly used word 'weak' in this context sounds particularly
paradoxical. It is, indeed, usually used as a synonymous for a
low-degree node with an additional "bridging" property, which has
resisted a quantitative formulation. We provide this definition
through Eq. \eqref{eq:CI}, according to which weak nodes are, de
facto, quite strong. Paraphrasing Granovetter's conundrum
\cite{granovetter}, Eq. \eqref{eq:CI} quantifies the ``strength of
weak nodes''.

The CI-algorithm scales as $\sim O(N \log N)$ by removing a finite
fraction of nodes at each step (Methods Section \ref{logn}).  This high
scalability allows us to find top influencers
in current big-data social media and optimal immunizators in
large-scale populations at the country level. The applications are
investigated next.

\begin{figure}[h]
\includegraphics[width=1\textwidth]{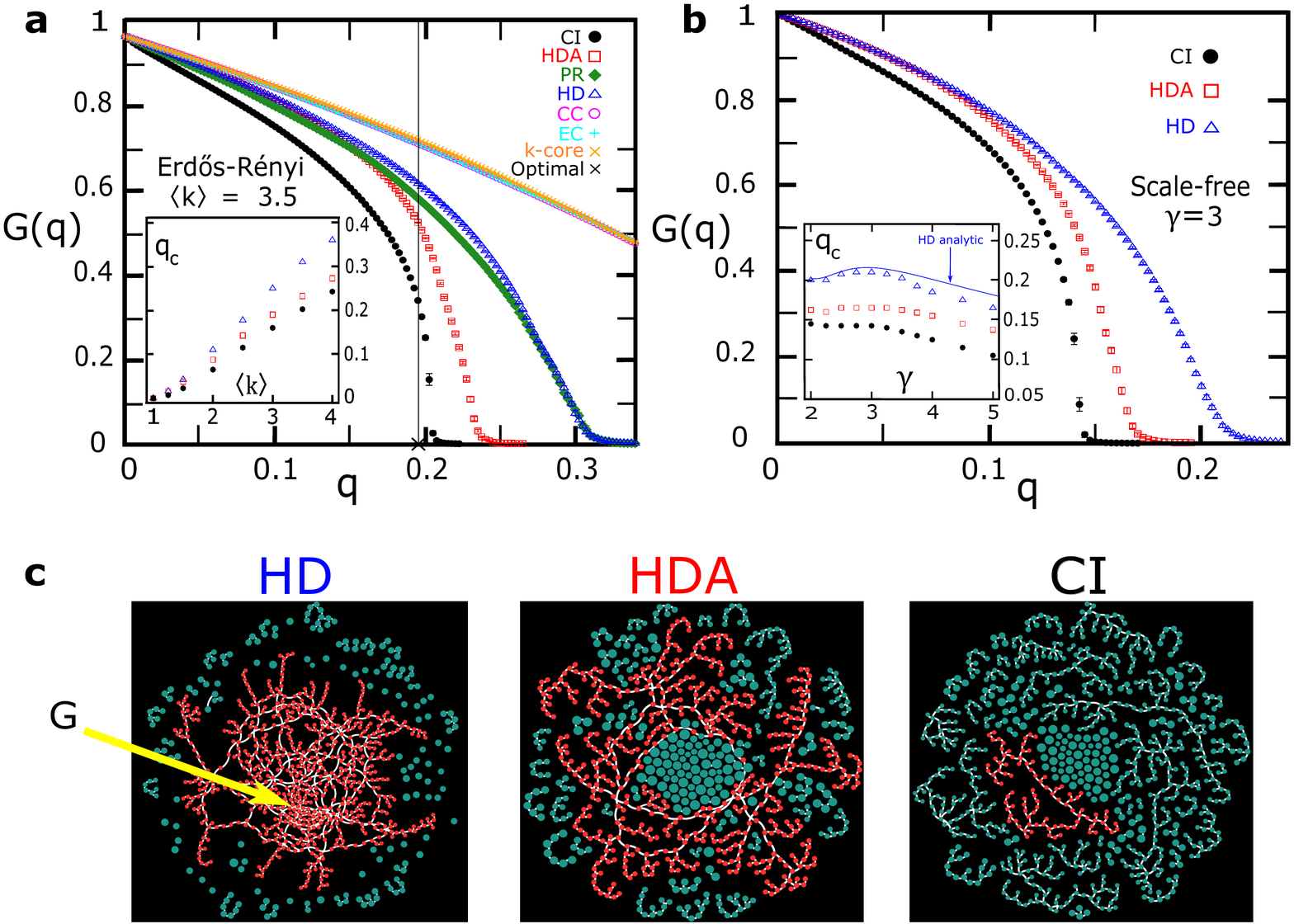} 
\caption{}
\label{fig:synthetic}
\end{figure}

Figure \ref{fig:synthetic}a shows the optimal threshold $q_c$ for random
Erd\"{o}s-R\'enyi (ER) network \cite{newman-book} (marked by the
vertical line) obtained by extrapolating the EO solution to infinite
size, $N\to\infty$, and $\ell \to \infty$ (Methods Section
\ref{sec:EO}).
In the same figure we compare the optimal threshold against the
heuristic centrality measures: high-degree (HD) \cite{barab},
high-degree adaptive (HDA), PageRank (PR) \cite{pagerank}, closeness
centrality (CC) \cite{freeman}, eigenvector centrality (EC)
\cite{freeman}, and k-core \cite{gallos} (see Methods Section
\ref{heuristics} for definitions).  Methods Section \ref{comparison}
and \ref{BP} show the comparison with the remaining heuristics
\cite{freeman,chen} and the Belief Propagation method of
\cite{zecchina2}, respectively, which have worst computational
complexity (and optimality), and cannot be applied to the network
sizes used here. Remarkably, at the optimal value $q_c$ predicted by
our theory, the best among the heuristic methods (HDA, PR and HD)
still predict a giant component $\sim 50\% - 60\%$ of the whole
original network. Furthermore, the influencer threshold predicted by
CI approximates very well the optimal one, and, notably, CI
outperforms the other strategies. Figure \ref{fig:synthetic}b compares
CI in scale-free (SF) network \cite{newman-book} against the best
heuristic methods, i.e. HDA and HD.
In all cases, CI produces smaller threshold and smaller giant
component (Fig. \ref{fig:synthetic}c).

\begin{figure}[h]
\includegraphics[width=1\textwidth]{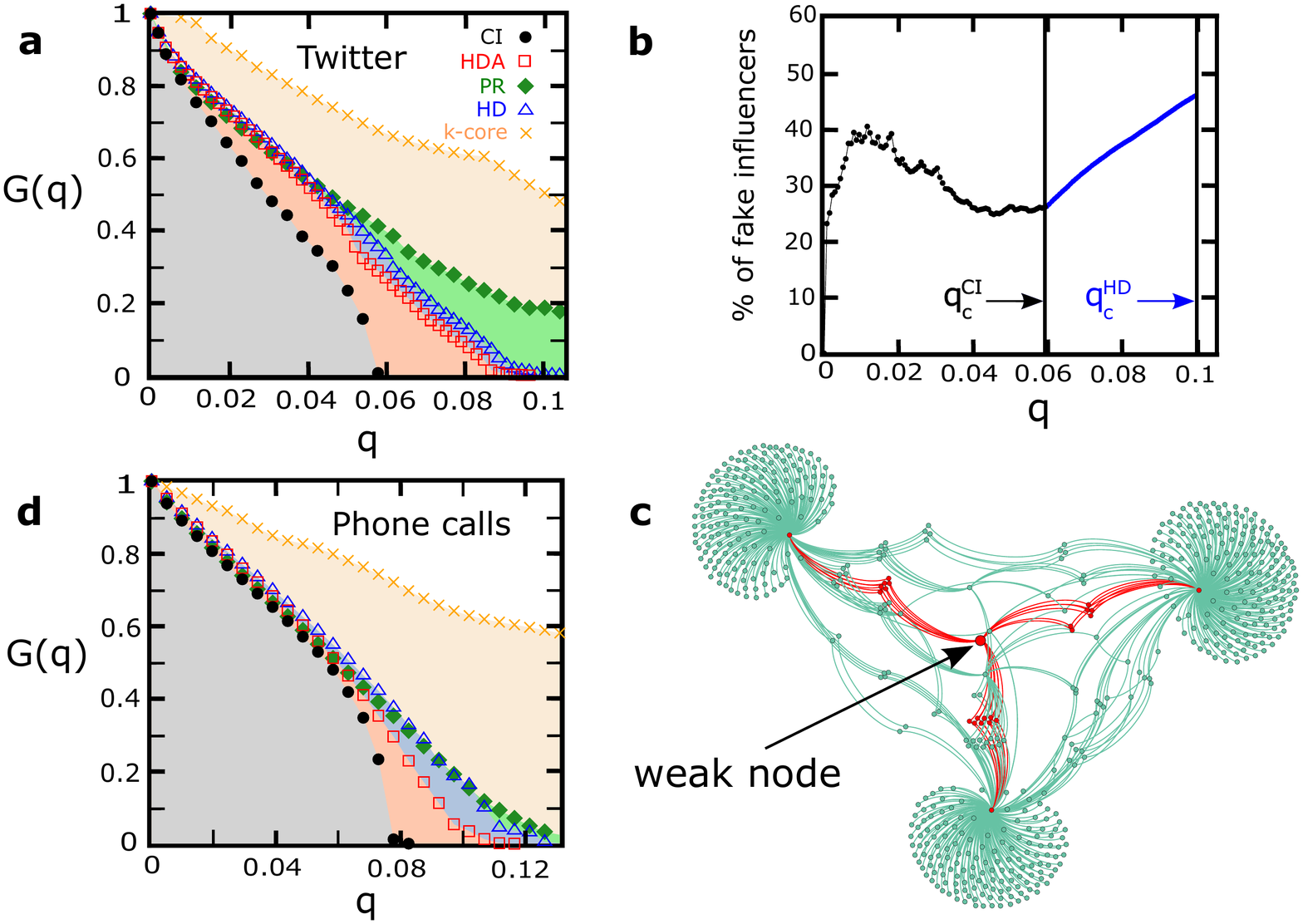} 
\caption{}
\label{fig:mobileMex} 
\end{figure}

As an example of information spreading network, we consider the web of
Twitter users (Methods Section \ref{twitter} \cite{pei}).  Figure
\ref{fig:mobileMex}a shows the giant component of Twitter when a
fraction $q$ of its influencers is removed following CI.  It is
surprising that a lot of Twitter users with a large number of contacts
have a mild influence on the network, as witnessed by the fact that,
when CI (at $\ell=5$) predicts a zero giant component (and so it
exhausts the number of optimal influencers), the scalable heuristic ranks
(HD, HDA, PR and k-core) still give a pretty big giant component of
the order of 30-70\% of the entire network, and, inevitably, find a
remarkably larger number of (fake) influencers which is 50\% larger
than that predicted by CI (Fig.  \ref{fig:mobileMex}b and Methods
Section \ref{twitter}). One cause for the poor performance of the
high-degree rank is that most of the hubs are clustered (rich-club
effect), which gives a mediocre importance to their contacts.  As a
consequence, hubs are outranked by nodes with lower degree surrounded
by coronas of hubs (shown in the detail of Fig. \ref{fig:mobileMex}c),
i.e. the weak-nodes predicted by the theory
(Fig. \ref{fig:explanation}e).

Finally, we simulate an immunization scheme on a personal contact
network built on the phone calls performed by $14$ million people in
Mexico (Methods Section \ref{mobile}).
Figure \ref{fig:mobileMex}d shows that our method saves a large amount
vaccines stockpile or, equivalently, find the smallest possible set of
people to quarantine outranking the scalable heuristics in large real
networks as well.  Thus, while the mapping of the influencer
identification problem onto optimal percolation is strictly valid for
locally tree-like random networks, our results may apply also for real
loopy networks, provided the density of loops is not excessively
large.

Our solution to the optimal influence problem shows its importance in
that it helps to unveil hitherto hidden relations between people, as
witnessed by the weak-node effect. This, in turn, is the byproduct of
a broader notion of influence, lifted from the individual
non-interacting point of view
\cite{freeman,pagerank,kleinberg,barab,cohen,chen,gallos,pei,pei2} to
the collective sphere: influence is an emergent property of
collectivity and top influencers arise from the optimization of the
complex interactions they stipulate.

{\bf Acknowledgments} 

This work was funded by NIH-NIGMS 1R21GM107641 and NSF-PoLS
PHY-1305476.  Additional support was provided by ARL.  We thank L. Bo,
S. Havlin and R. Mari for discussions and Grandata for providing the
data on mobile phone calls.

{\bf Author contributions}

Both authors contributed equally to the work presented in this paper.

{\bf Additional information}

The authors declare no competing financial interests. Supplementary
information accompanies this paper on www.nature.com/nature.
Correspondence and requests for materials should be addressed to
H.A.M. (hmakse@lev.ccny.cuny.edu).

\newpage

{\bf FIG. \ref{fig:explanation}.  Non-backtracking (NB) matrix and
  weak-nodes}. \textbf{a,} The largest eigenvalue $\lambda$ of $\Mia$
exemplified on a simple network. The optimal strategy for immunization
and spreading minimizes $\lambda$ by removing the minimum number of
nodes (optimal influencers) that destroys all the loops.  Left
panel: The action of the matrix $\Mia$ is on the directed edges of the
network. The entry
$\mathcal{M}_{2\to3,3\to5}=n_3\mathcal{B}_{2\to3,3\to5}=n_3$ encodes
node $3$'s occupancy ($n_3=1$) or vacancy ($n_3=0$). In this
particular case, the largest eigenvalue is $\lambda=1$.  Center panel:
Not-optimal removal of a leaf, $n_4=0$, which does not decrease
$\lambda$.  Right panel: Optimal removal of a loop, $n_3=0$, which
decreases $\lambda$ to zero.
\textbf{ b,} A NB walk is a random walk that is not allowed to return
back along the edge that it just traversed. We show a NB open walk
($\ell=3$), a NB closed walk with a tail ($\ell=4$), and a NB closed
walk with no tails ($\ell=5$). The NB walks are the building blocks of
the diagrammatic expansion to calculate $\lambda$.
\textbf{c,} Representation of the global minimum over $\mathbf{n}$ of
  the largest eigenvalue $\lambda$ of $\Mia$ vs $q$. When $q\ge q_c$,
  the minimum is at $\lambda=0$. Then, $G=0$ is stable (still,
  non-optimal configurations exist with $\lambda>1$ for which
  $G>0$). When $q< q_c$, the minimum of the largest eigenvalue is
  always $\lambda>1$, the solution $G=0$ is unstable, and then
  $G>0$. At the optimal percolation transition, the minimum is at
  $\mathbf{n}^*$ with $\lambda(\mathbf{n}^*,q_c) =1$. For $q=0$, we
  find $\lambda=\kappa-1$ ($\kappa=\la k^2\ra/\la k\ra$) which is the
  largest eigenvalue of $\Bia$ for random networks \cite{florant} with
  all nodes present $(n_i=1)$.
When $\lambda=1$, the giant component is reduced to a tree plus one
single loop (unicyclic graph), which is suddenly destroyed at the
transition $q_c$ to become a tree, causing the abrupt fall of
$\lambda$ to zero.  
 \textbf{d,}
  $\mathrm{Ball}(i,\ell)$ of radius $\ell$ around node $i$ is the set
  of nodes contained in the grey region and $\partial\mathrm{Ball}$ is
  the set of nodes on the boundary. The shortest path from $i$ to $j$
  is colored in red. 
 \textbf{e,} Example of a weak-node: a node with a small number of
  connections surrounded by hierarchical coronas of hubs at different
  $\ell$-levels.

{\bf FIG. \ref{fig:synthetic}.  Exact optimal solution and performance
  of CI in synthetic networks.} \textbf{a,} $G(q)$ in ER network
($N=2\times10^5$, $\la k\ra=3.5$, error bars are s.e.m. over $20$
realizations) for the true optimal solution with EO ($\times$), CI,
HDA, PR, HD, CC, EC and k-core.  The other methods are not scalable
and perform worst than HDA and are treated in Methods Sections
\ref{comparison} and \ref{BP}.  CI is close to the optimal
$q_c^{\mathrm{opt}}\sim 0.193$ obtained with EO in Methods Section
\ref{sec:EO}. Note that EO can estimate the extrapolated optimal value
of $q_c$, but it cannot provide the optimal configuration for large
systems.
Inset: $q_c$ (obtained at the peak of the
second largest cluster) for the three best methods vs $\la k\ra$.
 \textbf{b,} $G(q)$ for SF network with degree exponent $\gamma=3$,
maximum degree $k_{\mathrm{max}}=10^3$ and $N=2\times10^5$ 
(error bars are s.e.m. over $20$ realizations).
  Inset:
$q_c$ vs $\gamma$.  The continuous blue line is the HD analytical
result computed in Methods Section \ref{odd}.
 \textbf{c,} SF network
  with $\gamma=3$ after the removal of the $15\%$ of nodes, using the
  three methods.  CI produces a much reduced giant component (red
  nodes).

{\bf FIG. \ref{fig:mobileMex}.  Performance of CI in
large-scale real social networks.} \textbf{a,} Giant component $G(q)$
of Twitter users \cite{pei} ($N=469,013$) computed using CI, HDA, PR,
PR and k-core strategies (other heuristics have prohibitive running
times for this system size).
\textbf{b,} Percentage of fake influencers or false
    positives (PFI, Eq. S\ref{pfi}) in Twitter as a function of $q$,
  defined as the percentage of non-optimal influencers identified by
  HD algorithm in comparison with CI.  Below $q_c^{\mathrm{CI}}$,
  $\mathrm{PFI}$ reaches as much as $\sim40\%$ indicating the failure
  of HD in optimally finding the top influencers.  Indeed, to obtain
  $G=0$, HD has to remove a much larger number of fake influencers,
  which at $q_c^{\mathrm{HD}}$ reaches $\mathrm{PFI}\sim48\%$.
\textbf{c,} Example out of the many weak-nodes found in Twitter; the
crucial influencer missed by all heuristic strategies.
\textbf{d,} $G(q)$ for a social network of $1.4\times10^7$ mobile
phone users in Mexico
representing an example of big-data to test the scalability and
performance of the algorithm. CI immunizes this social network using
half a million less people than the best heuristic strategy (HDA),
saving $\sim$35\% of vaccine stockpile.

\clearpage 

{\bf Extended Data Fig. \ref{fig:pd}. HD high-degree threshold.} {\bf
  a,} HD influence threshold $q_c$ as a function of the degree
distribution exponent $\gamma$ of scale-free networks in the ensemble
with $k_{\rm max} = m N^{1/(\gamma-1)}$ and $N\to \infty$.  The curves
refer to different values of the minimum degree $m$: $1$ (red), $2$
(blue), $3$ (black). The fragility of SF networks (small $q_c$) is
notable for $m=1$, the case calculated in \cite{cohen}. In this case,
the network contains many leaves, and reduces to a star at $\gamma=2$
which is trivially destroyed by removing the only single hub,
explaining the general fragility in this case. Furthermore, in this
case, the network becomes a collection of dimers with $k=1$ when
$\gamma\to\infty$, which is still trivially fragile. This explains why
$q_c\to 0$ as $\gamma\to\infty$, as well.  Therefore, the fragility in
the case $m=1$ has its roots in these two limiting trivial
cases. Removing the leaves ($m=2$) results in a 2-core, which is
already more robust. For the 3-core $m=3$, $q_c\approx 0.4-0.5$
provides a quite robust network, and has the expected asymptotic limit
to a non-zero $q_c$ of a random regular graph with $k=3$ as
$\gamma\to\infty$, $q_c\to(k-2)/(k-1)=0.5$. Thus, SF networks become
robust in these more realistic cases and the search for other attack
strategies becomes even more important.
{\bf b,} HD influence threshold $q_c$ as a function of the degree
distribution exponent of scale-free networks with minimum degree $m=2$
in the ensemble where $k_{\rm max}$ is fixed and does not scale with
$N$.  The curves refer to different values of the cut-off
$k_{\mathrm{max}}$: $10^2$ (red), $10^3$ (green), $10^5$ (blue),
$10^8$ (magenta), and $k_{\mathrm{max}}=\infty$ (black), and show that
for typical $k_{\rm max}$ degree of $10^3$, for instance in social
networks, the network is fairly robust with $q_c\approx 0.2$ for all
$\gamma$.

{\bf  Extended Data Fig. \ref{fig:lambdaRS}.  RS estimation of the
   maximum eigenvalue.}  Main panel: the eigenvalue $\lambda_1^{\rm
   RS}(q)$ obtained by minimizing the energy function
 $\mathcal{E}(\mathbf{s})$ with the RS cavity method.  The curve was
 computed on a Erd\"{o}s-R\'enyi graph of $N=10,000$ nodes and average
 degree $\la k\ra=3.5$ and then averaged over 40 realizations of the
 network.  Inset: Comparison between the cavity method and extremal
 optimization for an ER graph of $\la k\ra=3.5$ and $N=128$. The
 curves are averaged over $200$ realizations (error bars are s.e.m.).

{\bf Extended Data Fig. \ref{fig:taufull}. EO estimation of the
  maximum eigenvalue.}  Eigenvalue $\lambda(q)$ obtained by minimizing
the energy function $\mathcal{E}(\mathbf{n})$ with $\tau$EO, plotted
as a function of the fraction of removed nodes $q$. The panels are for
different orders of the interactions. The curves in each panel refer
to different sizes of Erd\"{o}s-R\'enyi networks with average
connectivity $\la k\ra = 3.5$.  Each curve is an average over 200
instances (error bars are s.e.m.). The value $q_c$ where
$\lambda(q_c)=1$ is the threshold for a particular $N$ and many-body
interaction.

{\bf Extended Data Fig. \ref{fig:tau_pc}. Estimation of optimal
  threshold $q_c^{\rm opt}$ with EO.} {\bf a,} Critical threshold
$q_c$ as a function of the system size $N$ obtained with EO from from
Extended Data Fig. \ref{fig:taufull} of ER networks with $\la k\ra =
3.5$ and varying size. The curves refer to different orders of the
many-body interactions.  The data show a linear behaviour as a
function of $N^{-2/3}$, typical of spin glasses, for each many-body
interaction $\rho$. The extrapolated value $q_c^{\infty}(\rho)$ is
obtained at the $y$-intercept.
{\bf b,} Thermodynamical critical threshold $q_c^\infty(\rho)$ as a
function of the order of the interactions $\rho$ from {\bf a}. The
data scale linearly with $1/\rho$.  From the $y$-intercept of the linear
fit we obtain the thermodynamical limit of the infinite-body optimal
value $q_c^{\mathrm{opt}}= q_c^\infty(\rho\to\infty) = 0.192(9)$.

{\bf   Extended Data \ref{fig:allLevels}.  Comparison of the CI
    algorithm for different radius $\ell$ of the
    $\mathrm{Ball}(\ell)$}. We use $\ell = 1, 2, 3, 4, 5$, on a ER graph
  with average degree $\la k\ra = 3.5$ and $N=10^5$ (the average is
  taken over 20 realizations of the network, error bars are s.e.m.).
  For $\ell=3$ the performance is already practically
  indistinguishable from $\ell=4,5$.  The stability analysis we
  developed to minimize $q_c$ is strictly valid only when $G=0$, since
  the largest eigenvalue of the modified NB matrix controls the
  stability of the solution $G=0$, and not the stability of the
  solution $G>0$. In the region where $G>0$ we use a simple and fast
  procedure to minimize $G$ explained in Section \ref{neq}.  This
  explains why there is a small dependence on $\ell$ having a slightly
  larger $G$ for larger $\ell$, when $G>0$ in the region $q\approx
  0.15$.

{\bf  Extended Data Fig. \ref{fig:maxfrac}. Illustration of the
   algorithm used to minimize $G(q)$ for $q < q_c$. } Starting from
 the completely fragmented network at $q=q_c$, nodes are reinserted
 with the following criterion: each node is assigned and index $c(i)$
 given by the number of clusters it would join if it were reinserted
 in the network.  For example, the red node has $c(\mathrm{red})=2$,
 while the blue one has $c(\mathrm{blue})=3$. The node with the
 smallest $c(i)$ is reinserted in the network: in this case the red
 node.  Then the $c(i)$'s are recalculated and the new node with the
 smallest $c(i)$ is found and reinserted. The algorithm is repeated
 until all the removed nodes are reinserted in the network.

{\bf  Extended Data Fig. \ref{fig:decimation}.   Test of the decimation
   fraction}. Giant component as a function of the removed nodes using
 CI, for an ER network of $N=10^5$ nodes and average degree $\la k\ra
 = 3.5$.  The profiles of the curves are drawn for different
 percentages of nodes fixed at each step of the decimation algorithm.

{\bf  Extended Data Fig. \ref{fig:comparison}.  Comparison of the
   performance of CI, BC, and EGP}. We also include HD, HDA, EC, CC,
 k-core, and PR. We use a scale-free network with degree exponent
 $\gamma=2.5$, average degree $\la k\ra=4.68$, and $N=10^4$. We use
 the same parameters as in Ref. \cite{chen}.

{\bf Extended Data Fig. \ref{fig:ER_N200}. Comparison with BP.}  {\bf
  a,} Fraction of infected nodes $f$ as a function of the fraction of
immunized nodes $q$ in SIR from BP solution. We use a ER random graph
of $N=200$ nodes and average degree $\la k\ra=3.5$. The fraction of
initially infected nodes is $p=0.1$ and the inverse temperature
$\beta=3.0$.  The profiles are drawn for different values of the
transmission probability $w$: $0.4$ (red curve), $0.5$ (green), $0.6$
(blue), $0.7$ (magenta).  Also shown are the results of the fixed
density BP algorithm (open circles).
{\bf b,} Chemical potential $\mu$ as a function of the immunized nodes
$q$ from BP.  We use a ER random graph of $N=200$ nodes and average
  degree $\la k\ra=3.5$. The fraction of the initially infected nodes
  is $p=0.1$ and the inverse temperature $\beta=3.0$.  The profiles
  are drawn for different values of the transmission probability $w$:
  $0.4$ (red curve), $0.5$ (green), $0.6$ (blue), $0.7$
  (magenta). Also shown are the results of the fixed density BP
  algorithm (open circles) for the region where the chemical potential
  is non convex.
{\bf c,} Comparison between the giant components obtained with CI,
HDA, HD and BP. We use an ER network of $N=10^3$ and $\langle k
  \rangle = 3.5$.  We also show the solution of CI from
  Fig. \ref{fig:synthetic}a for $N=10^5$. We find in order of
  performance: CI, HDA, BP, and HD. (The average is taken over 20
  realizations of the network, error bars are s.e.m.)
{\bf d,} Comparison between the giant components obtained with CI,
HDA, HD and BPD. We use a SF network with degree exponent $\gamma =
  3.0$, minimum degree $\mathrm{k_{min}}=2$, and $N=10^4$ nodes.

{\bf Extended Data Fig. \ref{fig:comparisonBP_CI_fgamma}.  Fraction of
  infected nodes $f(q)$ as a function of the fraction of immunized
  nodes $q$ in SIR from BP.} We use the following parameters: initial
infected people $p=0.1$, and transmission probability $w=0.5$.  We use
an ER network of $N=10^3$ nodes and $\la k\ra = 3.5$.  We compare CI,
HDA and BP. All strategies give similar performance due to the large
value of the initial infection $p$ which washes out the optimization
performed by any sensible strategy, in agreement with the results of
\cite{zecchina2}, Fig 12a.

\newpage

\clearpage

\clearpage

\setcounter{figure}{0}


\begin{figure}[h]
\includegraphics[width=.75\textwidth]{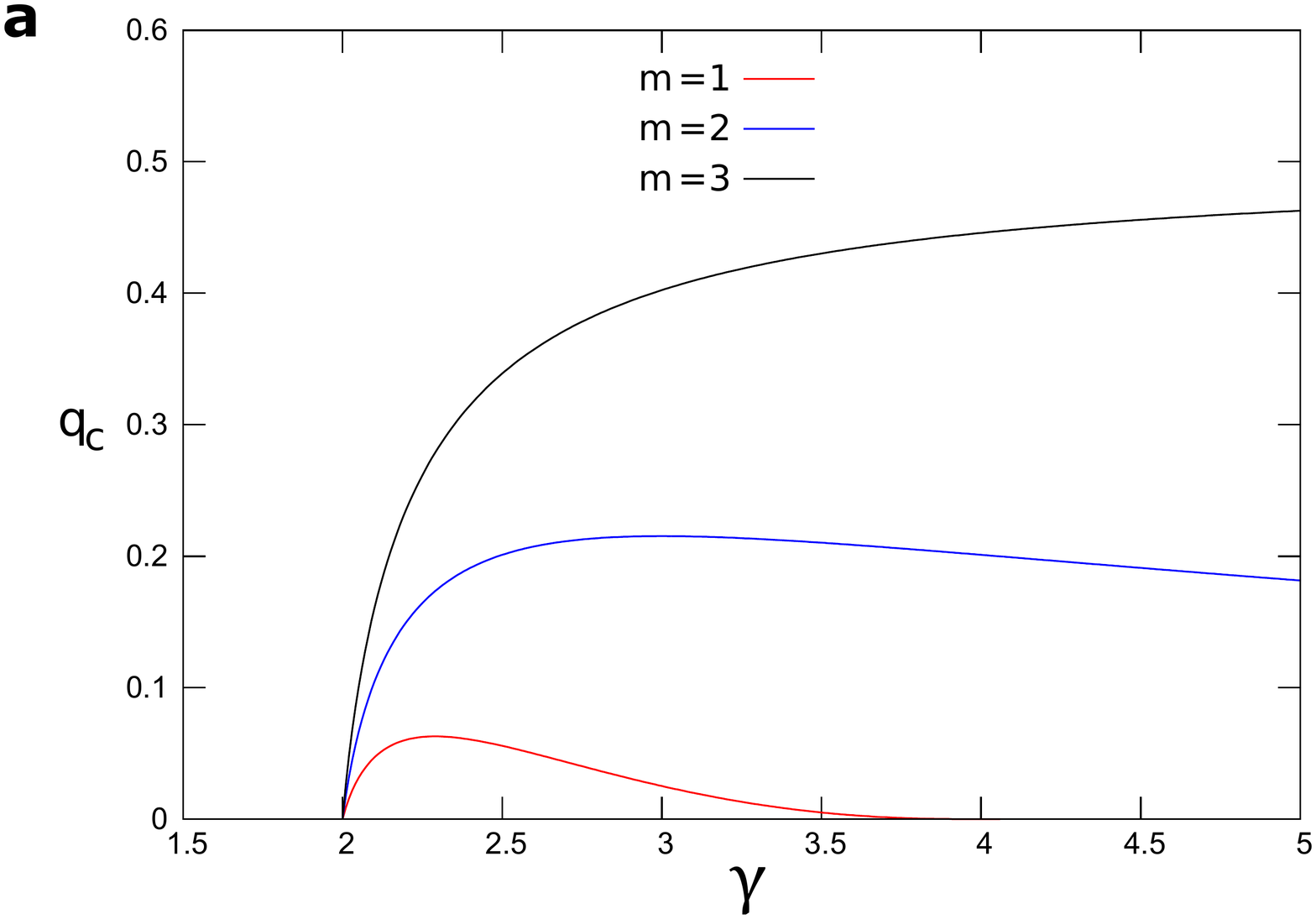}
\includegraphics[width=.75\textwidth]{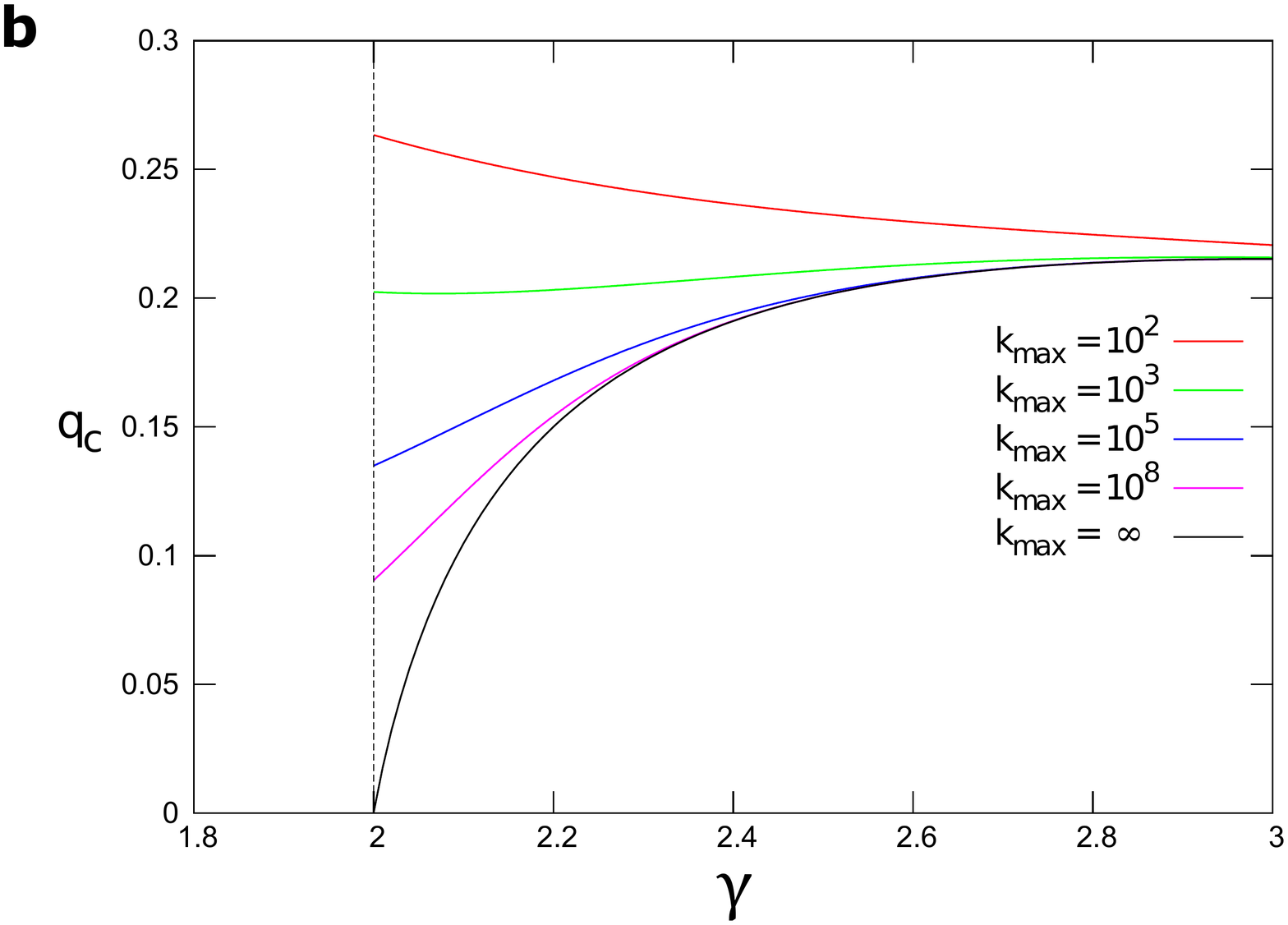} 
\caption{Extended Data}
\label{fig:pd} 
\end{figure}


\begin{figure}[h]
\includegraphics[width=.85\textwidth]{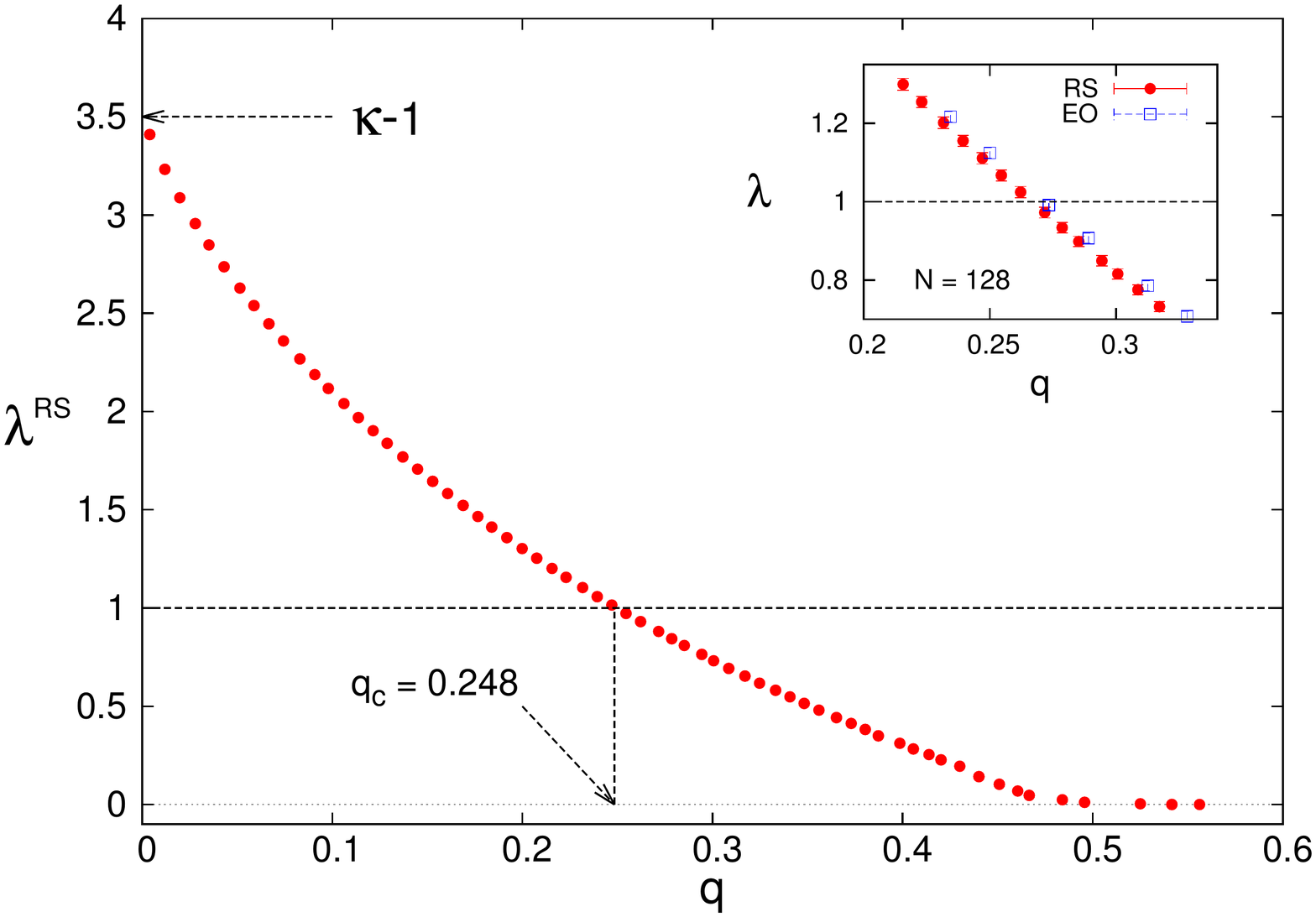} 
\caption{Extended Data}
\label{fig:lambdaRS} 
\end{figure}

\begin{figure}[h]
\includegraphics[width=1.0\textwidth]{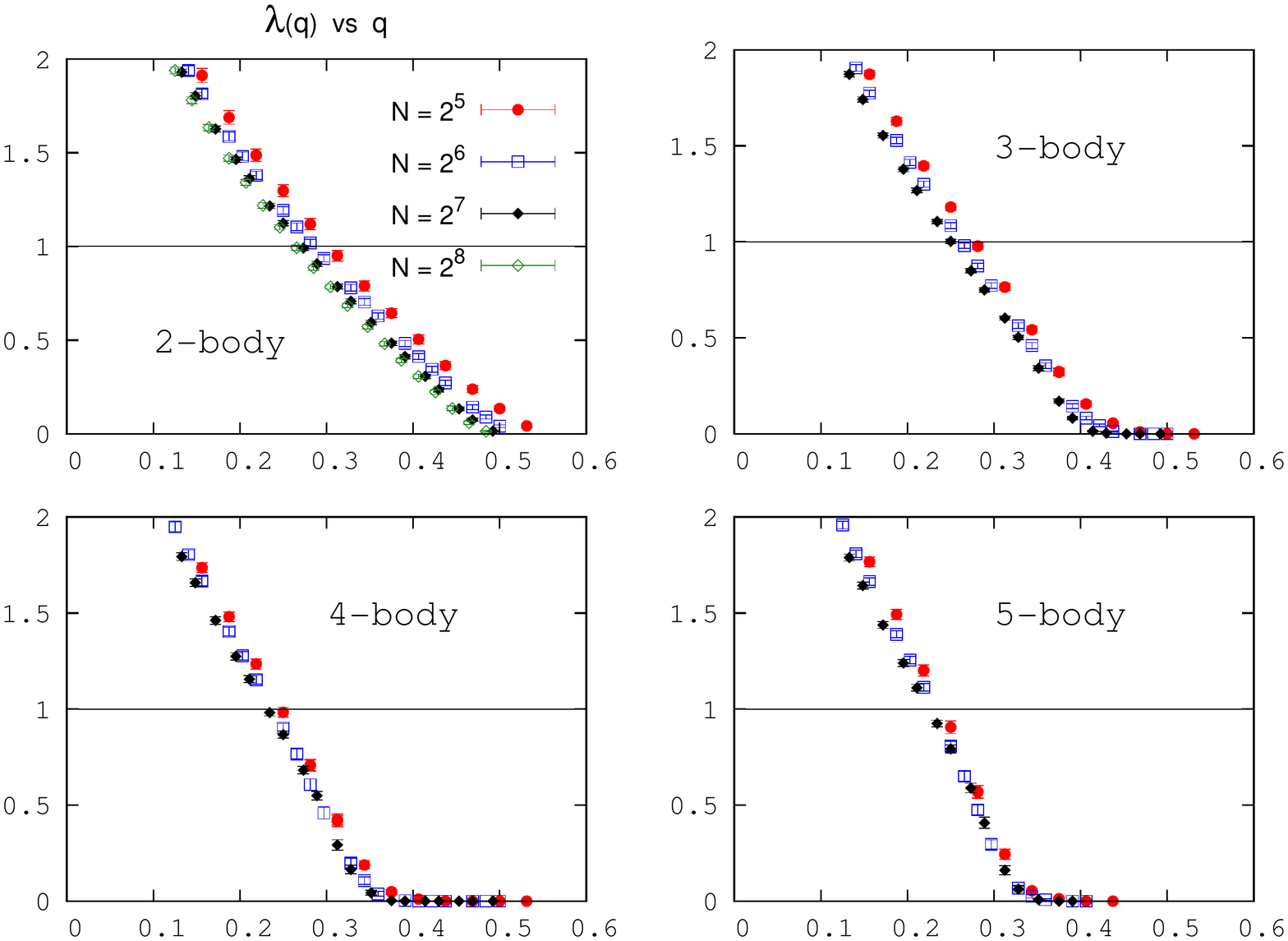} 
\caption{Extended Data}
\label{fig:taufull} 
\end{figure}

\begin{figure}[h]
\includegraphics[width=.75\textwidth]{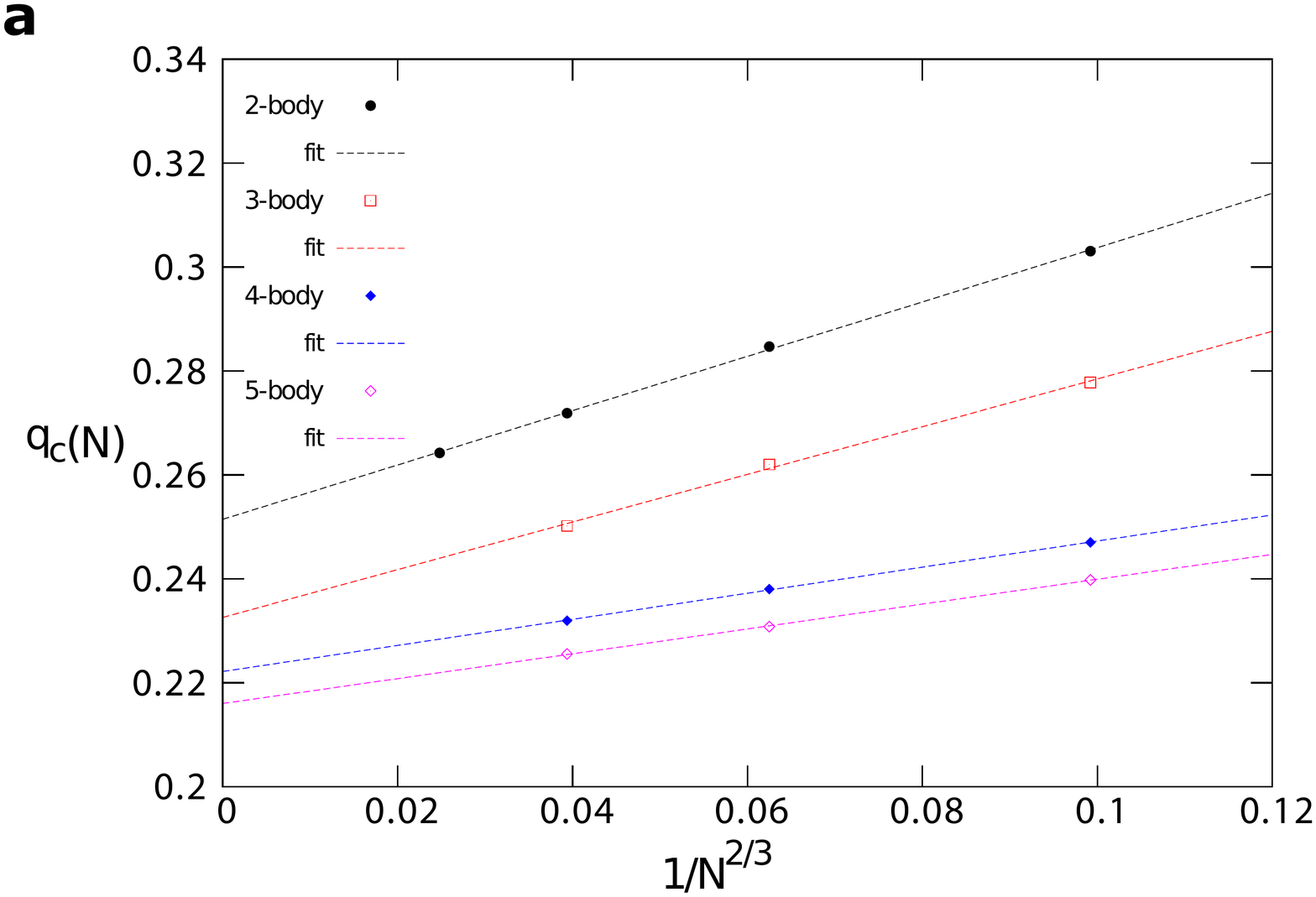} 
\includegraphics[width=0.75\textwidth]{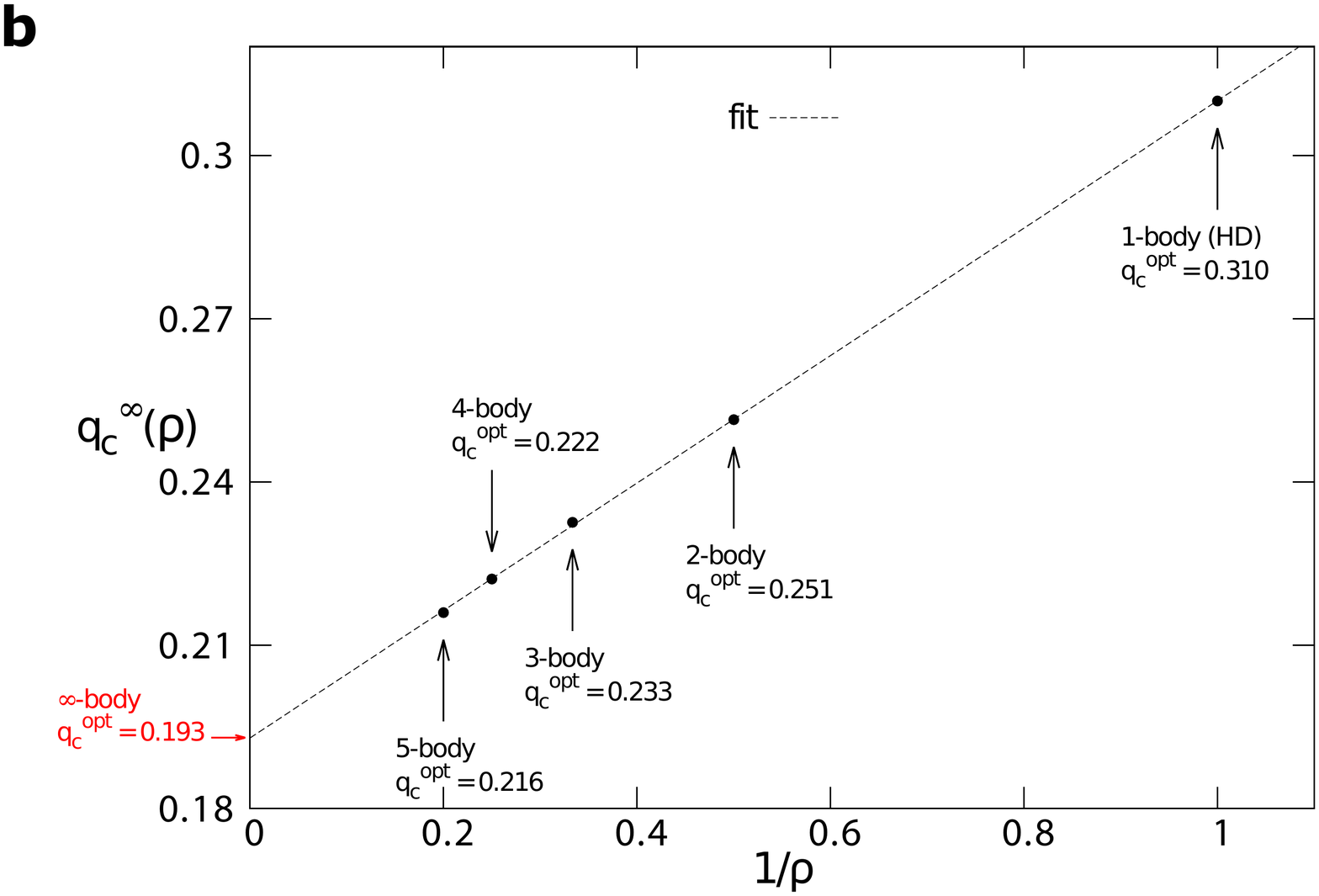} 
\caption{Extended Data}
\label{fig:tau_pc} 
\end{figure}


\begin{figure}[h]
\includegraphics[width=.75\textwidth]{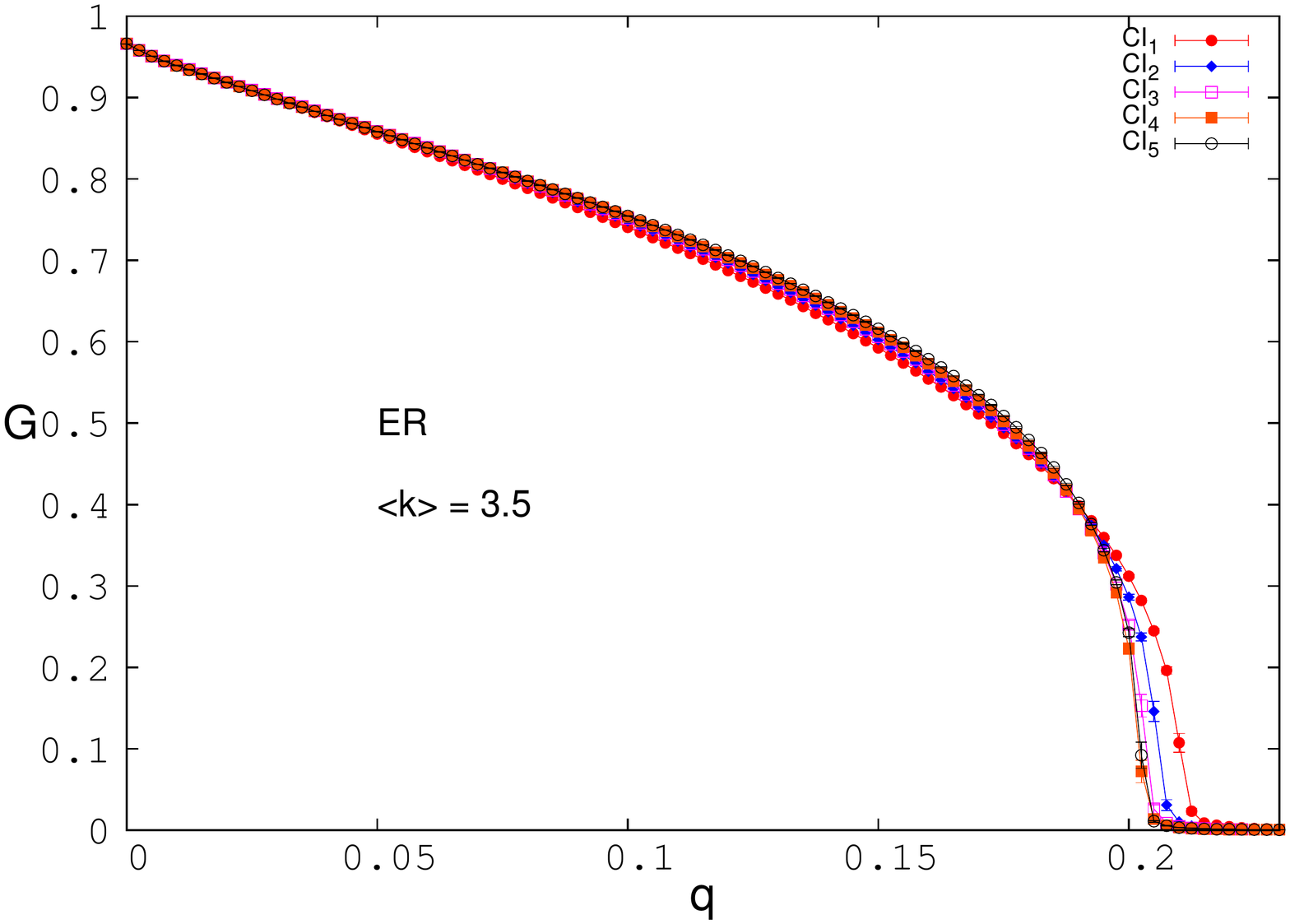} 
\caption{Extended Data}
\label{fig:allLevels} 
\end{figure}

\begin{figure}[h]
\includegraphics[width=.75\textwidth]{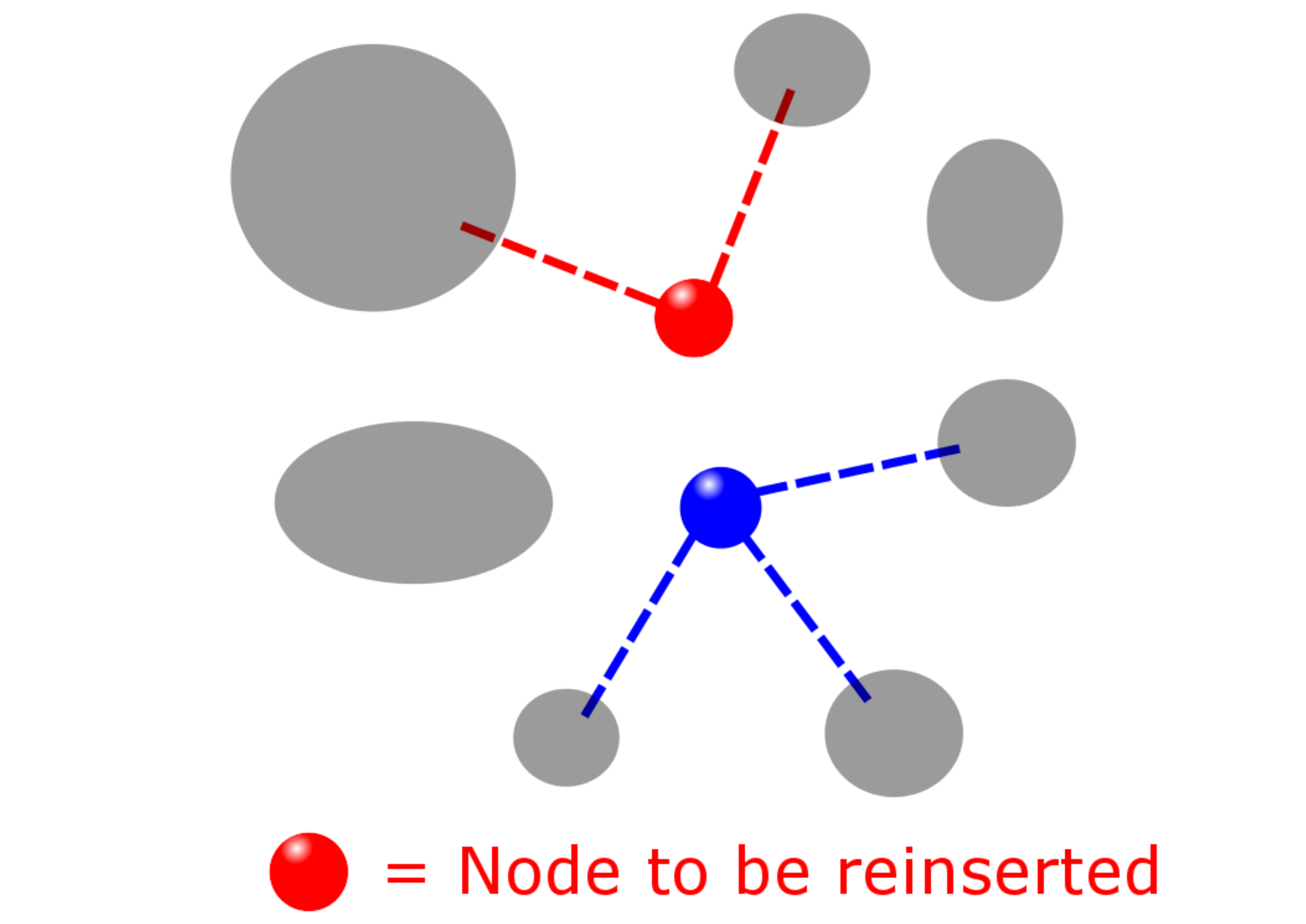}
\caption{Extended Data}
\label{fig:maxfrac}
\end{figure}

\begin{figure}[h]
\includegraphics[width=.75\textwidth]{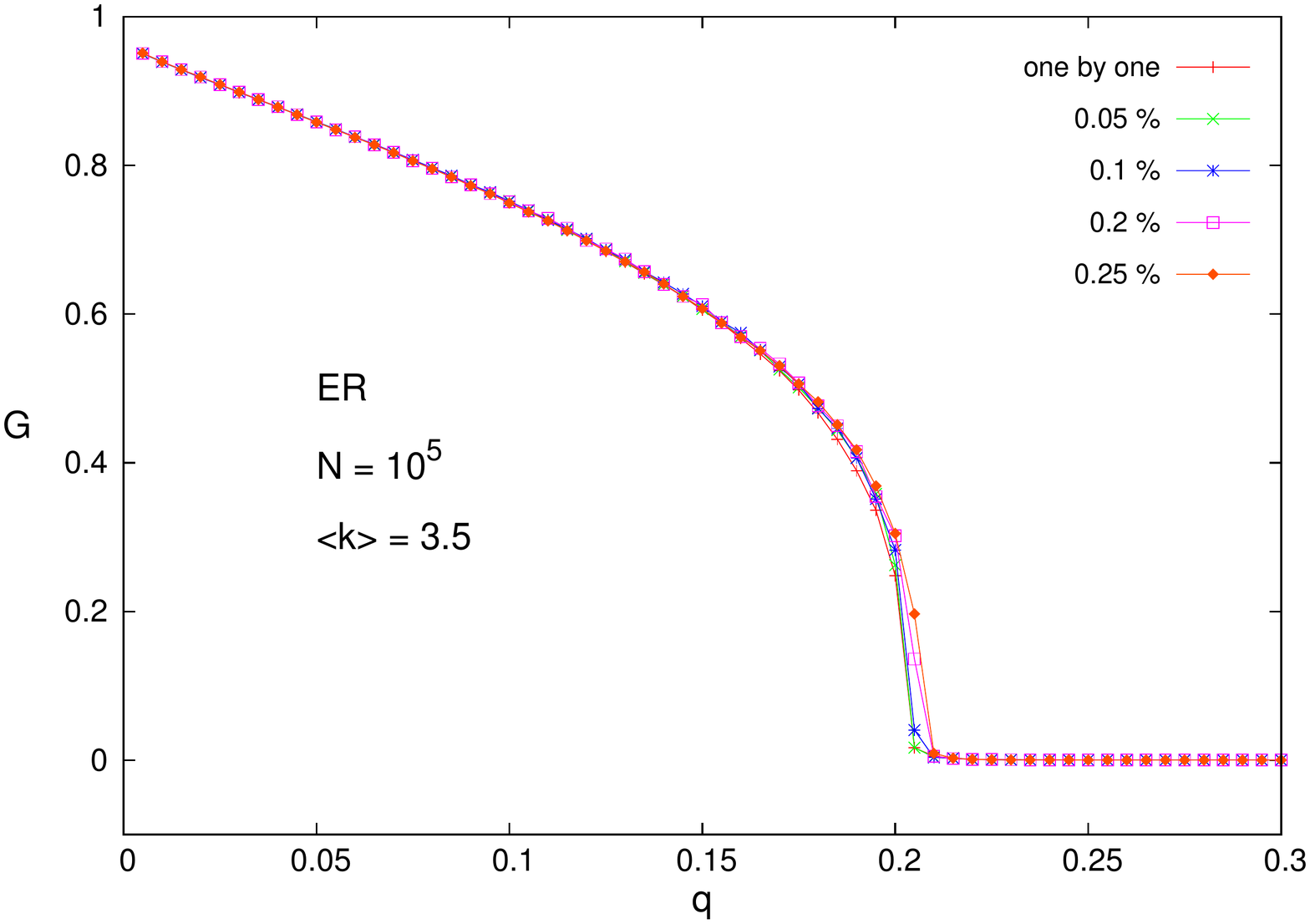}
\caption{Extended Data}
\label{fig:decimation}
\end{figure}

\begin{figure}[h]
\includegraphics[width=.75\textwidth]{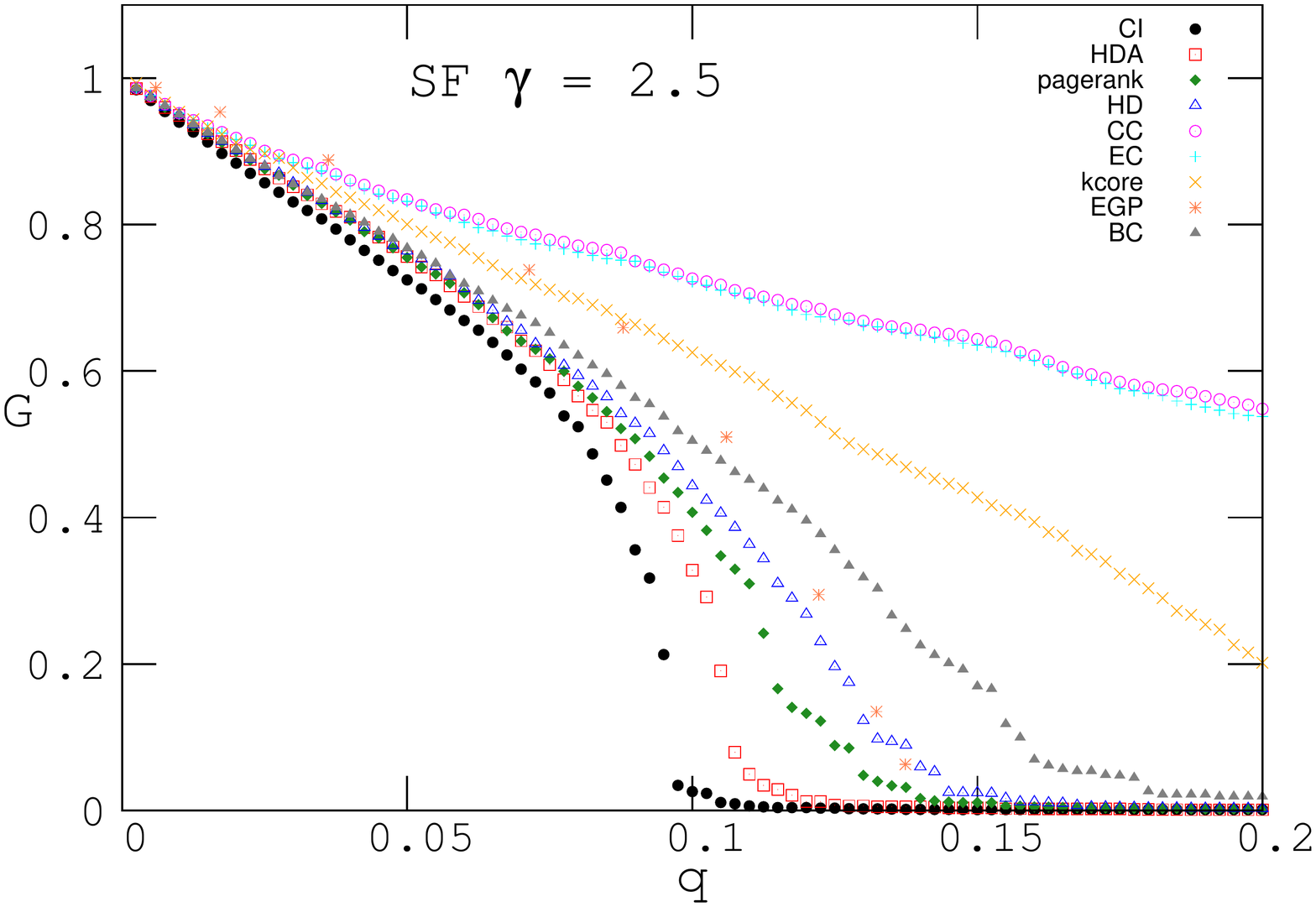} 
\caption{Extended Data}
\label{fig:comparison} 
\end{figure}

\begin{figure}[h]
\centerline{ \includegraphics[width=.5\textwidth]{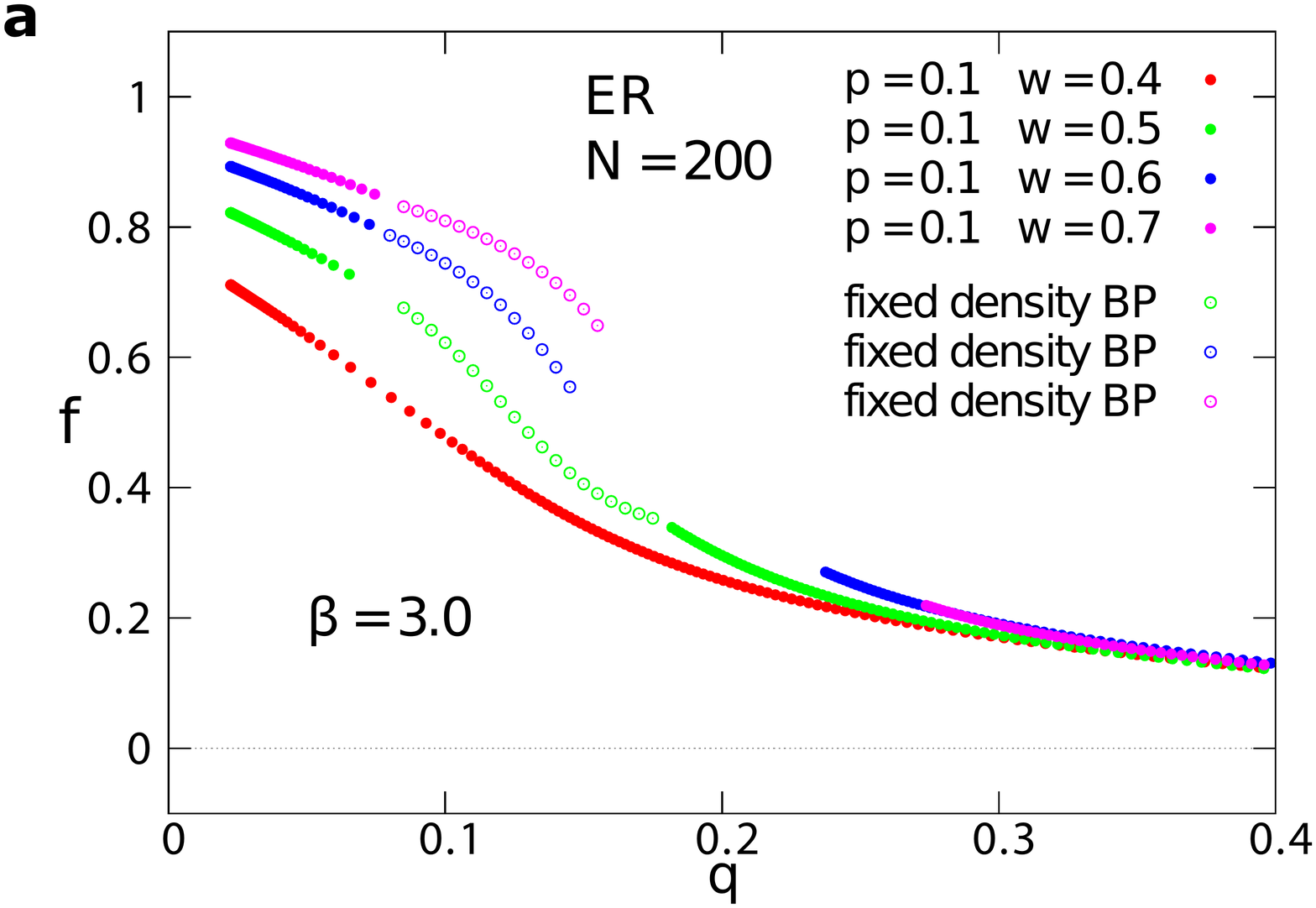}
\includegraphics[width=.5\textwidth]{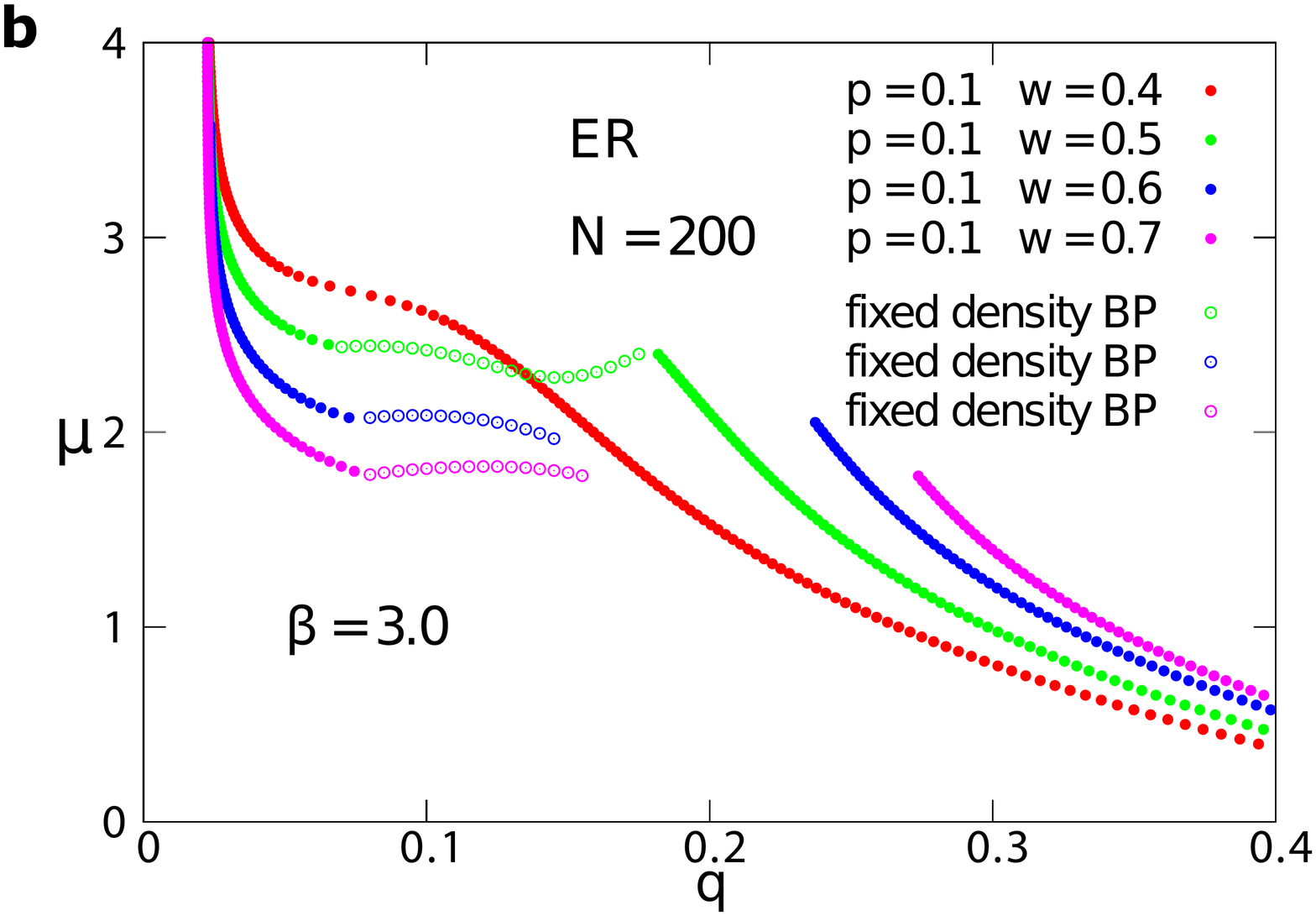}}
\centerline{  \includegraphics[width=.5\textwidth]{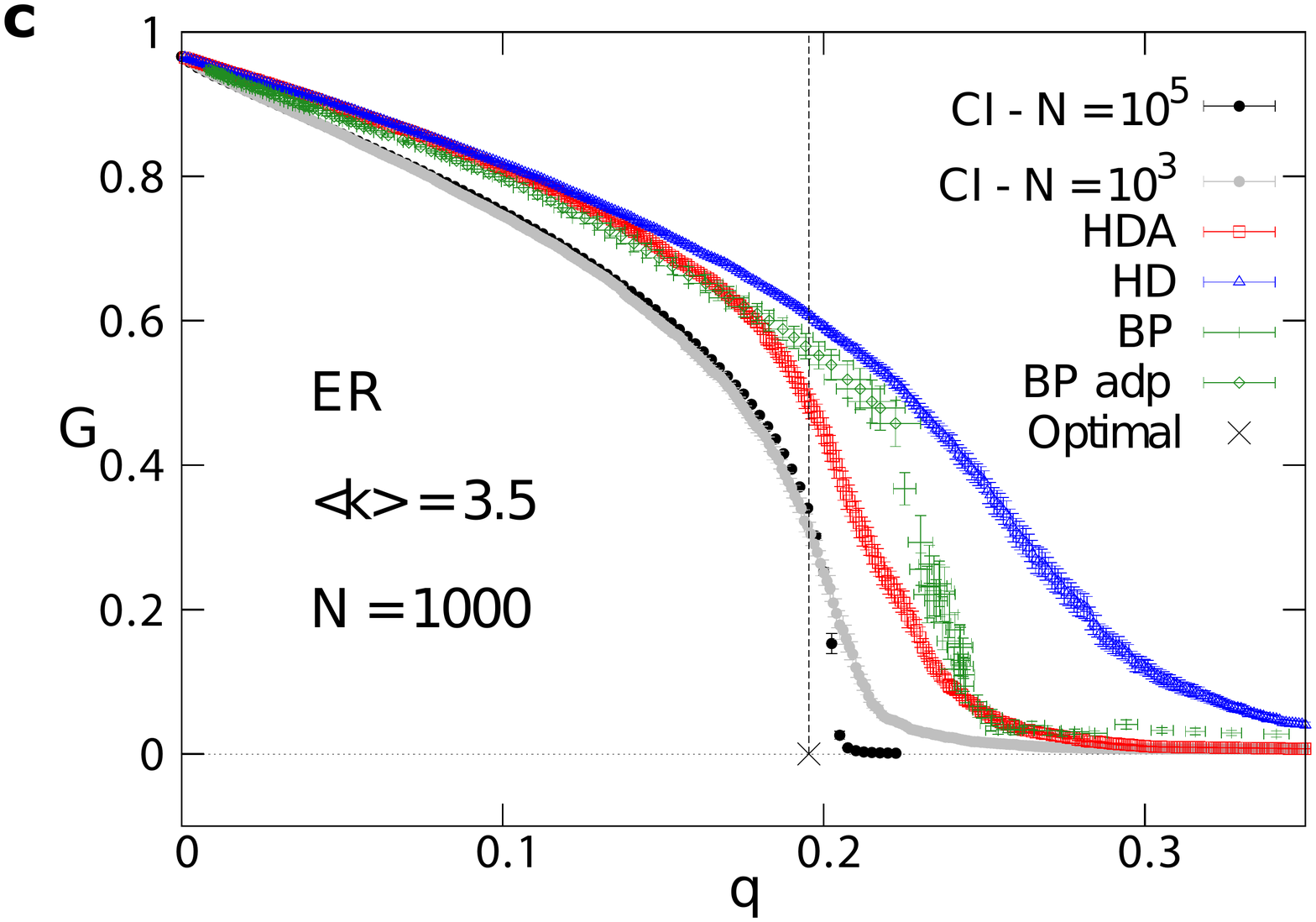}
  \includegraphics[width=.5\textwidth]{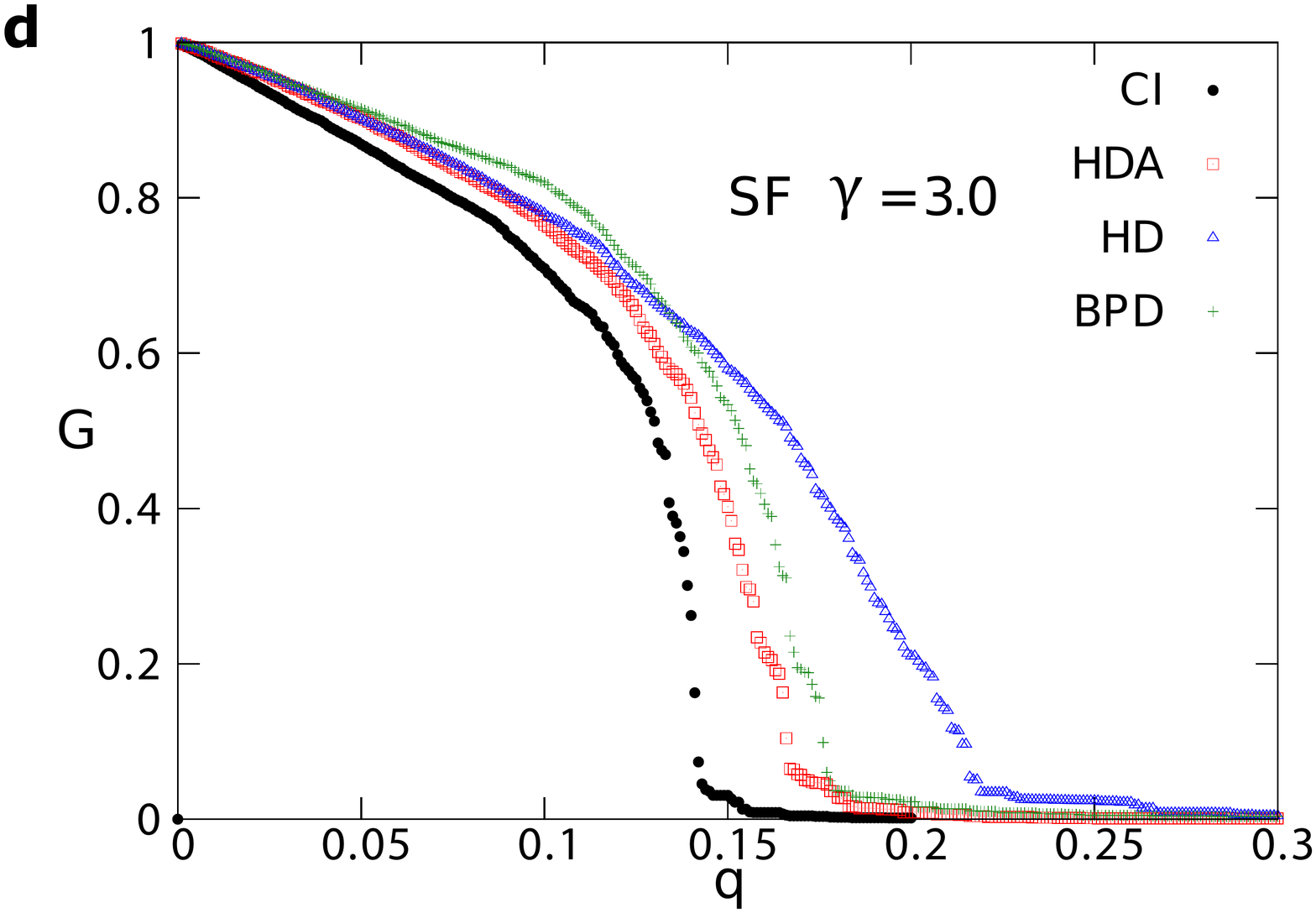}}
\caption{Extended Data}
\label{fig:ER_N200}
\end{figure}




\begin{figure}[h]
\includegraphics[width=.8\textwidth]{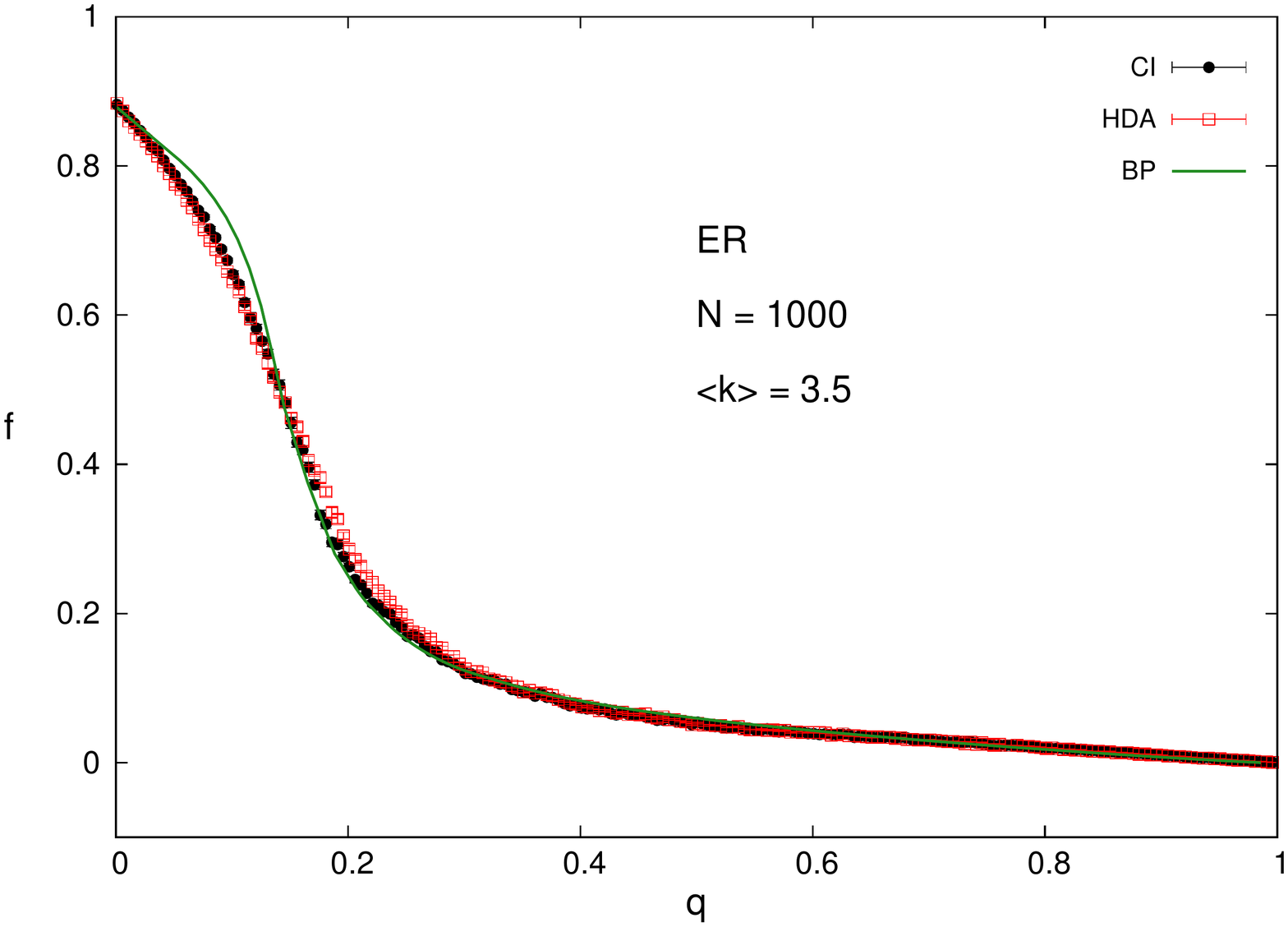}
\caption{Extended Data}
\label{fig:comparisonBP_CI_fgamma}
\end{figure}

\clearpage

\centerline{ \bf METHODS}

\centerline{\bf Influence maximization in complex networks through optimal percolation}

\medskip

\centerline{ \bf Flaviano Morone \& Hern\'an A. Makse}

\tableofcontents

\clearpage

\section{Heuristic methods used to identify 
  influential spreaders in complex networks}
\label{heuristics}

In this section we describe the existing heuristic algorithms in the
literature that have been used to identify influential spreaders and
superspreaders in networks. We use these heuristics to compare with
our collective theory of influence. A common feature of the heuristic
methods is that they are not designed from first principles and
therefore do not necessarily optimize an influence measure. Instead,
they are based on intuitive ideas about what is an
influencer. Besides, the heuristics constitute ranking of nodes
lacking the collective influence arising from considering all the
influencers at once. On the other hand, the theoretical framework for
maximizing the spread of influence of Kempe {\it et al.}  \cite{kempe}
and the Belief Propagation theory for the optimal immunization problem
of Altarelli {\it et al.}  \cite{zecchina2,semerjian} contain the necessary
optimization of influence. The greedy algorithm considered in
\cite{kempe} has prohibitive running time for all the networks
considered in our work. Detailed comparison with the Belief
Propagation is done in Methods Section \ref{BP}.

{\bf High-Degree (HD)} \cite{barab,pastor,cohen}.  In the HD method
nodes are ranked by degree, and sequentially removed starting from the
node of highest degree.  One of the limitations of this method is the
fact that hubs may form tightly-knit groups called``rich-clubs''
\cite{rich2,wasserman}.  Strategies based on high-degree will highly
rank these rich-club hubs.  On the other hand, an optimized scheme
will target only one of them to avoid overlap between the already
attacked areas in the network. {\bf High Degree Adaptive (HDA)} is the
adaptive version where the degree of the remaining nodes is recomputed
after each node removal.

{\bf PageRank (PR)} \cite{pagerank}.  This is the famous Google's
algorithm for ranking websites. It was proposed for first time in
\cite{pagerank} to ``condensing every page in the World Wide Web into
a single number, its PageRank. PageRank is a global ranking of all web
pages, regardless of their content, based solely on their location in
the Web's graph structure.'' PR can be thought of as the most
successful rank, ever. At its heart, it is another eigenvector
centrality. It computes the probability that, if someone follows links
on the web at random, performing a random walk of clicks, he/she
eventually hits your website. The higher this chance, the higher the
PR of the website. Therefore, sites that get linked more are
considered reputable, and, linking to other websites, they pass that
reputation along. Thus, the shortcoming with PR comes from the fact
that PR takes node's score into account when calculating other's
scores.  In other words, a high-PR site may confer a much higher score
to otherwise unpopular sites it happens to link.  Notice that in our
algorithm using the non-backtracking operator this problem is cured
nicely, since the influence is computed by "ignoring" the node you
come from.

{\bf K-core} \cite{gallos}.  K-core ranking is based on the k-shell
decomposition of the network.  Each node is assigned the k-shell
number, $k_S$, i.e. the order of the shell it belongs to.  In k-shell
decomposition, we first remove all nodes with degree $k= 1$ and
continue pruning the network iteratively until there is no node with
$k=1$. These removed nodes belong to the peripheric k-shell with index
$k_S=1$.  Similarly, the next k-shells are defined until all nodes are
pruned and we get to the kcore of the network. The rank based on kcore
produces good results in identifying single spreaders individually,
but has a poor performance for multiple spreaders (the case considered
in the present manuscript), because putting spreaders in the same
k-shell gives a marginal or null advantage, as recognized in
\cite{gallos}. That is, ranking the spreaders one by one, one may find
that the best of them are located in the core. However, when
considering the maximization of influence of all spreaders at the same
time, collective effects coming from interactions between the
spreaders via overlap of their spheres of influence are crucial: even
if the top spreader is in the core, the next spreader most probably
will not be in the core, because the core is already infected by the
first spreader.  Indeed, these interactions between spreaders are what
makes the problem hard to solve (NP-hard as shown in
\cite{kempe}). Thus, even if the best individually-ranked spreaders
might be located in the kcore, their collective influence is
determined by their full set of interactions. Therefore, for multiple
spreaders, the kshell ranking is not optimal, although, choosing core
nodes separated by a distance increases their optimality as already
shown in \cite{gallos}.

{\bf Eigenvector Centrality (EC)} \cite{ec}. It is the eigenvector
corresponding to the largest eigenvalue of the adjacency matrix.  Node
rank is the corresponding entry of the eigenvector. Nodes are removed
starting from the highest rank.  This method is not very powerful,
especially for the case of SF networks, where most of the weight may
be carried by few nodes (hubs), while the others have vanishingly
small weights, and thus they are not properly ranked.

{\bf Closeness Centrality (CC)} \cite{cc}. Closeness centrality
measures how close a vertex is to all other vertices in the graph.
More precisely CC at node $i$ is the inverse of the average distance
to all other nodes.  Nodes are ranked according to their CC from the
highest to the lowest score, and removed accordingly.  A property of
CC is that it tends to give high scores to individuals who are near
the center of local clusters (i.e. network communities), and hence it
over-allocates spreaders (or immunized nodes) next to each other.
Moreover, it comes with a high computational cost that prevents the
application to large networks.

Methods Section \ref{comparison} compares results with {\bf
  betweenness centrality} \cite{bc} and {\bf
  equal-graph-partitioning} \cite{chen} which present prohibitive
running times for the large-scale networks used here and present worst
performance than other heuristics.

\section{Collective Theory of Optimal Influence}
\label{theory}

Nodes forming complex networks play different roles, depending on the
process in which they participate \cite{newman-book}. Their inherent
strength and weakness emerge collectively from the pattern of
interactions with the other components.  Nonetheless, it is a common
practice to quantify node's importance in a network
\cite{newman-book,kleinberg}, for example social rank order
\cite{wasserman}, by individual node's attributes such as the amount
of its connections \cite{barab,cohen,pastor}, betweenness and
eigenvector centralities \cite{freeman}, or its closeness to the core
\cite{gallos}. This attitude is, nowadays, amplified by the augmented
reality provided by virtual social networks. This idea has also
permeated to other fields and it is common to strategies of
immunization \cite{pastor,chen}, viral spreading of information and
marketing in social media \cite{richardson1,pei}, as well as targeted
attack schemes to infrastructure networks \cite{barab,cohen}. However,
individual node ranking is an ambiguous definition of node's
influence, for it considers the influencers as isolated entities and
not in interaction with each other.
Yet, there has been an abundant production on the subject of
identifying most influential nodes and "superspreaders" using such
rankings \cite{freeman,pagerank,barab,cohen,chen,gallos,pei,pei2}.
The main problem is that all these methods do not optimize an
objective global function of influence. Instead, they are based on
assumptions about the importance of individual properties of the node
\cite{pei2} and, inevitably, they fail to take into account
the collective influence of the whole set of nodes.  As a consequence,
there is no guarantee of their performance.

On the other hand, a theoretical framework taking into account a
global maximization of influence was outlined by Kempe {\it et al.}
\cite{kempe} in the form of a discrete global optimization problem for
diffusion of information models such as the Linear Threshold Model
(LTM) of Granovetter \cite{LTM} and other variants \cite{watts}, whose
solution is proved to be NP-hard and is approximated by a greedy
algorithm \cite{kempe}. This approach makes leverage on a special
attribute of the function of influence to be optimized, called
submodularity, which express the decrease in the gain in the output of
the process after an increment of the input factors. This "diminishing
return" property is often lost in many optimization problems.  In
particular, submodularity does not apply, in general, to the giant
connected component in optimal percolation problem, which is the
function we minimize.  It should be said that for some LTMs, of the
type treated in \cite{kempe}, the objective function of influence to
be maximized can be proven to be submodular. However, this does not
hold true in general for other classes of LTMs, for example in the
case of a fixed choice of the thresholds, as explicitly stated in
\cite{kempe}, and which represents the type of problem studied in this
paper.  As a consequence even the greedy algorithm does not provide a
stable approximation to the optimal solution. Furthermore, greedy
searches are not scalable and therefore not applicable to current day
big data in social media and population immunization problems.  In our
case we face these difficulties.

Another pioneering approach to the problem of influence optimization
and immunization is outlined in Refs. \cite{zecchina1,zecchina2}.  In
particular, in Ref. \cite{zecchina2}, the authors use a very
interesting and principled method, based on Belief Propagation (BP),
to minimize the expected infection outbreak in an epidemic process
(modeled as susceptible-infected-recovered, SIR, or
susceptible-infected-susceptible, SIS) for a given number of immunized
nodes, and for a given choice of the parameters of the model, i.e.,
transmission probability $w$ and initial fraction of infected
individuals, $p$. While the method is able to find nearly optimal
solutions to the problem, it becomes unfeasible when $p\to0$, which
corresponds to the optimal influence problem treated here, because the
time complexity of the algorithm diverges as $p^{-3}$ for $p\to 0$.  A
full comparison between our theory and BP is performed in Methods
Section \ref{BP}.

From the theoretical standpoint, our main result is the discovery of a
method to map the problem of optimal influence onto the computation of
the minimal set of nodes that
minimizes the largest eigenvalue of the Non-Backtracking matrix of the
network. This operator has recently received a lot of attention thanks
to its high performance in the problem of community detection
\cite{florant,newman}. We show its formidable topological power in the
problem of optimal influence.  The problem we set up
is, in its most general formulation, intractably hard. We present a
perturbative solution along with a very fast algorithm, that we use to
simulate an optimized immunization/quarantine and superspreading
protocol on very large real networks.  The naive strategy,
corresponding to the lowest approximation, is given by the attack on
the high degree nodes. The first non-trivial attack strategy is
equivalent to find the ground state of a spin-glass like system, i.e.,
an anti-ferromagnetic Ising model with random bonds in a random
external field at fixed magnetization. Higher order approximations
produce superior performance compared to previous heuristic
strategies.  Furthermore, the algorithm is highly scalable with
running time $O(N \log N)$.

\subsection{Optimal Percolation}

A network is a set of $N$ nodes tied together by $M$ edges. The vector
$\mathbf{n}=(n_1,\dots,n_N)$ represents which node is present and
which one is removed. We adopt the convention that $n_i=1$ if node $i$
is present, and $n_i=0$ if node $i$ is removed (corresponding to an
influencer). The total fraction of removed nodes is denoted with $q$:
\be q=1-\frac{1}{N}\sum_{i=1}^{N}n_i\equiv1-\la n\ra\ .
\label{eq:unoccupied_sites}
\ee 

We call $G(q)$ the fraction of occupied sites belonging to the giant
(largest) connected component (in the limit $N\to\infty$ represents
the probability of existence of the giant component).  The optimal
percolation problem is finding the minimum fraction $q_c$ of nodes to
be removed such that $G(q_c)=0$: \be q_c\ =\ \min
\{q\in[0,1]:\ G(q)\ = 0\}\ .  \ee

For $q\geq q_c$, the network consists of a collection of clusters of
nodes whose sizes are subextensive.  Alternatively, for a fixed
fraction $q<q_c$, we search for the configuration of removed nodes
that provides the minimal non-zero giant connected component.

A given node $i$ can be disconnected from the giant component $G$,
either because it is directly removed, or because it is indirectly
detached by the removal of other nodes. In the former case $n_i=0$,
while in the second one $n_i=1$. Thus, we see that $n_i$ cannot tell
us whether node $i$ belongs to $G$ or not. We then need another
variable encoding the information that node $i$ belongs or not to
$G$. This variable is the probability of node $i$ to belong to the
giant connected component, $\nu_i$, and we agree to set $\nu_i=1$ if
$i\in G$, and $\nu_i=0$ otherwise. The fraction of nodes in the giant
component, upon removing $q$ of them from the network, is then given
by: \be G(q)\ = \ \frac{1}{N}\sum_{i=1}^N\ \nu_i\ .  \ee 

In principle, optimal percolation minimizes the giant component over
the configurations $\mathbf{n}$. However an explicit functional form
of $G(\mathbf{n})$ is not feasible. Our approach is to transform the
problem into the minimization over $\mathbf{n}$ of the largest
eigenvalue controlling the stability of the percolation solution,
which can be written explicitly in terms of $\mathbf{n}$.  We first
derive the relation between the vector
{\boldmath${\nu}$}$=(\nu_1,\dots,\nu_N)$ and the vector $\mathbf{n} =
(n_1,\dots,n_N)$. This can be easily done using a message passing
approach \cite{bianconi,lenka,montanari}.

Let us consider two connected nodes $i$ and $j$ and orient the
corresponding edge from $i$ to $j$. Now let us suppose to "virtually"
remove $j$ (create a ``cavity'' at $j$) from the network and ask
ourselves if node $i$ belongs to $G$ or not. This information can be
stored in an auxiliary quantity $\nu_{i\to j}=1,0$ representing the
probability of $i$ to belong to the giant connected component in the
absence of $j$.  The advantage of using the variables $\nu_{i\to j}$,
instead of $\nu_i$ is the fact that they satisfy a closed set of
equations. Clearly $\nu_{i\to j}=0$ if $n_i=0$. So the interesting
case is when $n_i=1$. Recalling that $j$ is momentarily absent from
the network, the chance that $i$ belongs to $G$ is determined by the
event "at least one among the neighbours of $i$ different from $j$,
belongs to $G$ when $i$ itself is virtually removed from the
network". For a locally tree-like network this statement can be
mathematically translated in the following message passing formula
\cite{bianconi,lenka}: \be \nu_{i\to j}\ =\ n_i\left[
  1\ -\ \prod_{k\in\partial i\setminus j}(1\ -\ \nu_{k\to i})
  \right]\ ,
\label{eq:CavityMess}
\ee where $\partial i\setminus j$ is set of nearest neighbours of $i$
minus $j$.  We can finally put back $j$ in the network and get the
real information $\nu_i$ as: \be \nu_i\ =\ n_i\left[
  1\ -\ \prod_{k\in\partial i}(1\ -\ \nu_{k\to i}) \right]\ .
\label{eq:RealMess}
\ee    

The system defined by Eq. \eqref{eq:CavityMess} always admits the
solution $\{ \nu_{i\to j}=0 \}$ for all $i\to j$ (regardless
  of the values of $n_i$), as can be verified by inspection.

As a consequence also $\{ \nu_i=0\}$ for all $i$, which in turn
gives $G=0$. This solution is stable, provided that the largest
eigenvalue of the linear operator represented by the $2M \times 2M$ 
matrix defined on the directed links $k\to \ell, i\to j$: 

\be \mathcal{M}_{k\to \ell, i\to j}\ =\ \frac{\partial \nu_{i\to
    j}}{\partial \nu_{k\to\ell}}\Big|_{\{\nu_{i\to j}=0\}} \ee is less
than one. We call $\lambda(\mathbf{n};q)$ the largest eigenvalue of
$\Mia$, which depends on the vector $\mathbf{n}$ and we add also a
parametric dependence on the fraction of removed nodes $q$. Thus, the
stability of a solution set $\mathbf{n}$ of $G=0$ is determined by the
condition $\lambda(\mathbf{n};q) < 1$.

For a fixed fraction $q$ there exist, in general, very many possible
configurations $\mathbf{n}$ that satisfy
Eq. \eqref{eq:unoccupied_sites}.  When $q<q_c$ each configuration
$\mathbf{n}$ gives $\lambda(\mathbf{n};q)>1$, since it is impossible
to find a set of nodes to remove such that $G(q)=0$, and this
corresponds to the instability of the solution $\{\nu_{i\to j}=0\}$
signaled by a value of $\lambda(\mathbf{n};q)$ larger than one.  On
the contrary, when $q>q_c$, we have two different possibilities: there
exist configurations $\mathbf{n}$ such that $\lambda(\mathbf{n};q)>1$,
which corresponds to nonoptimal node removals unable to destroy the
giant component ($G(q)>0$); on the other hand, there can be found
other configurations for which $\lambda(\mathbf{n};q)<1$, which
corresponds to a fragmented network with $G(q)=0$. As we approach
$q_c$ from above, $q\to q_c^+$, the number of configurations
$\mathbf{n}$ such that $\lambda(\mathbf{n};q)<1$ (and hence $G(q)=0$)
decreases and eventually vanishes at $q_c$. This situation is
exemplified in Fig. \ref{fig:explanation}c. 

We find that the matrix $\Mia$ is given in terms of the
Non-Backtracking (NB) matrix $\Bia$ \cite{hashimoto,nbt} for a
locally-tree like random network via the equation: \be
\mathcal{M}_{k\to \ell, i\to j}\ =\ n_i\mathcal{B}_{k\to \ell, i\to j}
\label{eq:modifiedNB}
\ee where
\be \mathcal{B}_{k\to \ell, i\to j} =
\begin{cases}
1\ \mathrm{if} \ \ell=i \ \mathrm{and}\  j\neq k\ ,\\
0\ \mathrm{otherwise}.
\end{cases}
\label{definition}
\ee
The modified NB operator $\hat{\mathcal{M}}$ is represented by a
$2M\times2M$ matrix on the $2M$ directed edges of the network. For
example, for the simple following graph with $N=6$ and $M=6$:
\begin{fmffile}{nb}
\begin{eqnarray}
\parbox{30mm}{
\begin{fmfgraph*}(50,40)
\fmfleft{l1}
\fmfright{r1,r2}
\fmfbottom{b1}
\fmf{plain, left=.5}{l1,v1,v2,v3,v1}
\fmf{plain, left=.5}{r1,v3}
\fmf{plain, left=.5}{r2,v2}
\fmfdotn{l}{1}
\fmfdotn{v}{3}
\fmfdotn{r}{2}
\fmflabel{$1$}{r2}
\fmflabel{$2$}{v2}
\fmflabel{$3$}{v1}
\fmflabel{$4$}{l1}
\fmflabel{$5$}{v3}
\fmflabel{$6$}{r1}
\end{fmfgraph*}
}\ \ \ \ \ \ \ \nonumber\\ 
\label{eq:nb}
\end{eqnarray}
\end{fmffile}
the corresponding $\hat{\mathcal{M}}$ matrix is a $12\times 12$ matrix
that reads (the associated non-backtracking matrix $\hat{\mathcal{B}}$
is obtained by setting $n_i=1$): \be
\begin{array}{ccccccccccccc}
 & 1\to2 & 2\to1 & 2\to3 & 2\to5 & 3\to2 & 3\to4 & 3\to5 & 4\to3 & 5\to2 & 5\to3 & 5\to6 & 6\to5\\
1\to2 & 0 & 0 & n_{2} & n_{2} & 0 & 0 & 0 & 0 & 0 & 0 & 0 & 0\\
2\to1 & 0 & 0 & 0 & 0 & 0 & 0 & 0 & 0 & 0 & 0 & 0 & 0\\
2\to3 & 0 & 0 & 0 & 0 & 0 & n_{3} & n_{3} & 0 & 0 & 0 & 0 & 0\\
2\to5 & 0 & 0 & 0 & 0 & 0 & 0 & 0 & 0 & 0 & n_{5} & n_{5} & 0\\
3\to2 & 0 & n_{2} & 0 & n_{2} & 0 & 0 & 0 & 0 & 0 & 0 & 0 & 0\\
3\to4 & 0 & 0 & 0 & 0 & 0 & 0 & 0 & 0 & 0 & 0 & 0 & 0\\
3\to5 & 0 & 0 & 0 & 0 & 0 & 0 & 0 & 0 & n_{5} & 0 & n_{5} & 0\\
4\to3 & 0 & 0 & 0 & 0 & n_{3} & 0 & n_{3} & 0 & 0 & 0 & 0 & 0\\
5\to2 & 0 & n_{2} & n_{2} & 0 & 0 & 0 & 0 & 0 & 0 & 0 & 0 & 0\\
5\to3 & 0 & 0 & 0 & 0 & n_{3} & n_{3} & 0 & 0 & 0 & 0 & 0 & 0\\
5\to6 & 0 & 0 & 0 & 0 & 0 & 0 & 0 & 0 & 0 & 0 & 0 & 0\\
6\to5 & 0 & 0 & 0 & 0 & 0 & 0 & 0 & 0 & n_{5} & n_{5} & 0 & 0
\end{array}
\ .
\ee
The largest eigenvalue of the NB matrix $\Bia$ is positive and
simple, as a consequence of the Perron-Frobenius theorem \cite{nbt},
and the corresponding eigenvector is such that all components are
positive. 

The NB matrix $\Bia$ has recently received a lot of attention
in context of detectability of communities in complex networks
\cite{florant,newman}. In that problem the interesting eigenvalue is
the second largest, since the corresponding eigenvector can be used to
label the nodes in different communities. This can be easily
understood for the case of two communities. Since the eigenvector
corresponding to the second largest eigenvalue has both positive and
negative components, one can assign to one community all the nodes
corresponding to the positive entries of the eigenvector, and to the
other community all the nodes labeled with the negative components
(the eigenvector corresponding to the largest eigenvalue does not work
to define communities since it has all positive components, so that it
is insensitive to the community structure of the network). This
clustering method can be also generalized to the case of multiple
(i.e. more than two) communities. This kind of community detection
protocol can be implemented also using other matrices, e.g. the
adjacency and the Laplacian matrices. What is crucial is the fact that
the NB operator has the optimal performance, in the sense that it is
able to detect communities down to the detectability threshold, while
the other spectral methods fail much before.

The NB matrix $\Bia$ also intervenes when one linearizes the belief
propagation equations, and it was first used to detect the location of 
the phase transition in Ref. \cite{Coja} for the 3-coloring problem.

In our problem we are interested in the largest eigenvalue of the
matrix $\Mia$ (not of $\Bia$), which is indeed a suitable modification
of the NB matrix via optimal removal of $n_i$. That is, $\Mia$ is a NB
matrix on a modified network where some nodes are removed in an
optimal way ($n_i=0$). According to the Perron-Frobenius theorem, the
largest eigenvalue of the matrix $\Bia$ is a strictly decreasing
function of $\Bia$, that is if $\Mia\leq\Bia$ (meaning that the
inequality holds entry by entry of the matrices) and $\Mia\neq\Bia$,
then $\lambda(\Mia)<\lambda(\Bia)$. In our case the matrix $\Mia$ can
be obtained from the matrix $\Bia$ by setting one or some of the
$n_i=0$.  Therefore the optimization problem for a given $q$ can be
rephrased as finding the optimal influencer configuration
$\mathbf{n}^*$ that minimize $\lambda(\mathbf{n};q)$ over all possible
configurations $\mathbf{n}$ satisfying $\la n \ra = 1 - q$.  Calling
$\lambda(\mathbf{n}^*; q)$ this minimum, we write: \be
\lambda(\mathbf{n}^*; q)\ \equiv\ \min_{\mathbf{n}:\la n\ra = 1-q}
\lambda(\mathbf{n};q)\ .  \ee The optimal threshold $q_c$ is the
solution of the equation: \be \lambda(\mathbf{n}^*; q_c)\ =\ 1.  \ee

Still, this equation is hard to solve, since there is no explicit
formula for $\lambda(\mathbf{n};q)$ as a function of $\mathbf{n}$. To
tackle this problem we propose a sequence of approximations to the
largest eigenvalue (and associated eigenvector) which is based on the
Power Method iterative scheme that we find quickly convergent to the
exact solution of the problem.  We stress that the
  optimal exact solution for $\ell\to \infty$ holds only under the
  assumption that the graph under consideration is locally-tree like.

Before to conclude this section, we notice that for the network
depicted in \eqref{eq:nb}, the modified NB matrix does not depend on
$n_1,n_4,n_6$, i.e., it does not depend on the variables outside the
loop. As a consequence, its largest eigenvalue does not depend on
those variables as well: $\lambda=\lambda(n_2,n_3,n_5)$.  Therefore,
removing the nodes at the end of the dangling edge does not reduce the
eigenvalue $\lambda$, which is one for the considered network:
$\lambda(1,1,1)=1$. On the contrary, by removing a node belonging to
the loop, e.g. nodes $2,3$ or $5$, the network becomes a tree, and the
eigenvalue is zero: $\lambda(0,1,1)=\lambda(1,0,1)=\lambda(1,1,0)=0$.
In general, the largest eigenvalue of a tree-network is equal to zero.
For networks with one loop (unicyclic graph), the largest eigenvalue
is equal to one, and for networks with many loops $\lambda$ is larger
than $1$. Thus we see that the result of minimizing the largest
eigenvalue of the modified NB matrix is a way to attack the loops in
the network.  When the eigenvalue reaches the critical value one
($\lambda=1$) the network consists of a single loop, that is suddenly
destroyed by the removal of a single node, causing the sharp drop of
the eigenvalue from one to zero. When the giant component is reduced
to a tree-like topology, it can be considered as completely
fragmented, since any tree can be destroyed with a subextensive number
of node removals.  Thus, our theory suggests that the best attack
strategy is to destroy the loops. This process is illustrated in
Fig. \ref{fig:explanation}c.

\subsection{Mapping of optimal immunization and spreading problems onto optimal percolation}
\label{mapping}

Here we map exactly the problems of optimal immunization and spreading
to the problem of minimizing the giant component of a network, ie,
optimal percolation.

In the immunization case, the quantity $\nu_i$ represents the
probability of node $i$ to be infected. Therefore minimizing the sum
$\sum_i\nu_i$ is equivalent to minimize the size of the disease
outbreak, which is obtained by optimally choosing the immunizator
nodes. These immunizators are exactly the nodes we need to remove to
fragment the network. Precisely, $n_i=0$ if node $i$ is an optimal
immunizator, and $n_i=1$ if not.  Note that we can also slightly
modify the equations to include the case of a transmission probability
of the disease $w$ smaller than one (the case treated explicitly in
our paper corresponds to $w=1$).  Including $w$, the main equation
reads: \be \nu_{i\to j}\ =\ n_i\left[1-\prod_{k\in\partial i\setminus
    j}(1-w\nu_{k\to i})\right]\ .
\label{eq:immunization_with_w}
\ee In this case the optimal immunizators can be still identified by
minimizing the largest eigenvalue of the modified NB operator $\Mia$.
The only difference is that the critical threshold $q_c$ is defined by
the following equation: \be \lambda(\mathbf{n}^*; q_c)\ =\ 1/w .  \ee
It should be noted, though, that the quantity $N^{-1}\sum_i\nu_i$,
which quantifies the total fraction of infected individuals, is
different from the giant component $G$ when $w < 1$.

The case of optimal spreaders is the dual of the optimal immunizators.
The optimized spreading problem \cite{kempe} consists of finding the
minimum number of nodes to "activate" in such a way that the
information percolates the network following, for instance, the Linear
Threshold Model (LTM) dynamics \cite{LTM,watts}.  The LTM simulates a
spreading of information process where an individual adopts an opinion
or information under ``peer pressure'', that is, it becomes activated
only after a given number of its neighbors are
\cite{LTM,watts,centola,korniss}.  The optimal influence threshold
represents the minimum fraction of spreader nodes we need to activate
to spread the information all over the network.  The mathematical
formulation of the problem is the following. Let us define $\nu_{i\to
  j}$ as the probability that, in the information spreading process,
node $i$ is eventually NOT activated in absence of node $j$. Moreover,
we assign to each node the number $n_i$, which, in this case, equals
zero, $n_i=0$, if $i$ is an initial spreader, and $n_i=1$ if not. We
note that in the case of immunization, $n_i=0$ corresponds to an
immunized node. Following the Linear Threshold Model (LTM) \cite{LTM}
of information spreading, in order for node $i$ to be activated in
absence of node $j$, the neighbouring nodes such that $k\in\partial
i\setminus j$ must be activated.  In the Linear Threshold Model
\cite{LTM,watts}, this situation corresponds to considering uniform
weights $w_{ij}=1$ for each edge $(i,j)$, and threshold of activation
$\theta_i=k_i-1$ for each node with degree $k_i$, such that a node $i$
becomes activated if the number of active neighbors is at least
$\theta_i$. The optimal spreading problem under the LTM is to find the
minimum set of initially activated spreaders, $q_c$, which will
percolate the information to the entire network, as defined in Kempe
{\it et al.} \cite{kempe}.  On the other limit, the optimal
immunizator problem corresponds to setting $\theta_i=1$, further
showing the dual relation between the immunization and spreading
problem.

The probabilities $\nu_{i\to j}$ for spreading can be computed
self-consistently through the following message passing equations
\cite{bianconi,lenka,montanari} in analogy to Eq. (\ref{eq:CavityMess}): \be
\nu_{i\to j}\ =\ n_i\left[1-\prod_{k\in\partial i\setminus
    j}(1-\nu_{k\to i})\right]\ .
\label{eq:spreader1}
\ee 

The total probability $\nu_i$ that $i$ is not activated is obtained
from Eq. (\ref{eq:spreader1}) by including the contribution of
$\nu_{j\to i}$ as well: 

\be \nu_{i}\ =\ n_i\left[1-\prod_{k\in\partial i}(1-\nu_{k\to
    i})\right]\ ,
\label{eq:spreader2}
\ee and the total fraction of nodes not activated in the spreading
process is: \be G\ =\ \frac{1}{N}\sum_{i=1}^{N}\nu_i\ .  \ee 

The optimization problem in spreading under the LTM consists in
minimizing the fraction of inactive nodes $G$, or equivalently,
maximizing the spreading over the network.  We notice the dual nature
of $G$ in immunization and spreading. In the former, $G$ represents
the fraction of inactive nodes, which we want to minimize to maximize
spreading. In the later, $G$ represents the connected component of
individuals who would get infected if the epidemic starts in a single
node. In this case, we also want to minimize $G$ to reduce the spread
of the epidemic. This is the basic reason why we are able to bring the
two problems under the same framework of optimal percolation.

In activated spreading, the minimization of $G$ is achieved by
optimally placing the initial spreaders $n_i=0$.  Equations
\eqref{eq:spreader1} and \eqref{eq:spreader2} are formally identical
to the equations of optimal influence \eqref{eq:CavityMess},
\eqref{eq:RealMess}. To summarize, in spreading, the giant component
$G$ represents the fraction of inactivated nodes. Therefore, by
minimizing the giant component in spreading we are effectively
maximizing the spreading of information. This completes the mapping
between the immunization and spreading problem to optimal percolation.

We notice that the mapping of the optimal LTM problem to optimal
percolation is done for a threshold $\theta_i=k_i-1$.  For
$\theta_i=1$, we recover the optimal immunization problem.  These two
problems can be solved by studying the local stability of the solution
$G=0$ in both cases. This is possible since at $q_c$ the transition is
of second order, and thus a local stability theory is applicable.  For
other intermediate values of the threshold $\theta_i=k_i-2, k_i-3,
\dots 2$, the transition at $q_c$ is of first order, of the type of
bootstrap percolation \cite{bootstrap}.  In these regimes, a local
stability criterion cannot be applied, and the optimal spreading
problem in these cases needs to be considered independently.  That is,
the largest eigenvalue of the NB operator is not guaranteed to provide
the optimal set when the transition is of first order.  Nevertheless,
we could expect that the proposed CI algorithm may work as well in
this regime of discontinuous transition. Another interesting
generalization of the present study is the case of randomly chosen
heterogeneous threshold $\theta_i$ in the LTM \cite{watts} or a fix
threshold \cite{korniss}.

A further interesting optimization problem is to find the optimal
spreaders under the SIR or SIS models \cite{gallos}. In general, this
problem cannot be translated into an optimal percolation problem,
because it does not have a transition from $G=0$ to $G>0$. Therefore,
it cannot be treated under our stability theory.  However, as for the
LTM with intermediate values of the threshold, the spreaders
identified by the CI algorithm could still be expected to be close to
optimal.

To conclude this section, we notice that for all optimization problems
on locally tree-like random networks that can be mapped onto an
optimal percolation problem, with a second order transition separating
the phases with $G=0$ and $G>0$, our theory holds true, and the CI
algorithm can be used accordingly.

In the next sections, we show that the optimal percolation set is
obtained as a (infinite) sequence of optimized ``attacks'' expressed
as successive approximations to the minimization of the largest
eigenvalue of the modified non-backtracking matrix. At the most
trivial level, we obtain the random attack of the network
corresponding to random percolation, Eq. S\ref{qran}. The zero-order
naive strategy corresponding to the lowest approximation of the theory
is given by the attack on the high degree nodes,
Eq. S\ref{eq:HDinfiniteCutoff}.  The first non-trivial collective
attack strategy is equivalent to find the ground state of a two-body
spin-glass like system, i.e., an anti-ferromagnetic Ising model with
random bonds in a random external field at fixed magnetization,
Eq. S\ref{eq:antiferromagnet}. Higher-order approximations consist of
increasing many-body problems, Eq. S\ref{eq:Costfunction2}, and
produce increasingly better performance compared to previous heuristic
strategies.

\subsection{  Limit of applicability of the
    theory of influence }
\label{limit}

The mapping of the optimal influence problem onto the optimal
percolation problem developed so far is strictly valid for locally
tree-like networks: in the message passing formulation
Eq. (\ref{eq:CavityMess}), the probabilities $\nu_{k\to i}$ are
assumed to be independent. This includes the thermodynamic limit of
the class of random networks of Erd\"{o}s-R\'enyi, scale-free networks
\cite{newman-book} and the configuration model (the maximally random
graphs with a given distribution \cite{configurational}) which are
locally tree-like and contains loops that grow as $\log N$
\cite{doro2}. Nonetheless, it is generally accepted, and confirmed by
many works, that results obtained for tree-like graphs apply quite
well also for loopy networks, provided the density of loops is not
excessively large \cite{zecchina2,florant,montanari}. Wherever the
number of such topological structures (loops) is abundant (e.g., to
finite dimensional lattices), the quality of the results obtained for
tree-like networks deteriorates. Indeed, the locally tree-like
approximation has been successfully used for a plethora of other
problems, like spin glass models on random graphs, coloring, matching,
bisection and maximum cut of graphs, and many other constraint
satisfaction problems \cite{montanari}.

\subsection{Random influence}
\label{zero}

The trivial case corresponds to random removal of nodes
(random percolation). It is obtained by taking the $n_i$ at random
from the distribution $P(\mathbf{n};q)$ such that the removal is
decoupled from the non-backtracking matrix:

 \be P(\mathbf{n};q)\ =\ \prod_{i=1}^{N}(1-q)^{n_i}q^{1-n_i}\ .  \ee
 Taking the expectation of the matrix $\mathcal{M}$
over $\mathbf{n}$, and exploiting the fact that $P(\mathbf{n};q)$ is factorized
over the sites, Eq. \eqref{eq:modifiedNB} becomes:
\be
\mathbb{E}_{\mathbf{n}}\ \mathcal{M}_{i\to j, k\to
  \ell}\ =\ (1-q)\mathcal{B}_{i\to j, k\to \ell}\ . 
\ee
Therefore, the eigenvalue $\lambda(\mathbf{n};q)$, averaged over $\mathbf{n}$,
is given by:
\begin{equation}
\lambda(q) = (1-q)\lambda_{\Bia},
\end{equation} 
where $\lambda_{\Bia}$ is the largest eigenvalue of the NB matrix
$\Bia$ for a random network. It is well known that the largest
eigenvalue of the NB matrix is equal to \cite{florant}:
\begin{equation}
\lambda_{\Bia}=\kappa-1,
\end{equation}
with $\kappa$ equals to the ratio of the first two moments of the
degree distribution:
\begin{equation}
\kappa=\la k^2\ra/\la k\ra.
\end{equation} 
The condition $\lambda(q_c)=1$ is nothing but the famous result for
random percolation \cite{bollobas} which has been obtained using the
NB matrix in \cite{lenka,radicchi}, i.e.:

\begin{equation}
q^{\rm ran}_c=1-(\kappa-1)^{-1}.
\label{qran}
\end{equation}

Thus, when $\mathbf{n}$ is decoupled with the NB
  matrix, the largest eigenvalue of NB captures the random percolation
  threshold for random networks. This result has been previously
  obtained in \cite{lenka} using similar ideas as used in the present
  derivation. The largest eigenvalue of the NB matrix in case of
  random removal of nodes is true only when the original graph is
  random. For a generic graph, the largest eigenvalue can be very
  different and not related to the first and second moment of the
  degree distribution.  On the other hand, the optimal threshold
arises by coupling the removal of nodes with the NB matrix, a case
that is treated next.

\subsection{Derivation of the main formula: cost energy function of influence, Eq. (\ref{eq:Costfunction})}
\label{mainformula}

Next, we derive Eq. (\ref{eq:Costfunction}) which
  holds only on very large locally tree-like graphs.  From now on we
omit $q$ in $\lambda(\mathbf{n};q)\equiv\lambda(\mathbf{n})$, which is
always kept fixed. For a given configuration $\mathbf{n}$, the
eigenvalue $\lambda(\mathbf{n})$ determines the growth rate of an
arbitrary nonzero vector $\mathbf{w}_0$ after $\ell$ iterations of the
matrix $\Mia$, provided that $\mathbf{w}_0$ has nonzero projection
onto the eigenvector corresponding to $\lambda(\mathbf{n})$. Denoting
with $\mathbf{w}_{\ell}(\mathbf{n})$ the vector at the $\ell$-th
iteration, \be \mathbf{w}_{\ell}(\mathbf{n})=\Mia^{\ell}\mathbf{w}_{0},
\ee we can write according to Power Method \cite{eigenvalue}:

\be \lambda(\mathbf{n})\ =\ \lim_{\ell\to\infty}\ \left[
  \frac{|\mathbf{w}_\ell(\mathbf{n})|}{|\mathbf{w}_0|}
  \right]^{1/\ell}\ , \ee where \be
|\mathbf{w}_\ell(\mathbf{n})|^2=\la\mathbf{w}_\ell(\mathbf{n})|\mathbf{w}_\ell(\mathbf{n})\ra=\la\mathbf{w}_0|(\Mia^{\ell})^{\dagger}\Mia^{\ell}|\mathbf{w}_0\ra.
\ee

For finite $\ell$ we define the $\ell$-dependent approximant
$\lambda_\ell(\mathbf{n})$ as: \be \lambda_\ell(\mathbf{n})=\left[
  \frac{|\mathbf{w}_\ell(\mathbf{n})|}{|\mathbf{w}_0|}
  \right]^{1/\ell}\ ,
\label{eq:approximant}
\ee so that we have:
\be
\lambda(\mathbf{n})\ =\ \lim_{\ell\to\infty}\lambda_\ell(\mathbf{n}).
\ee
We now derive the analytical expression of $\lambda_\ell(\mathbf{n})$.

We start by computing the first approximant $\ell=1$: \be
|\mathbf{w}_1(\mathbf{n})\ra=\Mia|\mathbf{w}_0\ra.  \ee In order to do
this, it is convenient to embed the matrix $\Mia$, whose dimension is
$2M\times2M$, in a larger space of dimension $N\times N\times N \times
N$. In this enlarged space $\Mia$ is given by: \be
\mathcal{M}_{ijk\ell}\ =
\ n_kA_{ij}A_{k\ell}\delta_{jk}(1-\delta_{i\ell})\ , \ee where each
index runs from $1$ to $N$: $i,j,k,\ell=1,\dots,N$, and $A_{ij}$ is
the adjacency matrix. Practically, we have represented $\Mia$ on the
nodes of the network, rather than on the directed edges. The Kronecker
deltas guarantee the non-backtracking nature of the NB walks.

As starting $2M$-dimensional vector $|\mathbf{w}_0\ra=|\mathbf{1}\ra$
in the space of links, we choose the vector with all components equal
one; the optimal solution is independent of this selection. The
components of the analogous $N\times N$ vector, $|\mathbf{w}_0\ra$ in
the enlarged space of nodes are given by $|w_0\ra_{ij}=A_{ij}$.

The right vector $|\mathbf{w}_1(\mathbf{n})\ra$ is computed as: \be
|w_1(\mathbf{n})\ra_{ij}=\sum_{k\ell}\mathcal{M}_{ijk\ell}\ |w_0\ra_{k\ell}\ =
n_jA_{ij}(k_j-1)\ , \ee while the left vector
$\la\mathbf{w}_1(\mathbf{n})|$ is given by: \be _{ij}\la
w_1(\mathbf{n})|=\sum_{k\ell}\ _{k\ell}\la w_0|\mathcal{M}_{k\ell
  ij}\ = n_iA_{ij}(k_i-1)\ .  \ee The factor $k_i-1$ will appear
frequently in the following, so it is worth to set the residual
degree:\be z_i\ \equiv k_i-1\ .  \ee

The norm $|\mathbf{w}_1(\mathbf{n})|^2$ is: 

\be
|\mathbf{w}_1(\mathbf{n})|^2\ = \sum_{ij}\ _{ij}\la
w_1(\mathbf{n})|w_1(\mathbf{n})\ra_{ij}
=\sum_{ij}A_{ij}z_iz_jn_in_j.
\label{norm1}  
\ee 

Since the norm of $|\mathbf{w}_0\ra$ is simply:
$|\mathbf{w}_0|^2=\sum_ik_i = 2 M$, we can write the final expression for
$\lambda_1(\mathbf{n})$ from Eq. (\ref{eq:approximant}) as: \be
\lambda_1(\mathbf{n})\ =\ \left[
  \frac{1}{2 M}\sum_{ij}A_{ij}z_iz_jn_in_j \right]^{1/2}.
\label{eq:2body}
\ee 

It is useful to give a graphical representation of the previous
formula in terms of a diagrammatic expansion, which becomes
indispensable for higher orders of the iteration. The interaction term
in the sum in the numerator on the r.h.s of Eq. (\ref{eq:2body}) can
be represented as
\begin{fmffile}{w1}
\begin{eqnarray}
A_{ij}(k_i-1)(k_j-1)n_in_j\ =\ \ \ \ \  \  &\ &\ 
\parbox{10mm}{
\begin{fmfgraph*}(15,3)
\fmfdotn{i}{1}
\fmfdotn{o}{1}
\fmfleft{i1,i2}
\fmfright{o1,o2}
\fmflabel{$n_i$}{i1}
\fmflabel{$n_j$}{o1}
\fmf{photon}{i2,i1}
\fmf{fermion}{i1,o1}
\fmf{photon}{o2,o1}
\fmflabel{$z_j$}{o2}
\fmflabel{$z_i$}{i2}
\end{fmfgraph*}
} 
\ \ \ \ \ \ \ \ \ 
\label{d1}
\end{eqnarray} 
\end{fmffile}
The meaning of the diagram is the following: each time a straight line
connects two sites $i$ and $j$, the variables $n_i$ and $n_j$ are
multiplied by each other (the meaning of the arrow is explained later
in Methods Section \ref{NBinterpretion} in terms of NB walks; for the
moment it can be thought of as an undirected line).  The wiggly lines
on the nodes means a multiplication by the factor $z_i=k_i-1$. The
diagram is then equal to $n_in_jz_iz_j$. The number of variables
appearing in the diagram gives the order of the interaction. In this
case, corresponds to a pair-wise interaction. Thus, the first
optimization order $\ell=1$ corresponds to a 2-body problem. We will
see that, in general, the $\ell$-order term in the approximant
describes a $2\ell$-body problem.

Let us compute for $\ell=2$, $|\mathbf{w}_2(\mathbf{n})\ra$: \be
|w_2(\mathbf{n})\ra_{ij}=\sum_{k\ell}\mathcal{M}_{ijk\ell}\ |w_1(\mathbf{n})\ra_{k\ell}\ =
n_jA_{ij}\sum_{\ell}A_{j\ell}n_{\ell}z_{\ell}(1-\delta_{i\ell}) , \ee
and also $\la\mathbf{w}_2(\mathbf{n})|$: \be _{ij}\la
w_2(\mathbf{n})|=\sum_{k\ell}\ _{k\ell}\la
w_1(\mathbf{n})|\mathcal{M}_{k\ell ij} =
n_iA_{ij}\sum_{k}A_{ik}n_{k}z_{k}(1-\delta_{kj})\ .  \ee The norm
$|\mathbf{w}_2(\mathbf{n})|^2$ is given by: \be
|\mathbf{w}_2(\mathbf{n})|^2=
\sum_{ijk\ell}A_{ij}A_{jk}A_{k\ell}(1-\delta_{ik})(1-\delta_{j\ell})z_iz_{\ell}
n_in_jn_kn_{\ell}.
\label{eq:4body}
\ee

There are two types of interactions in the sum on the r.h.s of
Eq. \eqref{eq:4body}: a $4$-body and a $3$-body interaction. The
graphical representation of the former is:

\begin{fmffile}{b4}
\begin{eqnarray}
 A_{ij}A_{jk}A_{k\ell}z_iz_{\ell}n_in_jn_kn_{\ell}=\ &\ \ \ \ \
\parbox{10mm}{
\begin{fmfgraph*}(35,7)
\fmfdotn{i}{1}
\fmfdotn{o}{1}
\fmfleft{i1,i2}
\fmfright{o1,o2}
\fmf{photon}{i2,i1}
\fmf{photon}{o2,o1}
\fmf{fermion}{i1,v1,v2,o1}
\fmfdotn{v}{2}
\fmfiv{l=$n_j$,l.a=50,l.d=-.30w}{c}
\fmfiv{l=$n_k$,l.a=132,l.d=-.30w}{c}
\fmflabel{$n_i$}{i1}
\fmflabel{$n_\ell$}{o1}
\fmflabel{$z_i$}{i2}
\fmflabel{$z_\ell$}{o2}
\end{fmfgraph*}
}\ \ \ \ \ \ \ \ \ \ \ \  \ \ \  
\label{4body}
\end{eqnarray}
\end{fmffile}

The diagram for the $3$-body interaction is:

\begin{fmffile}{b3x}
\begin{eqnarray}
 A_{ij}A_{jk}A_{ki}(z_i)^2n_in_jn_k=\ &\ \ \ \ \ \ \ \ \ \ \ \ \ 
\parbox{5mm}{
\begin{fmfgraph*}(20,20)
\fmftop{i1}
\fmfbottom{i2,i3}
\fmflabel{$n_i$}{i1}
\fmflabel{$n_j$}{i2}
\fmflabel{$n_k$}{i3}
\fmf{fermion}{i1,i2,i3,i1}
\fmffreeze
\fmfright{o1}
\fmfleft{o2}
\fmf{photon, left=.5}{i1,o1}
\fmf{photon,left=.5}{o2,i1}
\fmflabel{$z_i$}{o1}
\fmflabel{$z_i$}{o2}
\fmfdotn{i}{3}
\end{fmfgraph*}
} \ \ \ \ \ \ \ \ \ \ \ \ \ \ \ \ 
\label{3body}
\end{eqnarray}
\end{fmffile}

Then, the $\ell=2$ approximant of the eigenvalue,
$\lambda_2(\mathbf{n})$, is given by: \be \lambda_2(\mathbf{n})=
\left[ \frac{1}{2M}
  \sum_{ijk\ell}A_{ij}A_{jk}A_{k\ell}(1-\delta_{ik})(1-\delta_{j\ell})z_iz_{\ell}
  n_in_jn_kn_{\ell} \right]^{1/4}.  \ee

In the next order $\ell=3$, $|\mathbf{w}_3(\mathbf{n})|^2$ there
appears a term with $6$-body interactions:
\begin{fmffile}{b6}
\begin{eqnarray}
A_{ij}A_{jk}A_{k\ell}A_{\ell m}A_{mp}z_iz_{p}
n_in_jn_kn_{\ell}n_mn_p = \nonumber \\
=\parbox{10mm}{
\begin{fmfgraph*}(35,5)
\fmfdotn{i}{1}
\fmfdotn{o}{1}
\fmfleft{i1,i2}
\fmfright{o1,o2}
\fmf{photon}{i2,i1}
\fmf{photon}{o2,o1}
\fmf{fermion}{i1,v1,v2,v3,v4,o1}
\fmfdotn{v}{4}
\end{fmfgraph*}
}\ \ \ \ \ \ \ \ \ \ \ \ \  \ \ \ \ \ \ \ \ \ , \label{diagram:NBwalk3body}
\end{eqnarray}
\end{fmffile}
terms with $5$-body interactions:
\begin{fmffile}{b5}
\begin{eqnarray}
\label{eb5}
\ \ \ \ \ \ \ \  \ \ \ \ \ \  
\parbox{10mm}{
\begin{fmfgraph*}(20,8)
\fmfleft{i1}
\fmftop{i2}
\fmfbottom{i3}
\fmfright{o1,o2}
\fmf{fermion}{i1,v1,v2}
\fmf{fermion,left=-.4}{v2,o1}
\fmf{fermion,left=-1}{o1,o2}
\fmf{fermion,left=-.4}{o2,v2}
\fmffreeze
\fmf{photon, left=.5}{i1,i2}
\fmf{photon}{v2,i3}
\fmfdotn{i}{1}
\fmfdotn{v}{2}
\fmfdotn{o}{2}
\end{fmfgraph*}
}\ \ \ \ \ \ \ \ \ \ \ \ \ \ \ \ \ \ \ \ \ \ \ \nonumber 
\parbox{10mm}{
\begin{fmfgraph*}(20,8)
\fmfleft{i1,i2}
\fmfright{o1,o2}
\fmftop{f3}
\fmfbottom{f4}
\fmf{fermion}{i1,v1}
\fmf{fermion, left=-.5}{v1,o1,o2,v1}
\fmf{fermion}{v1,i2}
\fmffreeze
\fmf{photon, left=.5}{f4,i1}
\fmf{photon, left=-.5}{f3,i2}
\fmfdotn{i}{2}
\fmfdotn{v}{1}
\fmfdotn{o}{2}
\end{fmfgraph*}
}\ \ \ \ \ \ \ \ \ \ \ \ \ \ \ \ \ \ \ \ \ \nonumber \\ \nonumber
\\ &        \nonumber\\
\parbox{10mm}{
\begin{fmfgraph*}(20,8)
\fmfleft{i1}
\fmfright{o1,o2,o3}
\fmftop{i2}
\fmfbottom{i4}
\fmf{fermion}{i1,v1}
\fmf{fermion,left=-.5}{v1,o1}
\fmf{fermion,left=-.3}{o1,o2}
\fmf{fermion,left=-.3}{o2,o3}
\fmf{fermion,left=-.5}{o3,v1}
\fmffreeze
\fmf{photon, left=-.5}{i2,i1}
\fmf{photon}{i4,v1}
\fmfdotn{i}{1}
\fmfdotn{v}{1}
\fmfdotn{o}{3}
\end{fmfgraph*}
}\ \ \ \ \ \ \ \ \ \ \ \ \ \ \ \ \ \ \ \ \ \ \ \nonumber 
\parbox{10mm}{
\begin{fmfgraph*}(20,8)
\fmftop{i1}
\fmfright{o1,o2,o3}
\fmfleft{i2,i4}
\fmfbottom{b1}
\fmf{fermion, left=-.4}{i1,b1,o1}
\fmf{fermion, left=-.2}{o1,o2}
\fmf{fermion, left=-.2}{o2,o3}
\fmf{fermion, left=-.4}{o3,i1}
\fmffreeze
\fmf{photon,left=.25}{i2,i1}
\fmf{photon, left=.5}{i4,i1}
\fmfdotn{i}{1}
\fmfdotn{o}{3}
\fmfdotn{b}{1}
\end{fmfgraph*}
}\ \ \ \ \ \ \ \ \ \ \ \ \ \ \ \ \ \ \ \ \   \nonumber\\ \nonumber
\end{eqnarray}
\end{fmffile}
terms with $4$-body interactions:
\begin{fmffile}{bb4}
\begin{eqnarray}
\label{eb4}
\ \ \ \ \ \ \ \  \ \ \ \ \ \ 
\parbox{10mm}{
\begin{fmfgraph*}(20,8)
\fmfleft{i1}
\fmftop{i2}
\fmfbottom{i3}
\fmfright{o1,o2}
\fmf{fermion}{i1,v1}
\fmf{fermion, left=-.5}{v1,o1,o2,v1}
\fmffreeze
\fmf{photon, left=.5}{i1,i2}
\fmf{photon, left=-.5}{i3,o1}
\fmfdotn{i}{1}
\fmfdotn{v}{1}
\fmfdotn{o}{2}
\end{fmfgraph*}
}\ \ \ \ \ \ \ \ \ \ \ \ \ \ \ \ \ \ \ \ \ \ \ \nonumber
\parbox{10mm}{
\begin{fmfgraph*}(20,8)
\fmfleft{i1}
\fmftop{i2}
\fmfbottom{i3}
\fmfright{o1,o2}
\fmf{fermion}{i1,v1}
\fmf{fermion, left=.5}{v1,o2,o1,v1}
\fmffreeze
\fmf{photon, left=.5}{i3,i1}
\fmf{photon, left=-.5}{i2,i1}
\fmfdotn{i}{1}
\fmfdotn{v}{1}
\fmfdotn{o}{2}
\end{fmfgraph*}
}\ \ \ \ \ \ \ \ \ \ \ \ \ \ \ \ \ \ \ \ \ \ \ \nonumber
\\ &        \nonumber\\ \nonumber \\ 
\parbox{10mm}{
\begin{fmfgraph*}(16,7)
\fmfleft{i1,i2}
\fmfbottom{i3}
\fmftop{i4}
\fmfright{o1,o2}
\fmf{fermion, left=.5}{i1,i2,i4,i3,i1}
\fmffreeze
\fmf{photon}{o1,i3}
\fmf{photon}{o2,i4}
\fmfdotn{i}{4}
\end{fmfgraph*}
}\ \ \ \ \ \ \ \ \ \ \ \ \ \ \ \nonumber 
\parbox{10mm}{
\begin{fmfgraph*}(15,5)
\fmfbottom{i1,i2}
\fmftop{i3,i4}
\fmf{fermion, left=-.5}{i1,i2,i4,i3,i1}
\fmf{fermion}{i1,i4}
\fmfdotn{i}{4}
\fmffreeze
\fmftop{f1}
\fmfbottom{f2}
\fmf{photon}{f1,i1}
\fmf{photon}{f2,i4}
\end{fmfgraph*}
}\ \ \ \ \ \ \ \ \ \ \ \ \ \ \ \ \ \ \ \ \   
\end{eqnarray}
\end{fmffile}
and a term with $3$-body interaction:
\begin{fmffile}{bb3}
\begin{eqnarray}
\parbox{10mm}{
\begin{fmfgraph*}(20,10)
\fmfbottom{i1,i2}
\fmftop{i3}
\fmf{fermion}{i1,i2,i3,i1}
\fmfdotn{i}{3}
\fmffreeze
\fmfleft{o1}
\fmfright{o2}
\fmf{photon}{o1,i1}
\fmf{photon}{o2,i2}
\fmflabel{$i$}{i2}
\fmflabel{$j$}{i1}
\end{fmfgraph*}
}\ \ \ \ \ \ \ 
\label{d3}
\end{eqnarray}
\end{fmffile}

We see that the series expansion of the maximum eigenvalue can be
written in terms of a systematic diagrammatic expansion of increasing
levels of many-body interactions. The generalization is treated next.

\subsection{Interpretation in terms of NB walks and generalization to $2\ell$-body interactions}
\label{NBinterpretion}

Nodes entering in each type of interaction given by
$|\mathbf{w}_1(\mathbf{n})|^2$, $|\mathbf{w}_2(\mathbf{n})|^2$, and
$|\mathbf{w}_3(\mathbf{n})|^2$ are the same nodes visited by a
non-backtracking walk of length $1,3,5$, respectively (with the
possibility of traversing an edge multiple times). In general, the
diagrammatic expansion of the term $|\mathbf{w}_\ell(\mathbf{n})|^2$
will then contain all the possible graphs that can be built using the
nodes traversed by a NB-walk of length $2\ell-1$. This is the reason
why we put an arrow in the diagrammatic schematization of the
interaction terms: emphasizing the connection with NB walks.

The NB graphs in the expansion are built in the following way: (a) For
a given $\ell$ we construct all NB walks of $2\ell-1$ steps starting
at one node $i$ (with degree $k_i$) and ending in node $j$ (with
degree $k_j$).  The initial and ending point are indicated by a wiggly
line indicating their degree minus one.  (b) The initial and final node of the NB
walk do not need to be necessarily different.  (c) Loops are allowed in
the NB walk.  Thus, the shortest path between $i$ and $j$ might be
smaller than the $2\ell-1$ steps of the walk.  (d) We recall that the
condition for a NB random walk is only that it cannot come back
through the same link that it just came on, yet, it can visit the same
node several times.  (e) As well, the NB walks are allowed to travel
through the same links multiple times.  (f) The number of nodes of the
NB walk is the order of the interaction. (g) The dominant graph is
always a direct path (line) of length $2\ell-1$ and $2\ell$ nodes,
where the shortest path between the initial and ending node is
$2\ell-1$; we will see that all the other diagrams with loops are
negligible for sparse locally tree-like random graphs.

For instance, in the case $\ell=2$, we obtain two graphs representing
NB walks of length $2\ell-1=3$. The graph (\ref{4body}) represents a
NB walk of length 3 which traverses 4 distinct nodes, and thus it is a
4-body dominant interaction. The graph (\ref{3body}) also represents a
NB walk of length 3, but this time the NB walk starts and ends in the
same node $i$, rule (b) and (c), so this term contributes with a factor
$z_i^2$.

In the case $\ell=3$, the leading interaction is graph (\ref{eb5}) of
6-body interactions and 5 NB steps. The 3-body interaction (\ref{d3})
is a NB walk of 5 steps that starts at the node $i$ on the right and
traverses two links twice, rule (e), to end up in node $j$, resulting
in a triangular 3-body problem.

Next, we show that, in a sparse random network, all the NB graphs with
loops can be neglected to $O(1/N)$, in accordance with the assumption
of locally-tree like structure. Thus, the main equation
(\ref{eq:Costfunction}) takes into account only the leading
$2\ell$-body interaction for each $\ell$-level in the diagrammatic
expansion to $O(1/N)$. This situation is translated to the CI
algorithm where the ball is defined by a radius $\ell$ defined as the
shortest path between two nodes in the network.

It is important to note that, for locally tree-like graphs and for
very large system sizes $N$, the terms containing one or more loops
are suppressed by powers of $1/N$.  This can be understood through the
following argument.  Let us consider, for the sake of simplicity, an
ER random graph, where each edge is present with probability $z/N$,
where $z$ is the average degree (in ER the average degree is the same
as the average residual degree). The total number of loops in ER of
given length $\ell$ is given on average by (the final formula is valid
for general sparse random graphs \cite{configurational}, the quantity
$z$ being in general the average residual degree): \be
\mathcal{N}(\ell)\ =\ \left(\frac{z}{N}\right)^{\ell}\frac{1}{2\ell}\ N(N-1)\dots
(N-\ell+1) \sim \frac{z^{\ell}}{2\ell}\ , \ee where the factor
$1/2\ell$ comes from the fact that: i) anyone of the $\ell$ nodes in
the loop can be taken as starting point; ii) the loop can be traveled
in two directions.  Therefore, the probability $p(\ell)$ for a node to
belong to a loop of length $\ell$ is: \be p(\ell) =
\frac{\mathcal{N}(\ell)}{N} = \frac{z^{\ell}}{2\ell N}\ , \ee and thus
it is of order $O(1/N)$. The quantity $\mathcal{N}(\ell)$ enumerates
the number of non self-intersecting closed walks.  Self-intersecting
closed walks have a probability of order $O(1/N^2)$, since the
self-intersection is encountered, on average, in a fraction $1/N^2$ of
the total number of nodes. Thus, loops with many self-intersections
are suppressed by higher power of $1/N$.  This argument can be
generalized to random graph ensembles other than ER, provided they
have a locally tree-like structure in the limit $N\to\infty$.

As a consequence, for very large network, the leading term of
$|\mathbf{w}_\ell(\mathbf{n})|^2$ for a given $\ell$ (i.e., the one
containing a number of nodes exactly equal to $2\ell$, for instance
the diagram \eqref{diagram:NBwalk3body} in the expression of
$|\mathbf{w}_3(\mathbf{n})|^2$) is already a good approximation, which
becomes exact for $N\to\infty$.  However, for small networks all the
terms should be considered. In Methods Section \ref{sec:EO}, we will minimize
the eigenvalue $\lambda_\ell(\mathbf{n})$ by taking into account all
the possible interactions.

Since, in the limit $N\to\infty$, loops do not contribute to the
assessment of the norm $|\mathbf{w}_\ell(\mathbf{n})|$, the equation
for $|\mathbf{w}_\ell(\mathbf{n})|$ simplifies considerably, because
we can take into account only the terms with exactly $2\ell$-body
interactions; the remaining ones with loops decay as $1/N$ or
faster. 


The analytical expression for $|\mathbf{w}_1(\mathbf{n})|^2$ and $|\mathbf{w}_2(\mathbf{n})|^2$
given by Eqs. \eqref{norm1}, \eqref{eq:4body}
can be generalized to any 
$|\mathbf{w}_\ell(\mathbf{n})|^2$. We get
 \be
|\mathbf{w}_\ell(\mathbf{n})|^2=
\sum_{i_1i_2\dots i_{2\ell}}
A_{i_1i_2}A_{i_2i_3}\dots A_{i_{2\ell-1}i_{2\ell}}
(1-\delta_{i_1i_3})(1-\delta_{i_2i_4})\dots(1-\delta_{i_{2\ell-2} i_{2\ell}})
z_{i_1}z_{i_{2\ell}}
n_{i_1}n_{i_2}\dots n_{i_{2\ell}}.
\label{eq:w_general}
\ee Each term of the sum in Eq. \eqref{eq:w_general} can be associated
to a non backtracking walk of length $2\ell-1$, where edges may be
crossed multiple times.  To each node visited by the NB walk is
attached a variable $n_i$, and the extreme nodes of the walk have the
extra factors $z_{i_1}$ (starting point) and $z_{i_{2\ell}}$ (end
point).  When NB walks containing loops are neglected in
Eq. \eqref{eq:w_general}, the formula simplifies considerably, since
the only remaining walks are the ones of length $2\ell-1$ where each
one of the $2\ell$ nodes is visited only once. For example, let us
consider a loop free NB walk of length $2\ell-1$ starting from a given
node, say node $i$. It visits all nodes up to a distance $2\ell -1$,
and stops at the final node, say node $j$.  The total number of nodes
visited by this NB walk is $2\ell$.

Notice that, when loops are neglected, we can approximate the local
environment around any node by a tree, in line with the original
locally-tree like assumption of the whole approach. 
A simple way to implement the tree-like approximation is to consider 
only the NB walks of length $2\ell-1$ that start from a given node $i$
and end on nodes $j$, in such a way that these NB walks coincide with 
the shortest paths between $i$ and those nodes $j$. 
Moreover,
these NB walks contain the products $n_{i_1}n_{i_2}\dots n_{i_{2\ell}}$
with all $n_{i_\alpha}$ different from each other.  Hence, we can
finally write down the expression for the leading term of
$|\mathbf{w}_\ell(\mathbf{n})|^2$ as: \be
|\mathbf{w}_\ell(\mathbf{n})|^2 = \sum_{i=1}^{N} z_i
\sum_{j\in\partial\mathrm{Ball}(i,2\ell-1)}\left(\prod_{k\in\mathcal{P}_{2\ell-1}(i,j)}n_k\right)
z_j\ ,\label{eq:Costfunction2} \ee where $\mathrm{Ball}(i,\ell)$ is
the set of nodes inside a ball of radius $\ell$ around node $i$, where
the radius is defined taking the shortest path as the distance,
$\partial\mathrm{Ball}(i,\ell)$ is the frontier of the ball and
$\mathcal{P}_{\ell}(i,j)$ is the set of nodes belonging to the
shortest path of length $\ell$ connecting $i$ and $j$.  This is the
cost energy function of influence Eq. \eqref{eq:Costfunction} given in
the main text.

\subsection{Odd $(2\ell+1)$-body interactions}
\label{odd}

The energy (or cost) function Eq. \eqref{eq:Costfunction2} contains
only even $2\ell$-body interactions.  It is possible to interpolate
with odd interactions by considering an analogous Power Method
expansion of the eigenvalue in terms of the matrix elements
$\la\mathbf{w}_\ell(\mathbf{n})|\Mia|\mathbf{w}_\ell(\mathbf{n})\ra$.
Indeed, the eigenvalue $\lambda(\mathbf{n})$ can be also computed
using another series expansion in the Power Method \cite{eigenvalue}
as:

\be \lambda(\mathbf{n})\ =\ \lim_{\ell\to\infty}\ \left[
  \frac{\la\mathbf{w}_\ell(\mathbf{n})|\Mia|\mathbf{w}_\ell(\mathbf{n})\ra}
       {\la\mathbf{w}_0|\mathbf{w}_0\ra} \right]^{1/(2\ell+1)}\ .
\ee The explicit expression of
$\la\mathbf{w}_\ell(\mathbf{n})|\Mia|\mathbf{w}_\ell(\mathbf{n})\ra$
can be computed similarly to $|\mathbf{w}_\ell(\mathbf{n})|^2$,
following the same steps outlined in Methods Section
\ref{mainformula}.  As an example, for
$\la\mathbf{w}_1(\mathbf{n})|\Mia|\mathbf{w}_1(\mathbf{n})\ra$ we get:
\be \la\mathbf{w}_1(\mathbf{n})|\Mia|\mathbf{w}_1(\mathbf{n})\ra=
\sum_{ijk}A_{ij}A_{jk}(1-\delta_{ik})z_iz_k n_in_jn_k \ \ .  \ee The
asymptotic expression for $N\to \infty$ (i.e. the one neglecting
loops) is similar to Eq. \eqref{eq:Costfunction2}, and, for
$\ell\geq1$, reads: \be
\la\mathbf{w}_\ell(\mathbf{n})|\Mia|\mathbf{w}_\ell(\mathbf{n})\ra =
\sum_{i=1}^{N} z_i
\sum_{j\in\partial\mathrm{Ball}(i,2\ell)}\left(\prod_{k\in\mathcal{P}_{2\ell}(i,j)}n_k\right)
z_j\ ,
\label{eq:Costfunction3}
\ee
while for $\ell=0$ we find:
\be
\la\mathbf{w}_0|\Mia|\mathbf{w}_0\ra = 
\sum_{i=1}^{N} k_i(k_i-1)n_i\ .
\ee
We can introduce the equivalent of the 
approximant $\lambda_\ell(\mathbf{n})$ in Eq. \eqref{eq:approximant},
that we call $\lambda'_\ell(\mathbf{n})$:
\be \lambda'_\ell(\mathbf{n})\ =\  \left[
  \frac{\la\mathbf{w}_\ell(\mathbf{n})|\Mia|\mathbf{w}_\ell(\mathbf{n})\ra}
{\la\mathbf{w}_0|\mathbf{w}_0\ra} 
\right]^{1/(2\ell+1)}\ .
\ee

\subsection{HD attack at $\ell=0$ one-body problem}

It is interesting to consider the $\ell=0$ term,
$\lambda'_0(\mathbf{n})$, since it reproduces exactly the high-degree
(HD) strategy attacking the hubs calculated by Cohen {\it et al.}
\cite{cohen}. This term represents the one-body interaction where the
influencers are considered in isolation, only affected by the external
field, and therefore this strategy lacks the collective influence
effects found for $\ell\ge 2$. It reads: \be
\lambda'_0(\mathbf{n})\ =\ \frac{\la\mathbf{w}_0|\Mia|\mathbf{w}_0\ra}
        {\la\mathbf{w}_0|\mathbf{w}_0\ra}\ =\ \frac{\sum_i
          k_i(k_i-1)n_i}{\sum_i k_i}\ .
\label{highdegree}
\ee The sum on the numerator on the r.h.s represents the energy of a
gas of free particles in an external (site-dependent) field
$h_i=k_i(k_i-1)$. This field is always nonnegative
$h_i\geq0$. Therefore, in order to minimize $\lambda'_0(\mathbf{n})$
it is sufficient to sort the nodes according to the external field,
and then removing the ones corresponding to the highest $Nq$
fields. Since the field $h_i$ is monotonic increasing with the degree,
the minimization corresponds exactly to the removal of the high degree
nodes one by one. It is also easy to see that the stability condition
imposed on $\lambda'_0(\mathbf{n})$, \be \min_{\mathbf{n}:\la
  n\ra=1-q_c}\ \lambda'_0(\mathbf{n})\ =\ 1,
\label{eq:HD}
\ee gives exactly the threshold $q_c$ expected from HD as calculated
by Cohen {\it et al.} \cite{cohen}. Indeed, putting the expression for
$\lambda'_0(\mathbf{n})$ in Eq. \eqref{eq:HD}, we have, \be
\min_{\mathbf{n}:\la n\ra=1-q_c}\ \frac{\sum_i k_i(k_i-1)n_i}{\sum_i
  k_i}\ =\ 1\ .
\label{eq:HDcritic}
\ee As we said, the minimization of the numerator in the l.h.s. is
achieved by setting $n_i=0$ for the first $Nq_c$ highest degree node,
and $n_i=1$ for the remaining ones.  Therefore we can rewrite
Eq. \eqref{eq:HDcritic} as: \be \frac{1-q_c}{\la k\ra}\left( \la
k^2\ra' - \la k\ra' \right)\ =\ 1\ ,
\label{eq:HDaveraged}
\ee where the average $\la\cdot\ra'$ is performed using a modified
degree distribution $P'(k; q_c)$, which depends on $q_c$ and
represents the degree distribution in the network with the removed
hubs. To derive the explicit form of $P'(k; q_c)$, let us consider
first the relation between $q_c$ and the original $P(k)$. The fraction
$q_c$ of nodes to be removed is \be
\sum_{k=\zeta}^{\infty}P(k)\ =\ q_c\ , \ee where $\zeta$ is the lowest
degree of the removed nodes compatible with $q_c$.

The distribution $P'(k; q_c)$ is the degree distribution of the
remaining nodes (i.e. the ones for which $n_i=1$) and is given by: \be
P'(k; q_c)\ =\ \frac{1}{1-q_c}P(k)\theta(\zeta-k)\ .  \ee
 
We now solve Eq. \eqref{eq:HDcritic} in the case of scale free network
with degree distribution $P(k)\sim k^{-\gamma}$, degree exponent
$\gamma$, minimum degree $m$ and maximum degree $k_{\mathrm{max}}$.
We also work in the continuum limit, so that we can use integrals in
place of the sums.  The variable $\zeta$ as a function of $q_c$ reads:
\be q_c
=\ \int_{\zeta}^{k_{\mathrm{max}}}P(k)\ k\ dk\ =\ \frac{\zeta^{1-\gamma}-k_{\mathrm{max}}^{1-\gamma}}{m^{1-\gamma}-k_{\mathrm{max}}^{1-\gamma}}\ .
\ee

The average $\la k\ra'$ is given by: \be \la
k\ra'\ =\ \frac{1}{1-q_c}\int_{m}^{\zeta}P(k)\ k\ dk\ =\ \frac{1}{1-q_c}\frac{\gamma-1}{\gamma-2}\frac{m^{2-\gamma}-\zeta^{2-\gamma}}{m^{1-\gamma}-k_{\mathrm{max}}^{1-\gamma}},
\ee and $\la k^2\ra'$ by: \be \la
k^2\ra'\ =\ \frac{1}{1-q_c}\frac{\gamma-1}{\gamma-3}\frac{m^{3-\gamma}-\zeta^{3-\gamma}}{m^{1-\gamma}-k_{\mathrm{max}}^{1-\gamma}}.
\ee Putting these expressions in Eq. \eqref{eq:HDaveraged} we find the
following implicit equation for $q_c$: \be \frac{\gamma-2}{\gamma-3}
\frac{m^{3-\gamma}-\zeta^{3-\gamma}}{m^{2-\gamma}-k_{\mathrm{max}}^{2-\gamma}}
-
\frac{m^{2-\gamma}-\zeta^{2-\gamma}}{m^{2-\gamma}-k_{\mathrm{max}}^{2-\gamma}}=1\ .
\label{eq:SFgeneral}
\ee Equation \eqref{eq:SFgeneral} simplifies considerably when
$k_{\mathrm{max}}\to\infty$. Indeed (for $\gamma \geq 2$), we find the
same result as in \cite{cohen}:

\be \frac{\gamma-2}{\gamma-3}\left( m-
\frac{\zeta^{3-\gamma}}{m^{2-\gamma}}\right)
\ +\ \frac{\zeta^{2-\gamma}}{m^{2-\gamma}}=2\ .  \ee 

Considering the ensemble where $k_{\rm max} = m N^{1/(\gamma-1)}$, when
$N\to \infty$ then $k_{\mathrm{max}}\to\infty$, the relation between
$\zeta$ and $q_c$ is $\zeta=m\ q_c^{1/(1-\gamma)}$, and we finally
find: \be q_c^{(2-\gamma)/(1-\gamma)}= 2 +
\frac{\gamma-2}{\gamma-3}m\left(
q_c^{(3-\gamma)/(1-\gamma)}-1\right)\ ,
\label{eq:HDinfiniteCutoff}
\ee which is the known result for the HD attack \cite{cohen}. The
solution is shown in Extended Data Fig. \ref{fig:pd}a.

For $\gamma=2$, Eq. \eqref{eq:HDinfiniteCutoff} predicts a zero
critical $q_c$ for any $m$, and this is interpreted as an extreme
fragility of scale-free networks when the degree exponent is close to
$2$. Indeed, for $\gamma=2$, the network is essentially a star-graph,
which is trivially destroyed by removing the central hub.  This in
turn is a consequence of the fact that the natural cut-off
$k_{\mathrm{max}}$ diverges linearly with the system size for
$\gamma=2$, i.e., $k_{\mathrm{max}}\sim N$ (see Extended Data
Fig. \ref{fig:pd}a).

Nonetheless the situation changes a lot for other network ensembles,
where the cut-off can be very large but finite (which is the case of
all networks, both real and synthetic ones). Indeed,
Eq. (\ref{eq:SFgeneral}) for a finite cut-off $k_{\mathrm{max}}$ and
for $\gamma \to 2$ gives: \be \zeta\ = \ m + \log\left(\frac{\zeta
  k_{\mathrm{max}}}{m^2}\right)\ .  \ee By expressing $\zeta$ as a
function of $q_c$ via the equation: \be \zeta\ =
\ \frac{mk_{\mathrm{max}}}{m+q_c(k_{\mathrm{max}}-m)}\ , \ee we find
the equation for $q_c$ as a function of the cut-off
$k_{\mathrm{max}}$: \be
\frac{mk_{\mathrm{max}}}{m+q_c(k_{\mathrm{max}}-m)}\ =\ m +
\log\frac{k_{\mathrm{max}}^2}{m^2+mq_c(k_{\mathrm{max}}-m)}\ .
\label{eq:hdcutoff}
\ee The solution to Eq. \eqref{eq:hdcutoff} is shown in Extended Data
Fig. \ref{fig:pd}b. For $k_{\mathrm{max}}\to \infty $, the asymptotic
behaviour of the threshold $q_c$ is given by : \be
q_c\ \sim\ \frac{m}{\log(k_{\mathrm{max}})}\ \ \ \mathrm{for}\ \ k_{\mathrm{max}}\to\infty\ .
\ee Therefore, $q_c$ still vanishes when $k_{\rm max}\to\infty$ but as
the inverse of the logarithm of the cut-off. This very slow
convergence makes questionable the claim that scale free networks are
extremely fragile under hubs removal. Indeed, as can be seen in
Extended Data Fig. \ref{fig:pd}b, even for $k_{\mathrm{max}}$ of the
order of hundred millions, $q_c$ is still of order $0.1$. For more
realistic $k_{\rm max}=10^3$, which is typical of social networks,
$q_c\approx 0.2$ for all $\gamma$.  In these situations, the search
for other attack strategies becomes important.

\section{Optimization with the Cavity Method}
\label{cavity}

The eigenvalue $\lambda(\mathbf{n})$ can be minimized by using the
cavity method from spin glass theory \cite{MP} since we work with
sparse graphs.  The method can be applied in practice to the first
order approximation to the eigenvalue $\ell=1$, which is a pair-wise
interaction, i.e., Eq. S\ref{eq:2body}: \be
\lambda_1(\mathbf{n})\ =\ \frac{|\mathbf{w}_1(\mathbf{n})|}{|\mathbf{w}_0|}.
\ee For higher order many-body interactions, the cavity method becomes
much more involved and we will pursue other solving strategies.


The analytical Replica Symmetry (RS) solution obtained with the cavity
method for this pairwise model allows us to compare with the solution
given by EO. This is a very useful check of the correctness of the
problem solution, since both method are mathematically not rigorous.
An improvement over the RS solution can be obtained by applying the so
called 1step replica symmetry breaking cavity method (1RSB)
\cite{MP,montanari}, which should be compared as well with the EO
solution. This will be done in a future work. Here we only note that,
as far as the assessment of the ground state energy is concerned, the
difference between the RS and the 1RSB estimations is, typically, very
small (for example it is less than 2\% for spin glass models on random
graphs \cite{montanari}). We expect that the same scenario holds true
also for the model defined here.  Furthermore, a deeper physical
insight can be obtained when we recast the problem in terms of spin
glass theory as we show next.

The energy (cost) function of the problem is, Eq. S\ref{norm1}: \be
\mathcal{E}(\mathbf{n})\ \equiv\ |\mathbf{w}_1(\mathbf{n})|^2\ =\ \sum_{ij}A_{ij}(k_i-1)(k_j-1)n_in_j\ .
\ee The physics of this system is made more transparent when the
problem is formulated in terms of Ising spin variables $s_i=\pm1$. The
translation of the problem in the language of statistical mechanics
will turn the optimization problem to one of finding the ground
state of a spin-glass system.  The relation between $s_i$ and $n_i$ is
given by

\be s_i \equiv 2n_i-1.  \ee Note that the state $n_i=0$, meaning that
node $i$ is removed (influencer), corresponds to the spin down state
$s_i=-1$. On the other hand the state $n_i=1$ (node $i$ not removed)
corresponds to spin up $s_i=1$. Using these new variables the energy
function takes the more familiar form of an Ising model:

\be \mathcal{E}(\mathbf{s})\ =\ -\sum_{\la
  ij\ra}s_iJ_{ij}s_j\ -\ \sum_iH_is_i\ +\ \mathcal{C}\ ,
\label{eq:antiferromagnet}
\ee where the first sum on the r.h.s is over the pairs $\la i j\ra$ of
nearest neighbours sites in the network.  The coupling constants
$J_{ij}$ represent the interactions between the spins and they depend
on the details of the network. Explicitly they read: \be
J_{ij}\ =\ -\frac{1}{2} A_{ij} (k_i-1)(k_j-1)\ . 
\label{jij} \ee The local field
$H_i$ depends also on the topology and is given by: \be H_i\ =
\ -\frac{1}{2}k_i(k_i-1)(k_i^{nn}-1)\ , \ee where $k_i^{nn}$ is the
average nearest neighbors degree of vertex $i$, defined as: \be
k_i^{nn} \equiv k_i^{-1}\sum_{j\in\partial i}k_j.  \ee The
constant $\mathcal{C}$ does not depend on the spin variables and can
be ignored in the minimization problem (but must be included in the
evaluation of the energy $\mathcal{E}(\mathbf{s})$).

We introduce also the magnetization, \be m \equiv \sum_is_i/N, \ee of
the configuration $\mathbf{s}$, which is related to the number of
removed nodes $q$ by the equation \be m\equiv 1-2q.\ee Note that
$\mathcal{E}(m=-1) = 0$, and $\mathcal{E}(m = +1)=\la k\ra(\kappa -
1)^2$ (for uncorrelated networks), where $\kappa=\la k^2\ra/\la k\ra$.

The physical system defined by the energy function
\eqref{eq:antiferromagnet} is a disordered antiferromagnet
($J_{ij}\leq0$) in a random external magnetic field.  Remarkably, the
disorder in the model, $J_{ij}$, comes from the randomness in the
network via the adjacency matrix $A_{ij}$, even if for a given
instance of the problem both the couplings and the magnetic field are
deterministic (i.e. fixed by the topology of the underlying network
and the degree).

The problem we have to solve is tantamount to find the ground state of
the system with energy function \eqref{eq:antiferromagnet}. More
precisely, we want to find the ground state for a fixed value of the
magnetization $m$, which corresponds to keep fixed the number of
removed nodes $q$. This problem represents a quite novel system for
spin glass theory since the coupling $J_{ij}$ depends explicitly on
the contact network via Eq. S\ref{jij}.  This coupling between the
underlying network and the disorder is a quite unique feature of the
optimal percolation problem at the $\ell=1$ pair-wise level. The
problem becomes spin-glass when the system is forced to satisfy a given
magnetization, i.e., a given fraction of influencers, which is a
global constraint. This additional constraint of constant
magnetization represents a problem for the cavity method, since it
cannot be enforced locally. Nevertheless the problem can be
circumvented by introducing an external field $H$ that is chosen
self-consistently to fix the desired value of the magnetization as
done in \cite{sulc}.  In practice, we introduce the Legendre transform
of the energy function $\mathcal{E}(\mathbf{s})$, defined as \be
\begin{aligned}
E_H(\mathbf{s})\ &=\ \mathcal{E}(\mathbf{s})\ -\ MH\ ,\\
\frac{\partial E_H}{\partial H}\ &=\ -M\ ,
\end{aligned}
\ee where $M=\sum_is_i$ is the global magnetization.  At this point we
have to minimize the function $E_H(\mathbf{s})$ under the global
constraint defined by the equation $\partial E/\partial H = -M$.  This
can be done by properly gauging the external field $H$ during the
iteration of the cavity equations, as we will explain below.

The cavity equations for the system defined by the energy function
$E_H(\mathbf{s})$ are a set of equations for the cavity fields
$h_{i\to j}$ and the cavity bias $u_{i\to j}$, one for each directed
edge of the graph. These variables can be interpreted as messages
exchanged by the nodes: the bias $u_{i\to j}$ represents the incoming
message into node $j$ traveling along the edge connecting $i$ and
$j$; the field $h_{i\to j}$ is the outgoing message from node $i$
towards node $j$. Outgoing messages are computed from the incoming
ones in a self-consistent way. The cavity field $h_{i\to j}$
quantifies the tendency of spin $i$ to be $+1$ or $-1$, when the spin
on node $j$ has been pruned from the network (whence the name
cavity). The cavity bias $u_{i\to j}$ is determined by optimizing
between the interaction $J_{ij}$ and the cavity field $h_{i\to
  j}$. More precisely, the cavity equations for cavity bias and cavity
fields at zero temperature read: \be
\begin{aligned}
h_{i\to j}\ &=\ H(M) + H^i + \sum_{k\in\partial i\setminus j}u_{k\to i} 
\ ,\\
u_{k\to i}\ &=\ \frac{1}{2}\textrm{sign}(J_{ik} h_{k\to i})\min(|J_{ik}|, |h_{k\to i}|)\ .
\end{aligned}
\label{eq:cavity_eqs}
\ee Once a solution of the cavity equations \eqref{eq:cavity_eqs} has
been found, the total local effective field $h_i$ acting on spin $i$
can be computed through the formula: \be h_{i}\ =\ H(M) + H^i +
\sum_{k\in\partial i}u_{k\to i} \ee The main hypothesis of the
cavity method is the existence of a single pure state, which can be
rephrased as a hypothesis on the uniqueness of the solution of the
cavity equations \eqref{eq:cavity_eqs}. This is called the
replica-symmetric (RS) cavity method. It is also possible to
generalize the method to incorporate multiple solutions (the so called
cavity method at the level of 1step Replica Symmetry Breaking, 1RSB
\cite{MP}). The analysis of this second method will come in a
follow-up work. Here we limit ourselves to the RS cavity method.
  
The cavity equations \eqref{eq:cavity_eqs} can be interpreted as
updating rules for a message passing algorithm, and therefore they can
be solved iteratively starting from a random initial condition.  In
practice we add a "time" label $t$ to the cavity fields and we rewrite
Eqs. \eqref{eq:cavity_eqs} as dynamical message passing equations: \be
h_{i\to j}^{(t)}\ = H(M) + H^i + \sum_{k\in\partial i\setminus
  j}u_{k\to i}^{(t-1)}
\ .
\label{eq:dyn_cavity_eqs}
\ee Since the magnetization $M$ has to be kept fixed, the external
field $H(M)$ also needs to be updated at each step of the
iteration. Therefore, after a total update of the cavity fields
$h_{i\to j}^{(t)}$, the field $H(M)\equiv H^{(t)}(M)$ is recomputed by
solving the equation: \be \sum_i\textrm{sign}\left( H^{(t)}(M) + H^i +
\sum_{k\in\partial i}u_{k\to i}^{(t)} \right)\ =\ M.
\label{eq:H_field}
\ee

The solution of the Eqs. \eqref{eq:cavity_eqs} corresponds to the
fixed point of the map defined by Eqs. \eqref{eq:dyn_cavity_eqs} and
\eqref{eq:H_field}.

Once the solution of the cavity equations has been found, the RS
estimate of ground state energy $E_H^{\mathrm{RS}}$ is given by: \be
E_H^{\mathrm{RS}}\ =\ \sum_i\epsilon_H^i\ -\ \sum_{(ij)}\epsilon_{ij}\ ,
\ee where the site and link energies $\epsilon_H^i$ and
$\epsilon_{ij}$ are given, respectively, by \be
\begin{aligned}
\epsilon_H^i\ =\ -\max\left(
H(M) + H^i + \sum_{k\in\partial i}|h_{k\to i} + J_{ik}|, \right. \\ 
\left. -H(M) - H^i + \sum_{k\in\partial i}|h_{k\to i} - J_{ik}|
\right), \\
\epsilon_{ij}\ =\ -\max\left(
h_{i\to j}+J_{ij}+h_{j\to i},\ h_{i\to j}-J_{ij}-h_{j\to i},\right. \\
\left. -h_{i\to j}-J_{ij}+h_{j\to i},\ -h_{i\to j}+J_{ij}-h_{j\to i}
\right)\ .
\end{aligned}
\label{eq:energies_RS}
\ee

From the knowledge of the function $E_H^{\mathrm{RS}}$ we can compute
the energy $\mathcal{E}^{\mathrm{RS}}_M$ by inverting the Legendre
transform: \be \mathcal{E}^{\mathrm{RS}}_M\ =\ E_H^{\mathrm{RS}}
+\ MH(M)\ , \ee where $H(M)$ is the external field which produces the
desired value of the magnetization $M$, given by the fixed point of
Eq. \eqref{eq:H_field}.  The value of $\mathcal{E}^{\mathrm{RS}}_M$ is
all we need to compute the optimized eigenvalue. Since
$|\mathbf{w}_0|=\sqrt{N\la k\ra}$, we finally find: \be
\lambda_1^{\mathrm{RS}}(m)=\ \sqrt{\frac{\epsilon^{\mathrm{RS}}(m)}{\la
    k\ra}}\ ,
\label{eq:eigenvRS}
\ee
where $\epsilon^{\mathrm{RS}}(m) = \mathcal{E}^{\mathrm{RS}}_M/N $ is the 
(intensive) energy per spin. 

We observed that the message passing equations
\eqref{eq:dyn_cavity_eqs} never converge to a stable fixed point. This
is a consequence of the existence of very many solutions to the cavity
equations \eqref{eq:cavity_eqs}, which implies that the replica
symmetry is broken for this system. Nevertheless, the non-converged
messages can still be used to estimate the energy
$\epsilon^{\mathrm{RS}}(m)$.  In practice we run the algorithm for a
maximum number of $T_{\mathrm{max}}=10^6$ iteration. Then we use the
current value of the cavity fields $\{h_{i\to j}\}$ to compute the
energy of the system, and we average the energy on $T_{\mathrm{max}}$
more iterations.

In  Extended Data Fig. \ref{fig:lambdaRS} we show the optimized eigenvalue
$\lambda_1^{\mathrm{RS}}$ as a function of the removed nodes $q=1-2m$,
computed for an Erd\"{o}s-R\'enyi network with mean degree $\la k\ra =
3.5$ and size $N=10^4$. We find the transition point at this 2-body
interaction approximation to be $q_c=0.248$.

The continuous transition from the phase with
$\lambda_1^{\mathrm{RS}}=0$ to the phase with
$\lambda_1^{\mathrm{RS}}>0$ observed in Extended Data
Fig. \ref{fig:lambdaRS} is an artifact of the 2-body interaction
considered here. Indeed the real optimal eigenvalue (i.e. the
eigenvalue including $\infty$-body interactions) has to jump from zero
to one discontinuously at $q_c$ as depicted in
Fig. \ref{fig:explanation}c.  The reason is that the largest
eigenvalue of the non-backtracking matrix of a tree-graph is zero,
while the largest eigenvalue for a tree plus one single loop is
one. Adding more loops increases the eigenvalue.  Since there are no
other possible networks between a tree and a unicyclic graph, the
largest eigenvalue of the non-bactracking matrix cannot take values in
the interval $(0,1)$.  As a consequence it has to jump discontinuously
from 1 to zero at $q_c$.  In the 2-body approximation this jump is
smoothed by a continuous line. By considering higher order
interactions the eigenvalue would remain zero closer and closer to the
critical threshold $q_c$.  This is evident, for example, in the
problem with $3,4$ and $5$ interactions, that we solve using extremal
optimization next.  We expect that adding more and more interactions
the eigenvalue has several (continuous) transitions, when departing
from zero, for smaller and smaller values of $q$ and eventually for
interactions of infinite order it jumps discontinuously from zero to
one exactly at $q_c$.


In the inset of Extended Data Fig. \ref{fig:lambdaRS} we show also a
comparison between the RS cavity method and extremal optimization done
in the next section. As can be seen the difference is very small, and
we believe that they are actually very close to the true optimal
result.  The size of the ER network in this case is $N=128$, for which
EO gives the actual ground state. In favor of this conjecture we can
also say that the 1RSB estimate, which is in general more correct that
the RS one, is anyway very close to the latter, and would lie in
between of the two curves shown in the inset of Extended Data
Fig. \ref{fig:lambdaRS}.  The eigenvalue estimated with RS is anyway
slightly smaller. This is a typical feature of the RS cavity method,
in the sense that it gives a lower bound (and not an upper bound) to
the ground state energy of the system. On the contrary, EO provides an
upper bound to the optimal threshold. As the inset of Extended Data
Fig. \ref{fig:lambdaRS} shows, both the lower bound (RS) and the upper
bound (EO) are very close to each other, and therefore to the true
optimum.  Therefore, we observe that there can be different ways to
analytically assess the location of the optimal threshold.
Conversely, this does not mean that different methods are equally able
to find the actual configuration of optimal influencers, even if they
give similar analytical estimation for the threshold.  A more
important issue in any NP problem, perhaps the most relevant for any
practical purpose, is finding a scalable algorithm which approximates
the optimal configuration as better as possible and can be used for
very large networks.  This is the main reason why our CI algorithm was
designed for.

To conclude this section we want to observe that the antiferromagnetic 
nature of the optimal percolation problem is not totally unexpected.
Indeed, the antiferromagnetic interactions between nodes reflect the 
intuitive idea that immunizing contiguous nodes is less efficient 
than immunizing them in a staggered way.

\section{Minimization with Extremal Optimization (EO)}
\label{sec:EO}

In this section we describe another method for the minimization of the
eigenvalue $\lambda(\mathbf{n})$, called Extremal Optimization (EO)
\cite{EO}, which has the advantage, with respect to the cavity method,
to be easily implemented for higher $\ell$-order of
$|\mathbf{w}_\ell(\mathbf{n})|$. However, EO is still not scalable for
large networks, for which we will implement CI. Nevertheless, EO makes
use of the full energy function, including loops, and can be used to
extrapolate the solution for large networks where EO is not applicable
anymore.  Indeed, the EO algorithm is an efficient method to find
nearly optimal solutions.  It was used successfully to find the ground
state energy of spin glass models on random graphs, where it was shown
to be practically identical to the best available analytical
prediction \cite{EO}.  We believe that also in our case $\tau$-EO is
very close to the optimum when extrapolated to large systems. In
Extended Data Figs \ref{fig:taufull} and \ref{fig:tau_pc} we estimate
the optimal solution for large networks using a finite size scaling
analysis extrapolating to the infinite size limit and $\ell\to\infty$,
as explained next.

To explain the method in the simplest way let us consider again the
lowest non trivial approximation to the eigenvalue
$\lambda(\mathbf{n})
\sim\ \frac{|\mathbf{w}_1(\mathbf{n})|}{|\mathbf{w}_0(\mathbf{n})|}$
and the corresponding cost function $\mathcal{E}(\mathbf{n})$: \be
\mathcal{E}(\mathbf{n})\ \equiv\ |\mathbf{w}_1(\mathbf{n})|^2=\sum_{ij}A_{ij}(k_i-1)(k_j-1)n_in_j.
\label{eq:energyEO}
\ee

We now assign to each variable $n_i$ the fitness $b_i$:
\be
b_i\ =\ (k_i-1)\sum_{j}A_{ij}(k_j-1)n_j\ ,
\ee
so that we can rewrite the energy function \eqref{eq:energyEO} as
\be
\mathcal{E}(\mathbf{n})\ =\ \sum_i\ b_in_i\ .
\label{eq:Enefit}
\ee

Notice that this is similar to the form we adopt to define CI in
Eq. (\ref{eq:CI}). The CI-algorithm is an adaptive version
of the EO algorithm, in a sense that will be explained in
Sec. \ref{CI}. The EO algorithm being the exact minimization of the
largest eigenvalue of the NB matrix, which can only be achieved for
small systems.
 
Each node in the state $n_i=0$ (removed) gives zero contribution to
the energy, while nodes for which $n_i=1$ give a contribution equals
to their fitness.  Therefore, to minimize the energy
$\mathcal{E}(\mathbf{n})$ we have to find the set of nodes with the
lowest fitness, under the usual constraint $\sum_in_i=N(1-q)$.  Note
that the fitness $b_i$ of node $i$ depends on the states $n_j$ of its
neighbours $j$.

The aim of EO is to explore the space of states looking for the
configurations with the smallest fitness. Let now explain how it
works. For a fixed $q$, in the beginning the variables $n_i$ are
assigned at random into two sets: set $S_0$ containing the $qN$ nodes
to be removed $n_i=0$, and set $S_1$ containing $(1-q)N$ nodes with
$n_i=1$: \be
\begin{aligned}
S_1\ &\equiv\ \{i: n_i=1\}\ , \\
S_0\ &\equiv\ \{i: n_i=0\}\ , 
\end{aligned}
\ee with the constraints $|S_0|=Nq$, and $|S_0| + |S_1| = N$. The
initial separation of the nodes in these two groups is made
arbitrarily.  Then the fitness $b_i$ corresponding to this initial
configuration $C_0$ are evaluated and sorted separately for the two
groups.  The first move consists in exchanging the variable $n_i$ in
$S_1$ with the largest fitness, with the variable $n_j$ in $S_0$ with
the lowest one. In other words, we set $n_i=0$ and $n_j=1$.  This move
does not change the sizes of $S_1$ and $S_0$ and hence the global
constraint $\sum_in_i=N(1-q)$ remains satisfied. The energy function
$\eqref{eq:Enefit}$ corresponding to this new configuration $C_1$ is
evaluated and if $\mathcal{E}(C_1)<\mathcal{E}(C_0)$ the configuration
$C_1$ is stored together with the value of its energy. The process is
repeated by recomputing the new fitness and swapping the variable with
the highest value of $b_i$ from $S_1$ with the variable corresponding
to the lowest one in $S_0$. Note that the moves are accepted
unconditionally at each step and only the best configuration found so
far is saved. The algorithm is terminated after a maximum number of
iteration is reached.

What we have described so far is the basic EO algorithm. It can be
improved by introducing a tunable parameter, called $\tau$, so that we
will refer to it as $\tau$-EO. In this version of the algorithm, the
choice of the variables to be swapped is not performed
deterministically by selecting the ones with the largest and smallest
fitness, but, instead, they are picked up using a random
selection. This may look counterintuitive at first sight, since we
would not expect any improvement by randomizing the choice. Actually
this is not true, and an improvement can be achieved, provided that
the random rule is chosen judiciously.

In the $\tau$-EO algorithm, we sort the fitness in the two sets $S_0$
and $S_1$, in increasing order in $S_1$ and decreasing order in $S_0$:
\be
\begin{aligned}
&b_{\Pi(1)}\geq b_{\Pi(2)}\geq\cdots\geq b_{\Pi(|S_1|)}\ ,\\
&b_{\Lambda(1)}\leq b_{\Lambda(2)}\leq\cdots\leq b_{\Lambda(|S_0|)}\ ,
\end{aligned}
\ee where $\Pi$ and $\Lambda$ are two permutations of the labels $i$
of the variables in $S_1$ and $S_0$ respectively. The worst variable
in $S_1$ (the one with the highest fitness) is $n_{\Pi(1)}$, that we
want to change with the worst variable in $S_0$, i.e.,
$n_{\Lambda(1)}$ (the one with the lowest fitness). In the simple EO
algorithm this is exactly what we were doing: exchanging $n_{\Pi(1)}$
and $n_{\Lambda(1)}$. Now we rank the variables according to their
fitness. 

For the variables in $S_1$ the worst variable is $n_{\Pi(1)}$, which
has rank 1, while the best variable is $n_{\Pi(|S_1|)}$, which is of
rank $|S_1|$. For the variables in $S_0$ the variable of rank 1 is
$n_{\Lambda(1)}$, while $n_{\Lambda(|S_0|)}$ has rank $|S_0|$. Then we
consider the following probability distribution over the ranks $r$:
\be P(r)\ \propto\ r^{-\tau}\ ,
\label{eq:tauEO}
\ee with $r\in[1,|S_1|]$ or $r\in[1,|S_0|]$ for the ranks of variables
in $S_1$ or $S_0$, respectively. At each update, we draw two numbers
$r_1$ and $r_0$ from $P(r)$ and then we swap the variables
$n_{\Pi(r_1)}$ and $n_{\Lambda(r_0)}$.  Then the algorithm proceeds as
in the original EO. Note that for $\tau\to\infty$, we recover the
deterministic EO algorithm, swapping only the worst variables.  The
idea behind the choice of the scale free distribution $P(r)$ is to
ensure that no variable gets excluded from changing set, while giving
higher priority to the variables with worst fitness. The random
selection of the variables has the advantage, over the deterministic
process, to make possible global reconfigurations of the system, thus
climbing over the energy barriers and find better minima.
In our problem we found that a value of $\tau=1.7$ gives the best
results.  

As an application of this method, we perform the minimization of the
energy function $\mathcal{E}(\mathbf{n})$ on an Erd\"{o}s-R\'enyi
network with average degree $\la k\ra=3.5$. We considered different
system sizes $N=2^5,2^6,2^7,2^8$. For each size $N$ we took the
average of the ground state energy over 100 realizations. For each
instance we performed $N^3$ updates of the $\tau$EO routine. The
results for the eigenvalue $\lambda(q)=\sqrt{\mathcal{E}(q)/N\la
  k\ra}$ are shown in Extended Data Fig. \ref{fig:taufull}.

We observe that the finite size scaling of the optimal influence
threshold $q_c$, which is the solution of $\lambda(q_c)=1$, is
consistent with the scaling law: \be
q_c(N)\ =\ q_c(\infty)\ +\ AN^{-2/3}\ , \ee where $A$ is a coefficient
independent from the size $N$. This scaling form is the same of the
finite size correction to the thermodynamical energy density of a spin
glass system with pairwise interactions (e.g. the SK model) at and
below the de Almeida-Thouless line. Actually this scaling form is
observed also for other interesting thermodynamic quantities, like the
second cummulant of the overlap distribution \cite{parisi2}. In that
case the anomalous scaling (as opposed to the more natural $1/N$
correction expected from the central limit theorem) is due to the
existence of infinitely many zero modes, whose volume grows as
$N^{1/3}$.  In the language of our model, these zero modes represent
the infinitely many way to choose the set of optimal influencers.  It
would be very interesting then to interpret what kind of hidden
symmetry relates all these set of optimal influencers.

\subsection{$\tau$-EO with multibody interactions}

In the general case the energy function we want to minimize is: 
\be
\mathcal{E}(\mathbf{n})\ =\ |\mathbf{w}_\ell(\mathbf{n})|^2\ 
\label{eq:energyEOmultibody}
\ee which involves at most $2\ell$-body interactions. To treat systems
with at most a (odd) number of $(2\ell+1)$-body interactions, we
consider the energy function: \be
\mathcal{E}'(\mathbf{n})\ =\ \la\mathbf{w}_\ell(\mathbf{n})|\Mia|
\mathbf{w}_\ell(\mathbf{n})\ra\ .
\label{eq:energyEOmultibodyOdd}
\ee 

In order to apply the EO algorithm to systems with many-body
interactions, all that we have to do is to change the definition of
the fitness $b_i$.  For example, in the case of a system with $4$-body
interactions, described by $|\mathbf{w}_2(\mathbf{n})|^2$, we set
$b_i$ as: \be b_i=
\sum_{jk\ell}A_{ij}A_{jk}A_{k\ell}(1-\delta_{ik})(1-\delta_{j\ell})z_iz_{\ell}
n_jn_kn_{\ell}, \ee where $z_i=k_i-1$.

After that, the algorithm can be applied exactly in the same way as we
did for the system with two body interactions.  We use the algorithm
to minimize the energy function of a system with $3,4,5$-body
interactions as shown in Extended Data Fig. \ref{fig:taufull}.  Two
comments are in order. Firstly, we note that the eigenvalue is zero
for a larger interval of values of $q$ with respect to the one
computed in the system with $2$-body interactions. This observation
corroborates the idea that in the limit of infinitely many
interactions the eigenvalue jumps at $q_c$ from zero to one.

In Extended Data Fig. \ref{fig:tau_pc}a we show the threshold $q_c(N)$
as a function of the system size for different values of the order of
the many-body interactions $\rho$. The thermodynamic limit for each
many-body interaction $q_c^\infty(\rho)$ is obtained by $N\to \infty$
(the $y$-intercept in the figure).  The value at $\rho=1$ represents
the system with one-body interaction (equivalent to HD).  The value
$\rho=2$ represents the system with $2$-body interactions and energy
function $|\mathbf{w}_1(\mathbf{n})|^2$; $\rho=3$ corresponds to the
system with (at most) $3$-body interactions and energy
$\la\mathbf{w}_1(\mathbf{n})|\mathcal{M}|\mathbf{w}_1(\mathbf{n})\ra$,
as summarized in the following Table for the first three levels.
\\ \\
\begin{tabular}{|r|c|c|}
\hline
 &\ \  Even interactions\ \  &\ \  Odd interactions\ \  \\
\hline
\ \ Energy function\ \ &\ \ $\mathcal{E}_\ell({\bf n}) = |{\bf w}_\ell({\bf n})|^2$\ \ &\ \  $\mathcal{E}'_\ell({\bf n}) = \la{\bf w}_\ell({\bf n})|\hat{\mathcal{M}}|{\bf w}_\ell({\bf n})\ra$\ \ \\
\hline
\ \ Order of Interactions\ \  &\ \ $\rho = 2\ell$ \ \  &  $\rho = 2\ell+1$ \\
\hline
\ \ Leading diagram for $\ell=0$\ \  &
 & \ \ \ \ \ \ \ \ \ \ \ 
\begin{fmffile}{L0odd}
\parbox{10mm}{
\begin{fmfgraph*}(10,3)
\fmfbottom{i1}
\fmfdotn{i}{1}
\fmftop{o1}
\fmf{photon}{o1,i1}
\end{fmfgraph*}
}\end{fmffile}
one-body \\
\hline
\ \ Leading diagram for $\ell=1$\ \  &
\begin{fmffile}{L1even}
\parbox{10mm}{
\begin{fmfgraph*}(15,3)
\fmfdotn{i}{1}
\fmfdotn{o}{1}
\fmfleft{i1,i2}
\fmfright{o1,o2}
\fmf{photon}{i2,i1}
\fmf{fermion}{i1,o1}
\fmf{photon}{o2,o1}
\end{fmfgraph*}
} \end{fmffile}\ \ \ \ 
 two-body & \ \ \ \ \ \ \  
\begin{fmffile}{L1odd}
\parbox{10mm}{
\begin{fmfgraph*}(20,3)
\fmfdotn{i}{1}
\fmfdotn{o}{1}
\fmfleft{i1,i2}
\fmfright{o1,o2}
\fmf{photon}{i2,i1}
\fmf{photon}{o2,o1}
\fmf{fermion}{i1,v1,o1}
\fmfdotn{v}{1}
\end{fmfgraph*}
}\end{fmffile}\ \ \ \ \ \ \ \ \ \
three-body \\
\hline
\ \ Leading diagram for $\ell=2$\ \  &
\begin{fmffile}{L2even}
\parbox{10mm}{
\begin{fmfgraph*}(28,3)
\fmfdotn{i}{1}
\fmfdotn{o}{1}
\fmfleft{i1,i2}
\fmfright{o1,o2}
\fmf{photon}{i2,i1}
\fmf{photon}{o2,o1}
\fmf{fermion}{i1,v1,v2,o1}
\fmfdotn{v}{2}
\end{fmfgraph*}
}\end{fmffile} \ \ \ \ \ \ \ \ \ \ \ \ \ \  four-body & \ \ \ \ 
\begin{fmffile}{L2odd}
\parbox{10mm}{
\begin{fmfgraph*}(35,3)
\fmfdotn{i}{1}
\fmfdotn{o}{1}
\fmfleft{i1,i2}
\fmfright{o1,o2}
\fmf{photon}{i2,i1}
\fmf{photon}{o2,o1}
\fmf{fermion}{i1,v1,v2,v3,o1}
\fmfdotn{v}{3}
\end{fmfgraph*}
}\end{fmffile}\ \ \ \ \ \ \ \ \ \ \ \ \ \ \ \ \ \ five-body
 \\
\hline
\end{tabular}
\\
\\

In Extended Data Fig. \ref{fig:tau_pc}b, we extrapolate the infinite
size threshold $q_c^{\infty}(\rho)$ to the limit of $\infty$-body
interactions, i.e., for $\rho\to\infty$.  The scaling of
$q_c^{\infty}$ with $1/\rho$ is well consistent with a linear
behaviour.  We obtain the $\rho=\infty$ limit of
$q_c^{\infty}(\rho=\infty)\equiv q_c^{\mathrm{opt}}$ from a
least-squares fit.  For ER networks with average degree $\la k\ra=3.5$
studied here, we find $q_c^{\mathrm{opt}}=0.192(9)$. This is the value
of the optimal threshold shown in Fig. \ref{fig:synthetic}a in the
main text.

\section{CI Algorithm}
\label{CI}

We have shown so far that the problem of finding the optimal set of
influencers can be solved by minimizing the following cost function
which is the leading order approximation in $1/N$:

 \be E_\ell(\mathbf{n}) =
\sum_{i=1}^{N} z_i
\sum_{j\in\partial\mathrm{Ball}(i,\ell)}\left(\prod_{k\in\mathcal{P}_{\ell}(i,j)}n_k\right)
z_j\ ,\label{eq:CI_cost} \ee where $E_\ell(\mathbf{n}) =
|\mathbf{w}_{(\ell+1)/2}|^2$ for $\ell$ odd (corresponding to the
energy function $\mathcal{E}(\mathbf{n})$ in
Eq. S\ref{eq:energyEOmultibody}), and $E_\ell(\mathbf{n}) =
\la\mathbf{w}_{\ell/2}|\Mia|\mathbf{w}_{\ell/2}\ra$ for $\ell$ even
(corresponding to $\mathcal{E}'(\mathbf{n})$ in
Eq. S\ref{eq:energyEOmultibodyOdd}).  We recall that $z_i=k_i-1$.  We
define the collective influence strength at level $\ell$, of node $i$
as: \be \mathrm{CI}_{\ell}(i)\ =\ z_i
\sum_{j\in\partial\mathrm{Ball}(i,\ell)}\left(\prod_{k\in\mathcal{P}_{\ell}(i,j)}n_k\right)
z_j\ , \ee and we can rewrite Eq. \eqref{eq:CI_cost} as: \be
E_\ell(\mathbf{n})\ =\ \sum_{i=1}^{N} \mathrm{CI}_{\ell}(i)\ .  \ee
Notice that $\mathrm{CI}_{\ell}(i)$ is basically the same as the
fitness $b_i$ of the EO algorithm, and precisely:
$\mathrm{CI}_{\ell}(i)=b_in_i$.  A fast and efficient way to minimize
the cost function $E_\ell(\mathbf{n})$ is to adaptively remove the
nodes with the highest collective influence $\mathrm{CI}_{\ell}(i)$.
When all the nodes are present, corresponding to
$\mathbf{n}=\mathbf{1}$, $\mathrm{CI}_{\ell}(i)$ evaluates: \be
\mathrm{CI}_{\ell}(i)\ =\ z_i
\sum_{j\in\partial\mathrm{Ball}(i,\ell)}z_j\ .
\label{eq:cimaintext}
\ee 

This is the expression of $\mathrm{CI}_{\ell}(i)$ given in the main
text Eq. \eqref{eq:CI}.  By computing this quantity for each node, we
can find the one with the largest collective influence and then remove
it. We stress that the frontier of the Ball:
$\partial\mathrm{Ball}(i,\ell)$ consists of all the nodes $j$ that are
at a distance $\ell$ from $i$, the distance is measured as the minimum
path between $i$ and $j$. This definition is consistent with the fact
that we have neglected the NB walks with loops in the definition of
the energy functional Eq. S\ref{eq:Costfunction2} for large networks,
and therefore also in $\mathrm{CI}_{\ell}(i)$
Eq. S\ref{eq:cimaintext}, since they scale as $O(1/N)$ in random
networks as discussed in Section \ref{NBinterpretion}.

After the removal, the network consists of $N-1$ nodes, and we can
proceed as before, looking for the next node with the largest
$\mathrm{CI}_{\ell}$.  Since the removal of the first node changes the
degree of its neighbours, we need to decrease their degrees by one
before recomputing their $\mathrm{CI}_{\ell}$.  Removing one by one
the nodes according to this adaptive principle we can destroy the
network in a nearly optimal and very fast way. Besides, we can
significantly speed up the algorithm by decimating a finite fraction
of nodes at each step (see Section \ref{logn}).  The algorithm's
performance increases by using larger values of the radius $\ell$ of
the $\mathrm{Ball}(i,\ell)$. In Extended Data Fig. \ref{fig:allLevels}
we show the results for different values of $\ell$. We observe that
already for $\ell=3,4$ the algorithm reaches the top performance.

When $\ell$ becomes larger than the network diameter,
  then $\mathrm{CI}_\ell(i)=0$. In this situation different nodes are
  not distinguishable by the algorithm, and thus, the method is
  basically indistinguishable from a random one. Thus, the parameter
  $\ell$ should not exceed in practice the original network
  diameter. 
We also notice that dangling ends give zero contribution by CI, and
hence they are ignored by the algorithm. This is expected since
dangling ends should have zero influence in the network.

The CI algorithm Eq. (\ref{eq:CI}) is based on Eq.
  (\ref{eq:Costfunction}) which contains the many-body collective
  interactions that we refer to as ``collective influence''.  The CI
  algorithm incorporates the collective effects by considering the
  adaptive nature of the algorithm.  The adaptiveness of the CI
  algorithm, usually called decimation in the spin glass literature
  \cite{montanari}, is a collective way to select influential nodes,
  since the removal of each node depends heavily on the history of the
  process.


\subsection{Optimization for $G(q)\neq 0$}
\label{neq}

The theory we developed for the optimal fragmentation of networks
allows us to compute the optimal influence threshold $q_c$, i.e. the
smallest number of nodes to remove such that $G(q_c)=0$, together with
the corresponding configuration $\mathbf{n}^*$.

When $q<q_c$ the giant component is nonzero, a consequence of the fact
that the system of Eqs. \eqref{eq:CavityMess}
has another stable solution different from $\{\nu_{i\to j}\}$
identically zero.  Therefore, for $q<q_c$ the stability of the new
solution $G(q)\neq 0$ is no more controlled by the non-backtracking
operator, but a more complicated operator comes into play that depends
on the form of the solution itself.  To find the spectrum (or even the
largest eigenvalue) of this new matrix we have necessarily to know the
solution of the problem. This circumstance depauperates the method of
its power in $q<q_c$, since we need the solution to the problem to
solve the problem itself. In the regime $q>q_c$, where $G(q)=0$, this
solution can be easily guessed, as we did, but for $q<q_c$ no simple
ansatz can be adopted.

What can we do in the regime $q<q_c$ to minimize the size of the giant
component?

We know that the configuration $\mathbf{n}^*$ corresponds to a zero
giant component. Assuming that this configuration is the optimal one
(this hypothesis is not crucial in what follows and can be relaxed by
saying that $\mathbf{n}^*$ is the best approximation to the true
optimum), then the optimal trajectory in the configuration space,
starting from the point $\mathbf{n}=(1,1,1,\dots,1)$ corresponding to
$q=0$ and $G(0)=1$, must end up at the point $\mathbf{n}^*$ at $q=q_c$
where $G(q_c)=0$.

So far we know the final point $\mathbf{n}^*$ and we would like to
travel back the optimal trajectory up to the initial point
$\mathbf{n}=\mathbf{1}$.  In order to do that, let us suppose to
decrease infinitesimally the fraction of removed nodes $q$ from its
critical value $q_c$, that is $q=q_c-dq$ [$dq$ can be taken equal to
  $1/N$, so that $Ndq=O(1)$]. This amounts to explore a neighborhood
of the configuration $\mathbf{n}^*$, consisting of a collection of
configurations $\mathbf{n}'$ in which a number $Ndq$ of components
$n_i'=0$ is turned into $n_i'=1$. Here we are making the crucial
hypothesis that the optimal trajectory is continuous, in the sense
that in going from $q\to q-dq$ only a number $Ndq$ of components $n_i$
is changing state. Under this assumption the trajectory can be
followed adiabatically up to the point $\mathbf{n}=\mathbf{1}$.

Mathematically, this can be expressed by saying that the Hamming
distance between two neighbouring optimal configurations $\mathbf{x}$
and $\mathbf{y}$: $d(\mathbf{x},\mathbf{y})=\sum_{i=1}^{N}|x_i-y_i|$,
is equal to $d(\mathbf{x},\mathbf{y})=Ndq$.  This hypothesis may not
hold in the case where the optimal configuration $\mathbf{x}$
corresponding to $Nq$ removed nodes, and the one $\mathbf{y}$
corresponding to $N(q-dq)$ have an Hamming distance much larger than
$Ndq$. In this case the optimal trajectory has discontinuities,
jumping from one point to another which are not close to each
other. Physically this correspond to the fact that the optimal state
$\mathbf{y}$ cannot be obtained from the optimal state $\mathbf{x}$ by
flipping a finite number of components $n_i$, but requires a global
rearrangement of the system, which amounts to change the state of much
more variables $n_i$, whose number scales as $N^{\alpha}$ with
exponent $\alpha\in(0,1]$. This situation takes place in spin-glass
systems with Full-RSB thermodynamics, where this chaotic behaviour is
observed as a function of the temperature \cite{caos} (or as a
function of other control parameters like the bond strengths and the
magnetic field).  In that case, when the system is cooled from a
temperature $T$ to $T-dT$ (with $T$ below the critical point:
$T<T_c$), the Gibbs state corresponding to the higher temperature does
not survive after the cooling, but, instead, a completely new
equilibrium state appears at $T-dT$ (i.e. if we sample a typical
equilibrium configuration at temperature $T$, this will be very
distant, in the Hamming sense, from a configuration sampled at
$T-dT$).

It is highly plausible that the same situation (a chaotic trajectory)
is realized also in our problem. To keep things simple, we don't
explore this scenario, and analyze only the consequences deriving from
the hypothesis of a smooth optimal trajectory from $\mathbf{n}^*$ back
to $\mathbf{1}$. This approach will give us a very efficient
algorithm to minimize the giant component in all the interval
$q\in[0,q_c)$, at no additional computational cost. Therefore we take
  this performance as a practical justification of the main
  assumption, leaving to a future work the treatment of the more
  complicated chaotic scenario.

To take up the threads of our discussion, let us assume that optimal
configurations lie close to each other.  Knowing the optimal
configuration at the fraction $q$, we should be able to find the new
optimal one at $q-dq$ by changing the state of few variables, and
actually only one if we take $dq=1/N$. Practically we have to find the
new optimal configuration by changing the state of a single variable
from $0$ to $1$.
Practically we proceed in the following way.  At $q=q_c$, we have
$G(q_c)=0$. The corresponding configuration $n^*$ contains $Nq_c$
variables $n_i=0$ (removed nodes), and $N(1-q_c)$ variables $n_i=1$
(nodes that are present).  At this point we start to add back to the
network the nodes using the following algorithm.  We assign to each
removed node $n_i=0$ an index $c(i)$, which is calculated as the
number of clusters that node $i$ would join if it were reinserted in
the network, independently of their sizes.  Then we put back in the
network node $i^*$ such that $i^* = \mathrm{argmin}_{n_i: n_i=0}
c(i)$, by changing $n_i^*=0$ into $n_i^*=1$ (see Extended Data
Fig. \ref{fig:maxfrac}).  The idea behind this method is the fact that
we want to keep a maximally fragmented network after each reinsertion
of nodes.  We keep on reinserting nodes using the same criterion,
until no node is left for which $n_i=0$.  After each reinsertion the
indexes $c(i)$ are recalculated, and then the new node with minimum
$c(i)$ is chosen.

The running time of this algorithm is $O(MN\log N)$, where $M$ is the
number of edges.  Indeed, $O(M)$ operations are needed to assign the
indexes $c(i)$ and $O(N\log N)$ to sort them.  As we did for the case
of the main CI algorithm, the time complexity of this algorithm can be
reduced to $O(N\log N)$ (if $M=O(N)$), by reinserting a finite
fraction of nodes at time.

\subsection{Scalability of the CI algorithm}
\label{logn}



The time complexity needed to compute the quantity
$\mathrm{CI}_\ell(i)$ is proportional to the number of edges inside
the ball $\mathrm{Ball}(i,\ell)$. Since the radius $\ell$ is taken
finite, this calculation takes a time $O(1)$ for each node (even if
the prefactor increases with $\ell$). Thus, to compute the
$\mathrm{CI}_\ell(i)$ for all $i$ requires $O(N)$ operations. Sorting
the $\mathrm{CI}_\ell(i)$'s takes $O(N\log N)$.  The algorithm is
terminated when a number $Nq_c$ of nodes is removed. Therefore,
removing the nodes one-by-one, the total time complexity would be
$O(N^2\log N)$.  Actually we can keep the computational complexity to
$O(N\log N)$ without losing any performance, by simply removing a
finite fraction of nodes at each step (with a prefactor depending on
the percentage of nodes fixed at time).  In the next Section we
explore the performance of the CI algorithm for different
adaptive/decimation steps.


\subsection{ Effect of the percentage of fixed nodes during adaptive CI} 

In this section we show the performance of CI as a function of the
percentage of removed nodes at each step of the adaptive
algorithm. Indeed, removing a finite fraction of nodes at time reduces
the time complexity from $N^2\log N$ (corresponding to the one-by-one
removal) to $N\log N$.  In Extended Data Fig. \ref{fig:decimation} we
show the effect of the percentage of fixed nodes at each adaptive step
on an ER network of $N=10^5$ nodes and average degree $\la k\ra =
3.5$. As the figure shows, the performance of CI is practically
unaffected by the removal of up to $0.25\%$ of nodes at time
(i.e. $250$ nodes for the considered network) compared to the
one-by-one removal.

\section{Comparison with other heuristic methods}
\label{comparison}

In the main text Fig. \ref{fig:synthetic} we compare our solution with
heuristics: high-degree and high-degree adaptive \cite{barab,cohen},
PageRank \cite{pagerank}, kcore \cite{gallos}, eigenvector \cite{ec}
and closeness \cite{cc} centralities. We also compare in
Fig. \ref{fig:mobileMex}, for Twitter and Mobile Networks, CI with
HDA, HD, PR and k-core which are the only heuristics that are scalable
to these large-scale datasets. It remains to compare our results with
other popular heuristics which do not scale well with system size, and
therefore we use smaller systems of $10^4$ nodes: betweenness
centrality \cite{bc} and equal-graph-partitioning \cite{chen}.  We use
the same size and parameters of the scale-free network used in
\cite{chen}. The final comparison is with BP \cite{zecchina2} and it
will be done in the next section.

{\bf Betweenness centrality (BC) \cite{bc}.} Betweenness centrality of
node $i$ is the sum of the fraction of all-pairs shortest paths that
pass through $i$. BC is a very popular tool for network analysis,
which has applications in different fields, from community detection
to the human brain. However, it comes with a high computational cost
that prevents the examination of large graphs of interest.  The best
algorithm for BC computations has $O(NM)$ time complexity for
unweighted networks with $N$ nodes and $M$ vertices. It is not fast
enough, for example, to handle our 10$+$ million people network.
Extended Data Fig.  \ref{fig:comparison} shows its performance. It
does not outperform other centralities.

{\bf Equal-graph-partitioning (EGP) \cite{chen}.}  This method aims at
dividing the network in clusters of equal size.  It can behave well
for homogeneous networks, like random regular graph, where an equal
partition could be expected to destroy the network efficiently, but
loses a lot of performance for heterogeneous networks, like scale-free
networks, as we can see from Extended Data Fig. \ref{fig:comparison}.
Notice that we have used the same network parameters, size, and
definition of EGP as given in \cite{chen} in the comparison of
Extended Data Fig. \ref{fig:comparison}. In fact, we reproduce the
same curve and $q_c$ as found for EGP in \cite{chen}.


\section{Comparison with Belief Propagation algorithm of Altarelli {\it et al.} \cite{zecchina2}  } 
\label{BP}

The comparison with the Belief Propagation (BP) method proposed in
Ref. \cite{zecchina2} to optimally immunize a network deserves
particular care, because this method does not apply directly to the
problem we are treating here.  This is due to the fact that the
parameter $p$ in the work of Ref. \cite{zecchina2} (which is noted as
$q$ in \cite{zecchina2}) refers to the fraction of initially infected
individuals. In our work the fraction $p$ is assumed to be zero,
because in epidemic outbreaks the number of initiators of the epidemic
is very small, and typically of order $O(1)$ \cite{quarantine}.
For instance, Sierra Leone's explosion of Ebola cases in 2014 appeared
to stem from one traditional healer's funeral at which a single source
infected $14$ women; or the SARS outbreak in $2003$ started when one
doctor from China infected nine other guests in a Hong Kong hotel who
then spread the virus throughout the city and to Vietnam and Canada
(source- NY Times August 29, 2014, page A7, ``Outbreak in Sierra Leone
Is Tied to Single Funeral Where 14 Women Were Infected.''). Another
example is the patient zero-hypothesis in the AIDS epidemics
\cite{zeropatient}.

On the other hand the model of Ref. \cite{zecchina2} is valid for
$p>0$, in particular, the results of Ref. \cite{zecchina2} are
illustrated for $p=0.1$. The case $p=0.1$ would imply an epidemic
starting with $10\%$ of the entire population infected independently
at the same time. This would imply, for instance, 0.6 million people
in Sierra Leone spontaneously and independently being infected at the
same time, which would make any targeted immunization intervention
perform equally well in practice.
This result was shown by Ref. \cite{zecchina2} in Fig. 12a: when $p>0$
any reasonable targeted immunization method gives the same result for
the fraction of infected nodes vs immunized nodes. [Fig. 12a in
  \cite{zecchina2} treats the case of $p=10\%$, noted as $q=0.1$ in
  the notation of \cite{zecchina2}, and compares BP with greedy, HDA,
  eigenvector centrality and simulating annealing; all showing the
  same performance].

Therefore, the results of our paper are illustrated for $p = 0$.  That
being said, next, we compare our results with the BP algorithm in the
closest possible regime to ours when $p\to 0$, and also for $p=0.1$.
In the limit $p\to 0$, BP becomes unfeasible because the time
complexity of the BP algorithm diverges as $p^{-3}$ for $p\to 0$, as
we explain below.  The results are shown in Extended Data Figs
\ref{fig:ER_N200}c and \ref{fig:comparisonBP_CI_fgamma} and we
observe that BP does not perform better than CI. Furthermore, the poor
scalability of BP makes it prohibitive for the real networks of 10$+$
million people used in our work.

To perform a comparison, we need to briefly recall the approach of
Ref. \cite{zecchina2} and set the notation. The formulation of the
problem is based on the long time limit of the SIR dynamics, which is
described by the set of variables $\{\nu_i\}$, $i=1,\dots,N$, giving
the probability for each node to be infected after the epidemic
outbreak (in Ref. \cite{zecchina2} the variable $\nu_i$ is called $m_i$, 
but we prefer to use $\nu_i$ to make contact with our notation).
 These variables satisfy the following equations: \be
\nu_i\ = \ p + (1-p)\left[1-\prod_{k\in\partial i}(1-w\nu_{k\to
    i})\right]\ ,
\label{eq:qcav}
\ee where the parameter $p$ is the probability for node $i$ to be
initially infected; $w$ is the probability that a given neighbor $k$
of node $i$ transmits the disease to $i$; and the product on the
r.h.s. is over all neighbours $k$ of node $i$.  The variable $\nu_{k\to
  i}$ (named $m_{k\to i}$ in Ref. \cite{zecchina2}) is the probability that 
node $k$ is infected in a modified
network where node $i$ is absent. Each $\nu_{i\to j}$ is associated with
a directed edge $i\to j$ of the graph, and satisfies the following
equation: \be \nu_{i\to j}\ = \ p + (1-p)\left[1-\prod_{k\in\partial
    i\setminus j}(1-w\nu_{k\to i})\right]\ .
\label{eq:qtrue}
\ee To include the effect of immunization, the authors of Ref. \cite{zecchina2} 
 introduce a binary
variable $\s_i$ for each node $i$, taking values $\s_i=+1$ if node $i$
is immunized, and $\s_i=-1$ if not.  Equations \eqref{eq:qcav} and
\eqref{eq:qtrue} then become: \be
\begin{aligned}
\nu_i\ &= \ \frac{1-\s_i}{2}\left\{
p + (1-p)\left[1-\prod_{k\in\partial i}(1-w\nu_{k\to i})\right]
\right\} \equiv\ F_i(\s_i, \{\nu_{k\to i}\}_{k\in\partial i}) \ , \\
\nu_{i\to j}\ &= \ \frac{1-\s_i}{2}\left\{
p + (1-p)\left[1-\prod_{k\in\partial i\setminus j}(1-w\nu_{k\to i})\right]
\right\} \equiv\ F_{i\to j}(\s_i, \{\nu_{k\to i}\}_{k\in\partial i\setminus j})\ .
\end{aligned}
\label{eq:qcav_immuniz}
\ee To find the optimal immunization set, Ref. \cite{zecchina2}
minimizes the following cost (energy) function $E(\s, {\bf \nu})$: \be
E(\s, {\bf \nu})\ =\ \sum_{i=1}^N
\nu_i\ +\ \mu\sum_{i=1}^N\frac{1+\s_i}{2}\ \equiv\ \sum_{i=1}^N
e(\s_i,\nu_i)\ ,
\label{eq:costBP}
\ee where $\mu$ is a chemical potential controlling the fraction of
immunized nodes.  At this point, Ref. \cite{zecchina2} applies the
cavity method to estimate the single site marginal $P_i(\s_i)$, which
gives the probability that node $i$ is immunized. Approximating the
network with a tree rooted on node $i$, the authors of Ref. \cite{zecchina2}
derive the following equation to assess the probability distribution
$P_i(\s_i)$: \be P_i(\s_i)\ \simeq\ \int
\left(\prod_{k\in\partial i}\dd \nu_{k\to i}\dd \nu_{i\to k} Q_{k\to
  i}(\nu_{k\to i},\nu_{i\to k})\right) \mathrm{e}^{-\beta
  e(\s_i,\nu_i)} \prod_{k\in\partial i} \delta\left[\nu_{i\to k} -
  F_{i\to k}\right]\ , \ee where $\beta$ is the inverse temperature,
and the functions $Q_{k\to i}(\nu_{k\to i},\nu_{i\to k})$ satisfy the
following BP equations: \be Q_{i\to j}(\nu_{i\to j},\nu_{j\to
  i})\simeq\ \sum_{\s_i} \int \left(\prod_{k\in\partial i\setminus
  j}\dd \nu_{k\to i}\dd \nu_{i\to k} Q_{k\to i}(\nu_{k\to i},\nu_{i\to
  k})\right) \mathrm{e}^{-\beta e(\s_i,\nu_i)} \prod_{k\in\partial i}
\delta\left[\nu_{i\to k} - F_{i\to k}\right]\ .\label{eq:cav_zec}  \ee 

Next, we iterate the BP equations to perform a comparison with our
approach. These equations do not have an analytical solution, so
that, following Ref. \cite{zecchina2}, we solve them numerically by
discretizing the function $Q_{i\to j}(\nu_{i\to j},\nu_{j\to i})$ in a
number $\mathcal{N}_{bin}$ of bins. The computational cost to update
each message $Q_{i\to j}$ is of order $O(\mathcal{N}^{k_i-1})$, where
$k_i$ is the degree of node $i$.  This makes the algorithm practically
unfeasible on networks having nodes with large degree (think e.g. to
scale free graphs). To overcome this problem, the authors of
Ref. \cite{zecchina2} use a convolution trick, which reduces the
computational cost to $O((k_i-1)\mathcal{N}_{bin}^3)$. Using the
convolution method of Ref. \cite{zecchina2}, Eq. \eqref{eq:cav_zec}
reads: \be
\begin{aligned}
&Q_{i\to j}(\nu_{i\to j},\nu_{j\to i})\ \simeq\
\mathrm{e}^{-\beta\mu}\left[
\prod_{k\in\partial i\setminus j}\int
\dd \nu_{k\to i}\  
Q_{k\to i}(\nu_{k\to i},0)\right]\ 
 \delta(\nu_{i\to j})\ +\ \\
&\ +\ \frac{1}{1-p}\ 
M^{(k_i-1)}\left(\frac{1-\nu_{i\to j}}{1-p},\ (1-w\nu_{j\to i})\frac{1-\nu_{i\to j}}{1-p}\right)\ 
\mathrm{e}^{-\beta[1-(1-\nu_{i\to j})(1-w\nu_{j\to i})]}\ 
\Theta(\nu_{i\to j}-p)\ ,
\label{eq:BP_efficientForm}
\end{aligned}
\ee where the function $M^{(n)}(x,y)$ is defined iteratively by the
convolution: \be
\begin{aligned}
M^{(n)}(x,y) &= \int_0^1 \dd x_1\dd x_2\ \delta(x -
x_1x_2)\ M^{(n-1)}(x_1,y)\ M^{(1)}(x_2,y)\ ,\\ M^{(1)}(x,y) &=
\int_0^1 \dd \nu\ \delta[x -
  (1-w\nu)]\ Q\left(\nu,1-(1-p)\frac{y}{1-w\nu}\right)\ .
\end{aligned}
\label{eq:M_convolution}
\ee
In all the following numerical results we will always use the efficient form \eqref{eq:BP_efficientForm}
of the BP equations. 

From the knowledge of the functions $Q_{i\to j}(\nu_{i\to j},\nu_{j\to
  i})$, Ref. \cite{zecchina2} computes the probability distribution
$\mathcal{Q}_i(\nu_i)$ that node $i$ has been infected during the
epidemics [$\nu_i$ is defined in the first line of
  Eq. \eqref{eq:qcav_immuniz}], which is given by: \be
\mathcal{Q}_i(\nu_i)\ \simeq\ \mathrm{e}^{-\beta\mu}\left[
  \prod_{k\in\partial i}\int \dd \nu_{k\to i}\ Q_{k\to i}(\nu_{k\to
    i},0)\right]\ \delta(\nu_{i})\ +\ \frac{\mathrm{e}^{-\beta
    \nu_i}}{1-p}\ M^{(k_i)}\left(\frac{1-\nu_i}{1-p},\ \frac{1-\nu_i}{1-p}\right)\ \Theta(\nu_i-p)\ .
\ee Moreover, they estimate the single spin marginal $P_i(\s_i)$ as:
\be P_i(\s_i)\ \simeq\ \mathrm{e}^{-\beta\mu}\left[
  \prod_{k\in\partial i}\int \dd \nu_{k\to i}\ Q_{k\to i}(\nu_{k\to
    i},0)\right]\ \delta(\s-1)\ +\ \left[\int\dd
  x\ \mathrm{e}^{-\beta[1-(1-p)x]}\ M^{(k_i)}(x,x)\right]\ \delta(\s+1)\ .
\ee Once the authors of Ref. \cite{zecchina2} obtained the probability
distributions $\mathcal{Q}_i(\nu_i)$ and $P_i(\s_i)$, they can compute
the average fraction of infected nodes $f$: \be f =
\frac{1}{N}\sum_{i=1}^{N}\ \int \dd
\nu_i\ \nu_i\ \mathcal{Q}_i(\nu_i)\ ,
\label{eq:infected}
\ee
and the average fraction of immunized nodes $q$:
\be
q =\ \frac{1}{N}\sum_{i=1}^N\frac{1+\la\s_i\ra}{2}\ =\ 
\frac{1}{N}\sum_{i=1}^N\frac{1}{2}\left(1+\sum_{\s_i}\ \s_i\ P_i(\s_i)\right)\ .
\label{eq:immunized}
\ee

\subsection{BP adaptive}
\label{sec:BPadap}

Before we compare BP with our method, we need to illustrate the BP
method on a ER network to clarify some technical issues. We consider a
small ER random graph of $N=200$ nodes, where BP can be studied, and
average degree $\la k\ra=3.5$.  We use the following values of the
parameters: fraction of initially infected nodes $p=0.1$, inverse
temperature $\beta = 3.0$, and transmission probabilities $w=0.4, 0.5,
0.6, 0.7$.  The results are shown in Extended Data
Fig. \ref{fig:ER_N200}a, where we plot the fraction of infected nodes
$f$ versus the fraction of immunized nodes $q$.  As already noticed in
Ref. \cite{zecchina2}, we observe that, while for $w=0.4$ the curve is
continuous in the whole range of values of $q$, for larger values of
$w$ the curves get interrupted at a certain value of $q$.  This is due
to the fact (as mentioned by the authors of Ref. \cite{zecchina2})
that the free-energy is non-convex in that region of values of $q$,
and the chemical potential is flat as shown in Extended Data
Fig. \ref{fig:ER_N200}b. Therefore, all values of $q$ in that
region cannot be explored using the normal BP method.  Physically, the
fact that the thermodynamical potential becomes non-convex is the
signature of a phase transition happening at a certain value of $w$.
To overcome this problem, the authors of Ref. \cite{zecchina2} suggest
the following technique. One adds an extra magnetic field $H$ to the
energy function $E(\s, {\bf \nu})$ in Eq. \eqref{eq:costBP}, which is
then adjusted at each update of the BP equations to keep fixed the
value of immunized nodes $q$.  We implemented this adaptive BP method
(called 'fixed density BP' in Extended Data Fig. \ref{fig:ER_N200}a),
and we found that the missing part of the curve can be effectively
reconstructed for some values of $w$ larger than $w=0.4$. Nonetheless,
for even bigger values of $w$, we found that the missing part of the
curve cannot be fully reconstructed, since the adaptive algorithm does
not converge anymore. Usually the non-convergence of the BP algorithm
is associated to the existence of a phase transition (different from
the aforementioned one), marking the limit of validity of the
replica-symmetric cavity method. We then expect that for those values
of $q$, where reconstruction is impossible, a different BP method has
to be used, in order to deal with the phenomenon of replica symmetry
breaking.

In the next Section we compare BP with CI in two different settings.
The first case is the closest one to the regime where CI is defined
(i.e. when $p\to 0$ and $w\to 1$). The second type of comparison
is the case where $p>0$ and the BP method can be used for all values of
$q\in[0,1]$.  


\subsection{Comparison}
\subsubsection{First comparison}

Here, we compare BP in the closest possible regime to CI, i.e. for
$p\to 0$ and $w\to 1$.  Solving numerically the BP equations requires
to discretize the functions $Q_{i\to j}(\nu_{i\to j},\nu_{j\to i})$ in
a number $\mathcal{N}_{bin}$ of bins of the order
$\mathcal{N}_{bin}\sim 1/p$, in order to have good numerical accuracy,
because the smallest possible non-zero value assumed by $\nu_i$ is
$\nu_i=p$, as stated in Ref. \cite{zecchina2}.  The BP running time is
of order $O(M\mathcal{N}_{bin}^{3})$, $M$ being the number of edges in
the graph. The factor $\mathcal{N}_{bin}^{3}$ comes from the
computation of the function $M^{(n)}(x,y)$ in
Eq. \eqref{eq:M_convolution}, that requires a double integration over
$x_1$ and $x_2$ (giving a factor $\mathcal{N}_{bin}^{2}$), for each
value of $y$ (giving an extra factor $\mathcal{N}_{bin}$).  Since
$\mathcal{N}_{bin}\sim 1/p$, the BP running time diverges as $p^{-3}$
for $p\to 0$.  This is the reason why we cannot use BP directly for
$p=0$.  So, we set $p=0.01$, as small as possible, and $w=0.99$, as
close to 1 as possible, in the BP algorithm. Moreover, we choose a
quite high value of the inverse temperature $\beta = 10$, close enough
to the zero temperature limit.  Note that for this value of
$p=10^{-2}$, the number of bins needed for good numerical resolution
is of the order of $\mathcal{N}_{bin} \sim 10^2$, which introduces a
prefactor in the computational cost of the algorithm already of order
$10^6$.

We compare CI and BP on a small ER network of $N=10^3$ nodes and
average degree $\la k\ra=3.5$, where BP can be run efficiently to do a
study over the parameter space.  Since for those values of $p$ and $w$
we cannot compute the full curve $f(q)$ (for the reasons explained in
Section \ref{sec:BPadap}), we compare the giant component found by BP
and the one obtained with CI (notice that $f$ coincides with the giant
component $G$ in the limits $p\to 0$ and $w\to 1$).  In order to
choose which nodes have to be removed according to BP, we use the
following criterion: we run BP and we assign to each node the value of
the sign of its magnetization: $\mathrm{sign}(\la \s_i\ra)$ (the value
of the inverse temperature we chose, $\beta = 10$, is sufficiently
high for the spins $\s_i$ to be highly polarized). Then, node $i$ is
removed if $\mathrm{sign}(\la \s_i\ra)=1$, and it is not if
$\mathrm{sign}(\la \s_i\ra)=-1$.  When BP does not converge, we stop
the algorithm after a maximum number of iterations and we use the
unconverged marginals to assign the magnetizations. In this way we can
draw the full curve $G(q)$ even if BP does not converge. The result of
the comparison is shown in Extended Data Fig. \ref{fig:ER_N200}c,
where we can see that BP is not better than CI, and performs slightly
worst than the HDA method.





To conclude this section, we mention two other versions of the BP
algorithm.  The first one is developed in Ref. \cite{semerjian}. The
technique used in Ref. \cite{semerjian} is the same BP technique as
the one introduced by Altarelli {\it et al.}
\cite{zecchina1,zecchina2}.  From the analytical point of view,
Ref. \cite{semerjian}  improves the lower bound on the threshold
$q_c$ by considering the effects of 1 step replica symmetry breaking
(1RSB), obtaining slightly larger lower bounds than those predicted by
the replica symmetry (RS) approach of Altarelli {\it et al.}:
$q_c^{\rm RS} \le q_c^{\rm 1RSB}$, or $\theta_{\rm min,0} \le
\theta_{\rm min,1}$, respectively in the notation of
Ref. \cite{semerjian}.  Hence the lower bound in Ref. \cite{semerjian}
is larger than the one obtained by Altarelli {\it et al.}

The second variant of the BP algorithm is used in Refs.
\cite{zhuo,zhuo2} for solving the undirected feedback vertex set
problem.  This algorithm, named Belief Propagation Guided Decimation
(BPD), improves the time complexity of the BP approach of
\cite{zecchina1,zecchina2,semerjian} and can be tested in SF
networks. In Extended Data Fig. \ref{fig:ER_N200}d we compare the BPD
with CI algorithm where we find evidence of the best performance of
CI.

\subsubsection{Second comparison}

In this section we compare CI and BP in a different way; this time in
the case where BP is well defined and CI is not, for parameter values
$p\neq 0$ and $w\neq 1$. Thus, we use $p=0.1$ and $w=0.5$. So, this
second comparison represents the opposite situation with respect to
the previous one.  We compare the two methods in the following way. We
use BP to compute directly the fraction of infected nodes $f(q)$ as a
function of the fraction $q$ of immunized nodes by means of
Eqs. \eqref{eq:infected}--\eqref{eq:immunized}.  To compare against
CI, we have to simulate explicitly the SIR process, since we cannot
estimate directly the $f(q)$. More precisely, we first identify the
immunized nodes with CI, and then we run the SIR algorithm to obtain
the final fraction of infected individuals $f(q)$.  

The result of the comparison is reported in Extended Data
Fig. \ref{fig:comparisonBP_CI_fgamma}, for an ER network of $N=10^3$
nodes and average degree $\la k\ra = 3.5$. The values of the initially
infected individuals $p$ and the transmission probability $w$ are
$p=0.1$ and $w=0.5$. The value of the inverse temperature $\beta$ used
in the BP algorithm is $\beta =10$ (for the portion of the curve where
the adaptive BP algorithm is needed, we chose the lowest possible
temperature such that the algorithm converges).  As the figure shows,
there is little difference between BP and CI, with CI slightly better
for small $q$.  Moreover we checked that even using HDA gives more or
less the same results as BP and CI, as the authors of
Ref. \cite{zecchina2} also show in Fig. 12a of their work. Therefore,
in the case when $p>0$ (meaning that a finite fraction of the entire
network is already infected from the very beginning of the epidemic
outbreak), any reasonable targeted immunization technique gives the
same result. That is, the optimization achieved by any method is
washed out by the large number of already infected people, and all
strategies perform equally well.  On the contrary, in the case when
$p=0$, i.e. when the epidemic is initiated by a superspreader event
$O(1)$, different strategies behave very differently, with CI being
the best so far.

We notice, en passant, that the analytical BP estimation of $f(q)$
gives a lower bound on the actual $f(q)$. That is, if we used the same
procedure as for CI, by first identifying the immunized nodes and then
computing the fraction of infected ones trough the outcome of the SIR
process, the resulting curve $f(q)$ would lie above the analytical BP
estimation.


Finally, we note that EO estimates the optimal numerical value of the
threshold $q_c$ as a numerical extrapolation to $N\to \infty$ and
$\ell\to\infty$.  While EO can estimate this threshold accurately
(providing an upper bound very close to the real optimum), it cannot
provide the actual optimal configuration $\mathbf{n}^*$ for large
system sizes. This is of course a general feature due to the
NP-hardness of the problem.

Indeed, the EO method we use to estimate the value of optimal
threshold for ER random graphs in Extended Data Fig. \ref{fig:tau_pc}b
may not be the only way to assess analytically that result. Indeed,
there are other methods to approximate the location of the optimal
threshold, which can provide lower or upper bounds. For instance, the
BP (or cavity) method investigated above writes down approximate
self-consistent equations for the optimization problem, that are
solved iteratively to get an estimation of the optimal
threshold. Often, the BP equations do not converge (as a consequence
of the NP-hardness of the problem), but an attitude has gained a
foothold in the statistical physics community, which amounts to ignore
convergence problems and use anyway an unconverged solution as an
estimation of the optimal threshold.  Indeed, in all cases where this
approach has been pursued, it has been shown that the BP analytical
prediction provides a lower bound to the optimal threshold. On the
contrary, the EO method employed in our work provides an upper bound
to the optimal threshold. Therefore, different analytical methods can
give, indeed, predictions which are close to each other and, hence,
close to the optimal value of the threshold.

Furthermore, it may not be impossible to find the exact analytical
value of the threshold even if the problem is NP, as in the case of
the Sherrington-Kirkpatrick model for spin glasses \cite{montanari},
where the Parisi ansatz provides the correct solution.  We also notice
that analytical solutions are based on the analysis of the most
probable case in general, but not for a specific instance of the
problem. Indeed, not every NP-complete problem can be analysed in
this way. Some problems do not permit a discussion based on the most
probable case. A random chosen satisfiability problem, for example,
is almost always easy to solve, because a random sequence of symbols
almost always does not make sense.

In our problem of optimal percolation, even though the numerical value
of the threshold could be known exactly with EO or other method, the
main problem remains open: finding an optimal configuration that is as
close to the minimal as possible in the large system size. The most
relevant challenge for practical applications of NP problems is not to
estimate theoretically the value of the threshold $q_c$, but to find a
scalable algorithm (for realistic applications should be at most $O(N
\log N)$) which is able to approximate as close as possible a real
optimal configuration $\mathbf{n}^*$ at $q_c$.

Our algorithmic solution to this NP problem is then CI: a scalable
algorithm $\sim O(N \log N)$ that contains the physics of the optimal
configuration, and it is necessarily an approximation to the true
optimum; being a $O(N\log N)$ algorithm it cannot give the optimal
solution unless P = NP.  Thus, proper benchmarking does not compare
the analytical value of the threshold $q_c$.  Benchmarking should be
carried out by comparing the optimal configurations with the
corresponding giant components for large size networks, which should
be at least of the order of $10^7+$ nodes (as we have done in Fig
\ref{fig:mobileMex}d), showing an improvement both in the running time
and efficiency.

\section{A new paradigm of influence in social media: Twitter}
\label{twitter}

In the next two sections we show that the performance of our method is
confirmed in two real networks. We study two prototypical examples of
real networks: Twitter web and a social network derived from phone
calls. The former is used to test our theory as a new paradigm of
influence, while the second can be used to design an immunization
protocol in the case of an epidemic outbreak.

We have paved the way to explore the consequences of our theory in 
real networks, where the assumption of tree-like structure that is the
basis of our theory is not necessarily satisfied.
The reason to be interested in such a kind of problem is
that it is manifestly in the interest of man's communal existence to
understand how people increase their influence when they tie one
another. The critical question is to what extent one can define a
measure for influence solely on the basis of social contact network.
The answer might be hard to find, but, at the same time, one cannot
deny that the network itself mirrors the mutual relations of users,
and hence it must contain information about their influence. The
resulting network-based influence estimation can always be
supplemented by measures of activity and engagement. 

With this caveat in mind, our optimal percolation theory uncovers the
optimal influencers in social media. In this context, the measure of
node-influence in social media is the drop in the size of the giant
cluster which would happen if the node in question were removed. Such
a measure of influence is related to the ability to spread the news
to the largest portion of the network as shown by our mapping of the
maximal spreading problem in LTM (with $\theta_i=k_i-1$) to optimal
percolation. We test this idea in Twitter, next.

Twitter is the online social networking and microblogging service that
has gained worldwide popularity. Here we use the dataset of
approximately 16 million tweets sampled between January 23rd and
February 8th, 2011 and publically shared by Twitter (\url{http://trec.
  nist.gov/data/tweets/}) (also available at \url{http://jamlab.org},
see Ref. \cite{pei} for more details). The natural way to get the
social network is to extract the follower network through Twitter
API. Unfortunately, due to the access rate limit of Twitter API, it is
impossible to obtain the full information of the follower network in a
reasonable time. Furthermore, many of the follower links are not
active. To approximate the social network, we use an alternative way -
the mention network \cite{pei}. In contrast to the normal tweets,
mentions are tweets containing @username and usually include personal
conversations or references. In fact, the mention links have stronger
strength of ties than follower links. Therefore, the mention network
can be viewed as a stronger version of interactions between Twitter
users. In the mention network, if user $i$ mentions user $j$ in
his/her tweets, there exists a link from $i$ to $j$.  In order to
better represent the social contacts, we also add to the network the
retweet relations from the tweets. A retweet (RT @username)
corresponds to content forward with the specified user as the nominal
source. If user $i$ retweets a tweet of user $j$, then a contact is
established between $j$ and $i$. We then consider all links to be
undirected.  In this way, the social network of Twitter is
constructed. The resulting network has $N=469,013$ nodes and $M =
913,457$ edges.

We measure the collective influence of a group of nodes as the drop in
the size of the giant component which would happen if the nodes in
question were removed. The results are shown in
Fig. \ref{fig:mobileMex}a, showing the better performance of CI in
comparison with HDA, PR, HD and k-core. The other heuristics and BP
cannot be run in this large dataset.

In Fig. \ref{fig:mobileMex}b we plot the percentage of fake
influencers (PFI) or false positives as a function of the fraction of
removed nodes $q$. This quantity is defined with respect to the HD
method, and represents the amount of different influencers between HD
and CI.  More precisely, we call $S_{\mathrm{CI}}(q)$ the set of
influencers (i.e. removed nodes) found by CI at a given value of $q$:
\be S_{\mathrm{CI}}(q)=\{x_{\mathrm{CI}}^1, x_{\mathrm{CI}}^2, \dots,
x_{\mathrm{CI}}^{Nq}\}\ , \ee and $S_{\mathrm{HD}}(q)$ the
corresponding vector for HD. Moreover we denote by $|S(q)| = Nq$ the
size of the set. Notice that $|S_{\mathrm{CI}}(q)|$ is upper bounded
by $Nq_c^{\mathrm{CI}}$, i.e. $|S_{\mathrm{CI}}(q)|\leq
Nq_c^{\mathrm{CI}}$, where $q_c^{\mathrm{CI}}$ is the influence
threshold obtained with CI. Indeed $Nq_c^{\mathrm{CI}}$ is the maximum
number of influencers. Analogously, $|S_{\mathrm{HD}}(q)|$ is upper
bounded by $Nq_c^{\mathrm{HD}}$.

We define $\mathrm{PFI}(q)$ as: \be \mathrm{PFI}(q) = 100\times
\left[1- \frac{|S_{\mathrm{CI}}(q)\cap S_{\mathrm{HD}}(q)|}{Nq}
  \right]\ .
\label{pfi}
\ee

In other words, we measure the percentage of different nodes removed
by CI and HD. As shown in Fig. \ref{fig:mobileMex}b, the PFI at the
critical threshold of CI is $\mathrm{PFI}(q_c^{\mathrm{CI}})\sim26\%$,
meaning that HD misses roughly $1/4$ of the total number of (real,
i.e. optimal) influencers.  As a consequence the giant component for
HD is still very large, $G_{\mathrm{HD}}(q_c^{\mathrm{CI}})\sim 0.37$,
and hence, HD needs to keep on removing nodes to fragment completely
the network.  This comes at the price of including a large number of
fake influencers at the end of the process $q_c^{\mathrm{HD}}$, where
$\mathrm{PFI}(q_c^{\mathrm{HD}})\sim48\%$.

In the same way, if one knew the true (optimal) configuration of
influencer, one could analogously define the fake CI influencers which
do not overlap with the optimal ones. Actually, the obtained nearly
optimal set by the CI method has an unknown overlap with the true
optimal solution.  On the other hand, the impossibility to find such
an optimal set (because of the aforementioned prohibitive running
time), makes the CI influencer set the natural substitute for the
optimal one, and, hence, the reference set for studying the overlap
with node sets identified by other methods.

\section{Halting epidemics:
Mobile phone call network}

\label{mobile}

While medicine has made solid advances in the isolation of new
vaccines for an increasing number of diseases, and may expect to make
still greater ones, no certain claim can be established for a
corresponding advance in preventive immunization.

It is of deep social importance to have a fast and optimal
intervention strategies when new outbreaks of disease break
out. Prevention methods are still limited for various reasons: many
virus are responsible of diseases of animals that can be transmitted
to humans and thus causing epidemics. It is difficult, if not
impossible, to control the populations of vectors and natural
reservoirs, or predict what changes in the environment can favor the
epidemics. The development of new drugs is usually not the solution
to the problem.

It is generally accepted that an efficient way to fight epidemic
diseases is the execution of immunization protocols and fast
quarantine procedures, together with the spread of the knowledge of
these dangers and the efforts to remove the environmental causes that
favor them \cite{quarantine}.  Probably a certain percentage of
diseases will always remain undefeated, but if only one can succeed in
reducing to a minority the majority that is today vulnerable, one will
have accomplished a great deal, perhaps indeed everything that can be
accomplished.  In this situation is highly desirable to have a guiding
strategy which enables to select who must be vaccinated or put in
quarantine. Our theory offers a protocol of selection, the closest one
to the optimal. This result is important because immunization doses
can be limited or very expensive in practice, and without an optimal
distribution these resources can go inadvertently to waste.

To investigate the applicability of CI to an immunization/quarantine
scheme in a real large-scale social network we consider a social
contact network built from the mobile phone calls between people in
Mexico. Data has been provided by GranData.

A mobile phone call social network reflects people's interactions in
social lives, and is generally accepted as a proxy of a human contact
network. For example, the mobile phone network from Mexico can help us
to design effective immunization strategies, by identifying the most
relevant social contacts among people.  The disease spreads through
direct contacts of infected people and proximity and mobility data
from mobile phone networks can serve as a proxy of human movements and
possible spreading patterns in human contact networks.

In order to build the network, we put a link
between two people if there is a reciprocal exchange of phone calls
between them in a observation window of three months (i.e. a call in
both directions), and the number of such reciprocal calls is larger
than or equal to three.  This criterion gives us a network of
$N=14,346,653$ nodes, with an average degree $\la k\ra = 3.53$ and a
maximum degree $k_{\mathrm{max}}=419$.  The result of the CI
algorithm, compared to HD and HDA, is shown in
Fig. \ref{fig:mobileMex}d.

The phone call network is the prototype of big-data, where a scalable
(i.e.  almost linear) algorithm is mandatory. Indeed, the size of this
network already rules out many heuristic methods with quadratic (or
larger) running time (CC, EC, BC, and EGP) and also BP.  From the
perspective of performance, CI is better by a very good margin.
Indeed, it fragments the network using about $500,000$ people less
than the best heuristic strategy (HDA) implying a saving of the same
number of vaccines in a hypothetical immunization campaign. Moreover,
when CI gives a zero giant component, HDA gives still $G\sim0.3$,
i.e. a connected network of $\sim 4\times10^6$ people. This result,
together with the result on Twitter, indicates that, although the
theory has been developed for a locally-tree like network, in real
networks with loops the CI-algorithm performs quite well as well.


\newpage

\begin{thebibliography}{99}


\bibitem{richardson1} Domingos P. \& Richardson, M. Mining
  knowledge-sharing sites for viral marketing. {\it Proc. 8th ACM
      SIGKDD Intl. Conf. on Knowledge Discovery and Data Mining},
    p61-70 (2002).

\bibitem{pastor} Pastor-Satorras R. \&, Vespignani, A. Epidemic spreading
  in scale-free networks. {\it Phys. Rev. Lett.} {\bf 86}, 3200-3203
  (2001).

\bibitem{newman-perco} Newman, M. E. J. Spread of epidemic disease on
  networks.  {\it Phys. Rev. E} {\bf 66}, 016128 (2002).

\bibitem{kempe} Kempe, D., Kleinberg, J. \& Tardos, E. Maximizing the
  spread of influence through a social network. {\it Proc. 9th ACM
    SIGKDD Intl. Conf. on Knowledge Discovery and Data Mining},
  p137-143 (2003). doi: 10.1145/956750.956769


\bibitem{newman-book} Newman, M. E. J. {\it Networks: An Introduction}
  (Oxford University Press, Oxford, 2010).


\bibitem{freeman} Freeman, L. C. Centrality in social networks:
  conceptual clarification. {\it Social Networks} {\bf 1}, 215-239
  (1979).

\bibitem{pagerank} Brin, S. \& Page, L. The anatomy of a large-scale
  hypertextual web search engine. {\it Comput. Networks ISDN} {\bf
    30}, 107-117 (1998).

\bibitem{kleinberg} Kleinberg, J. Authoritative sources in a
  hyperlinked environment. {\it Proc. 9th ACM-SIAM Symposium on
    Discrete Algorithms} (1998). Extended version in {\it J. of ACM}
  {\bf 46}, 604-632 (1999).

\bibitem{barab} Albert, R., Jeong, H. \& Barab\'asi, A.-L. Error and
  attack tolerance of complex networks. {\it Nature} {\bf 406}, 378-382
  (2000).


\bibitem{cohen} Cohen, R., Erez, K., ben-Avraham \& D., Havlin, S.
  Breakdown of the Internet under intentional attack. {\it
    Phys. Rev. Lett.} {\bf 86}, 3682-3685 (2001).

\bibitem{chen} Chen, Y., Paul, G., Havlin, S., Liljeros, F. \&
 Stanley, H. E. Finding a better immunization strategy.  {\it
    Phys. Rev. Lett.} {\bf 101}, 058701 (2008).


\bibitem{gallos} Kitsak, M., Gallos, L. K., Havlin, S., Liljeros, F.,
  Muchnik, L., Stanley, H. E. \& Makse, H. A. Identification of
  influential spreaders in complex networks. {\it Nature Phys.} {\bf
    6}, 888-893 (2010).

\bibitem{zecchina1} Altarelli, F., Braunstein, A., Dall'Asta, L. \&
  Zecchina, R. Optimizing spread dynamics on graphs by message
  passing. {\it J. Stat. Mech.} P09011 (2013).


\bibitem{zecchina2} Altarelli, F., Braunstein, A., Dall'Asta, L.,
  Wakeling, J. R. \& Zecchina, R. Containing epidemic outbreaks by
  message-passing techniques.  {\it Phys. Rev. X} {\bf 4}, 021024
  (2014).

\bibitem{hashimoto} Hashimoto, K. Zeta functions of
  finite graphs and representations of p-adic groups. {\it
    Adv. Stud. Pure Math.} {\bf 15}, 211-280 (1989).

\bibitem{Coja} Coja-Oghlan, A., Mossel, E. \& Vilenchik, D. A spectral
  approach to analyzing Belief Propagation for 3-Coloring.  {\it
    Combinatorics, Probability and Computing} {\bf 8}, 881-912 (2009).

\bibitem{LTM} Granovetter, M. Threshold models of collective
  behavior. {\it Am. J. Sociol.}  {\bf83}, 1420-1443 (1978).


\bibitem{watts} Watts, D. J. A simple model of global cascades on
  random networks.  {\it Proc. Natl. Acad. Sci. U.S.A.} {\bf 99},
  5766-5771 (2002).

\bibitem{pei} Pei, S., Muchnik, L., Andrade, J. S. Jr., Zheng, Z. \&
  Makse, H. A.  Searching for superspreaders of information in
  real-world social media. {\it Sci. Rep.} {\bf 4}, 5547 (2014).

\bibitem{pei2} Pei, S. \& Makse, H. A. Spreading dynamics in complex
  networks, {\it J. Stat. Mech.} P12002 (2013).

\bibitem{bollobas} Bollob\'as, B. \& Riordan, O. {\it
  Percolation} (Cambridge Univ. Press, Cambridge, 2006).

\bibitem{bianconi}
 Bianconi, G. \& Dorogovtsev, S. N. Multiple percolation transitions 
in a configuration model of network of networks.
{\it Phys. Rev. E} {\bf 89}, 062814 (2014).

\bibitem{lenka} Karrer, B., Newman, M. E. J. \& Zdeborov\'a, L.
  Percolation on sparse networks. {\it Phys. Rev. Lett.} {\bf 113},
  208702 (2014).



\bibitem{nbt} Angel, O., Friedman,  J. \& Hoory, S. The non-backtracking
  spectrum of the universal cover of a graph. 
{\it Trans. Amer. Math. Soc.} {\bf 367} 4287-4318 (2015).



\bibitem{florant} Krzakala, F., Moore, C., Mossel, E., Neeman, J.,
  Sly, A., Zdeborov\'a, L. \& Zhang, P.  Spectral redemption in
  clustering sparse networks.  {\it Proc. Natl. Acad. Sci. U.S.A.}
  {\bf 110}, 20935-20940 (2013).

\bibitem{newman} Newman, M. E. J. Spectral methods for community
  detection and graph partitioning. {\it Phys. Rev. E} {\bf 88},
  042822 (2013).

\bibitem{radicchi} Radicchi, F. Predicting percolation thresholds in
  networks. {\it Phys. Rev. E} {\bf 91}, 010801(R) (2015).

\bibitem{MP} M\'ezard, M. \& Parisi, G. The cavity method at zero
  temperature. {\it J. Stat. Phys.} {\bf 111}, 1-34 (2003).

\bibitem{EO} Boettcher, S. \& Percus, A. G. Optimization with extremal
  dynamics.  {\it Phys. Rev. Lett.} {\bf 86}, 5211-5214 (2001).


\bibitem{granovetter} Granovetter, M. The strength of weak ties. {\it
  Am. J. Sociol.}  {\bf 78}, 1360-1380 (1973).


\newcounter{firstbib}
\setcounter{firstbib}{\value{NAT@ctr}}
\end{thebibliography}

\begin{thebibliography}{99}
\setcounter{NAT@ctr}{\value{firstbib}}


\bibitem{rich2} Colizza, C., Flammini, A., Serrano, M. A. \& Vespignani, A.
  Detecting rich-club ordering in complex networks.  {\it Nature
    Phys.} {\bf 2}, 110-115 (2006).

\bibitem{wasserman} Wasserman, S. \& Faust, K. {\it Social Network
  Analysis} (Cambridge Univ. Press, Cambridge, 1994).

\bibitem{ec} Straffin, P. D. Linear algebra in geography: eigenvectors
  of networks. {\it Mathematics Magazine} {\bf 53}, 269-276 (1980).

\bibitem{cc} Bavelas, A. Communication patterns in tasks oriented
  groups. {\it J. Acoust. Soc. Am.} {\bf 22}, 271-282 (1950).

\bibitem{bc} Freeman, L. C. A set of measures of centrality based on
  betweenness. {\it Sociometry} {\bf 40}, 35-41 (1977).

\bibitem{montanari} M\'ezard, M. \& Montanari, A.  {\it Information,
  Physics, and Computation} (Oxford University Press, USA, 2009).

\bibitem{centola} Centola, D., Eguiluz, V. \& Macy, M. W. Cascade
  dynamics of complex propagation. {\it Physica A: Statistical Mechanics
  and its Applications} {\bf 374}, 449-456 (2007).

\bibitem{korniss} Singh, P., Sreenivasan, S., Szymanski, B. K. \&
  Korniss, G. Threshold-limited spreading in social networks with
  multiple initiators. {\it Sci. Rep.} {\bf 3}, 2330 (2013).

\bibitem{bootstrap} Baxter, G. J., Dorogovtsev, S. N., Goltsev,
  A. V. \& Mendes, J. F. F. Bootstrap percolation on complex networks.
  {\it Phys. Rev. E} {\bf 82}, 011103 (2010).

\bibitem{configurational} Wormald, N. C. The asymptotic connectivity
  of labelled regular graphs. {\it J. Combinatorial Theory B} {\bf
    31}, 156-167 (1981).

\bibitem{doro2} Dorogovtsev, S. N., Mendes, J. F. F. \& Samukhin, A. N.
      Metric structure of random networks. {\it Nucl. Phys. B}
     {\bf 653}, 307-338 (2003).

\bibitem{eigenvalue} Bhatia, N. P. \& Szeg\"o, G. P. {\it Stability theory of 
dynamical systems} (Springer-Verlag, Berlin Heidelberg, 2002).

\bibitem{sulc} \~Sulc, P. \& Zdeborov\'a, L. Belief propagation for
  graph partitioning. {\it J. Phys. A: Math. Theor.}  {\bf 43}, 285003
  (2010).

\bibitem{parisi2} Parisi, G., Ritort, F. \& Slanina, F. Several
  results on the finite-size corrections in the
  Sherrington-Kirkpatrick spin-glass model. {\it J. Phys. A:
    Math. Gen.}  {\bf 26}, 3775-3790 (1993).


\bibitem{caos} Bray, A. J. \& Moore, M. A. Chaotic nature of the
  spin-glass phase.  {\it Phys. Rev. Lett.} {\bf 58}, 57-60 (1987).

\bibitem{quarantine} Ferguson, N. M., {\it et al.} Strategies for
  containing an emerging influenza pandemic in Southeast Asia.  {\it
    Nature} {\bf 437}, 209-214 (2005).


\bibitem{zeropatient} Auerbach, D. M., Darrow, W. W., Jaffe, H. W. \& Curran,
  J. W. Cluster of cases of the acquired immune deficiency
  syndrome. Patients linked by sexual contact. {\it Am. J. Med.} {\bf
    76}, 487-492.  (1984).

\bibitem{semerjian} Guggiola, A., Semerjian, G. Minimal contagious sets
in random regular graphs. {\it J. Stat. Phys.} {\bf 158}, 300-358. (2015). 

\bibitem{zhuo} Zhou, H.-J. Spin glass approach to the 
feedback vertex set. {\it Eur. Phys. J. B} {\bf 86}, 455-463 (2013). 

\bibitem{zhuo2} Qin, S.-M. \& Zhou, H.-J. Solving the undirected 
feedback vertex set problem by local search. 
{\it Eur. Phys. J. B} {\bf 87}, 273 (2014). 




\end{thebibliography}
\end{document}